\documentclass[traditabstract]{aa}
\usepackage{txfonts}
\usepackage{natbib}
\usepackage{graphicx}
\usepackage{amssymb}
\usepackage{epsfig}
\usepackage{rotating}
\usepackage{lscape}
\usepackage{color, ulem}

\graphicspath{{figures/}}
\newcommand{\ud}{\mathrm{d}}
\newcommand\T{\rule{0pt}{2.6ex}}
\newcommand\B{\rule[-1.2ex]{0pt}{0pt}}

\begin{document}
\def\sizex{16.0 cm}
\def\bigx{10.0 cm}
\def\smallerxsize{7.0 cm}
\def\smallxsize{10.0 cm}
\def\smallysize{12.0 cm}

\title {Galaxy clustering in the CFHTLS-Wide: the changing
  relationship between galaxies and haloes since
  $z\sim1.2$\thanks{Based on observations obtained with
    MegaPrime/MegaCam, a joint project of CFHT and CEA/DAPNIA, at the
    Canada-France-Hawaii Telescope (CFHT) which is operated by the
    National Research Council (NRC) of Canada, the Institut National
    des Sciences de l'Univers of the Centre National de la Recherche
    Scientifique (CNRS) of France, and the University of Hawaii. This
    work is based in part on data products produced at TERAPIX and the
    Canadian Astronomy Data Centre as part of the Canada-France-Hawaii
    Telescope Legacy Survey, a collaborative project of NRC and
    CNRS.}}  \offprints {J.\ Coupon} \date{Received date / Accepted
  date} \titlerunning{Galaxy clustering in the CFHTLS Wide}
\authorrunning{Coupon et al.}  \author{
  J. Coupon\inst{1,2}\fnmsep\thanks{\email{coupon@asiaa.sinica.edu.tw}},
  M. Kilbinger\inst{3,4},
  H. J. McCracken\inst{5}, 
  O. Ilbert\inst{6},
  S. Arnouts\inst{7},
  Y. Mellier\inst{5},\\
  U. Abbas\inst{8},
  S. de la Torre\inst{9},
  Y. Goranova\inst{5}, 
  P. Hudelot\inst{5},
  J.-P. Kneib\inst{6},
  and O. Le F\`evre\inst{6}
}

\institute{
  Astronomical Institute, Graduate School of Science, Tohoku
  University,  Sendai 980-8578, Japan \and
  Institute of Astronomy and Astrophysics, Academia
  Sinica, P.O. Box 23-141, Taipei 10617, Taiwan \and 
  Excellence Cluster Universe, Technische Universit\"at M\"unchen, Boltzmannstr. 2, 85748 Garching,
  Germany \and
  Universit\"ats-Sternwarte M\"unchen, Scheinerstr. 1, 81679
  M\"unchen, Germany \and
  Institut d'Astrophysique de Paris, UMR7095 CNRS, Universit\'e Pierre
  et Marie Curie, 98 bis Boulevard Arago, 75014 Paris, France \and 
  Laboratoire d'Astrophysique de Marseille (UMR 6110), CNRS-Universit\'e
  de Provence, 38, rue Fr\'ed\'eric Joliot-Curie, 13388 Marseille Cedex
  13, France \and 
  Canada-France-Hawaii Telescope, 65--1238 Mamalahoa Highway, Kamuela,
  HI 9674 \and
  INAF-Osservatorio Astronomico di Torino, 10025 Pino Torinese, Italy \and
  SUPA, Institute for Astronomy, University of Edinburgh, Royal
  Observatory, Blackford Hill, Edinburgh EH9 3HJ, UK
}

\abstract%
{ It has become increasingly apparent that studying how dark matter
  haloes are populated by galaxies can provide new insights into
  galaxy formation and evolution. In this paper, we present a detailed
  investigation of the changing relationship between galaxies and the
  dark matter haloes they inhabit from $z \sim 1.2$ to the present
  day. We do this by comparing precise galaxy clustering measurements
  over 133~$\deg^2$ of the ``Wide'' component of the
  Canada-France-Hawaii Telescope Legacy Survey (CFHTLS) with
  predictions of an analytic halo occupation distribution (HOD) model
  where the number of galaxies in each halo depends only on the halo
  mass. Starting from a parent catalogue of $\sim3\times10^6$ galaxies
  at $i'_{\rm AB}<22.5$ we use accurate photometric redshifts
  calibrated using $\sim10^4$ spectroscopic redshifts to create a
  series of type-selected volume-limited samples covering $0.2 < z <
  1.2$.  Our principal result, based on clustering measurements in
  these samples, is a robust determination of the luminosity-to-halo
  mass ratio and its dependence on redshift and galaxy type. For the
  full sample, this reaches a peak at low redshifts of $M^{\rm
    peak}_{\rm h}=4.5\times10^{11} h^{-1} M_\odot$ and moves towards
  higher halo masses at higher redshifts. For redder galaxies the peak
  is at higher halo masses and does not evolve significantly over the
  entire redshift range of our survey. We also consider the evolution
  of bias, average halo mass and the fraction of satellites as a
  function of redshift and luminosity.  Our observed growth of a
  factor of $\sim 2$ in satellite fraction between $z\sim1$ and
  $z\sim0$ is testament to the limited role that galaxy merging plays
  in galaxy evolution for $\sim10^{12} h^{-1} M_\odot$ mass haloes at
  $z<1$.  Qualitatively, our observations are consistent with a
  picture in which red galaxies in massive haloes have already
  accumulated most of their stellar mass by $z\sim1$ and subsequently
  undergo little evolution until the present day.  The observed
  movement of the peak location for the full galaxy population is
  consistent with the bulk of star-formation activity migrating from
  higher mass haloes at high redshifts to lower mass haloes at lower
  redshifts.}


\maketitle

\section{Introduction}
\label{sec:intro}

In our current paradigm of galaxy formation, haloes of dark matter
grow from tiny imperfections in the early Universe. Against the
background (at the present day) of an accelerating Universe,
structures grow ``hierarchically'': small haloes form first and merge
to build up larger ones, resulting in a complex filamentary network
where the most massive haloes lie at the nodes of the cosmic web
\citep{2006Natur.440.1137S}.  Galaxies form as baryons fall into the
centres of dark matter haloes and the inflow of cold gas provides the
fuel for star formation
\citep{1978MNRAS.183..341W,1980MNRAS.193..189F}. While this scenario
broadly reproduces the observed galaxy distribution over a large range
of scales and cosmic time, several open questions remain concerning
which processes regulate star formation inside haloes and how this
leads to the Hubble sequence observed at the present day.

These physical processes which drive galaxy formation and regulate
star formation are expected to correlate closely to the mass of dark
matter haloes which host galaxies. Therefore, a profitable avenue to
pursue to better understand the physics of galaxy formation is to
determine how the relationship between the stellar mass $M_{\rm star}$
and the dark matter halo mass $M_{\rm h}$ changes over the lifetime of
the Universe. In the local Universe, abundance-matching studies have
shown that at $z<0.1$ the stellar-to-halo mass ratio (SHMR, $M_{\rm
  star}/M_{\rm h}$) reaches a maximum of a few percent at $M_{\rm h}
\sim 10^{12}M_\odot$, much smaller than the universal baryon fraction
\citep{2010MNRAS.404.1111G}. This indicates that the conversion
efficiency of baryons to luminous galaxies is highest at these halo
masses. In less massive haloes, supernovae winds are responsible for
gas reheating, which reduces star formation and flattens the galaxy
luminosity function at the faint end \citep{2003ApJ...599...38B}.  In
more massive haloes, feedback from active galactic nuclei
\citep{2006ApJ...652..864H,2008MNRAS.391..481S} can ``quench'' star
formation, leading to formation of the ``red sequence'' of passively
evolving galaxies \citep{2006MNRAS.370..645B,2006MNRAS.365...11C}. The
relative importance of these different processes as well as their
dependence on environment, redshift and galaxy type is still very much
an open question.

Ideally we would like to make a detailed comparison between the
luminous content of dark matter haloes and the halo masses themselves
as a function of redshift.  Several methodologies exist to estimate
halo masses: for example, using satellite kinematics
\citep{2011MNRAS.410..210M}, X-ray temperatures
\citep{2006PhR...427....1P,2009Natur.460..213C}, or gravitational
lensing \citep{2006MNRAS.371L..60H,2006MNRAS.368..715M,
  2010ApJ...724..511A,2011arXiv1104.0928L}. However, the range in
redshifts over which these techniques can be applied is limited, and
amassing large samples is not always easy.

At the price of making some assumptions concerning the profiles of
dark matter haloes and their evolution in number density, another way
of inferring halo masses is from the observed galaxy clustering and
mass functions.  Usually, one makes the simplifying assumption that
the number of galaxies in a given dark matter halo only depends on the
halo mass, an assumption which has been verified with $N$-body
simulations
\citep{2003ApJ...593....1B,2005ApJ...629..625B,2002ApJ...575..587B,
  2004ApJ...609...35K}. Furthermore, the galaxy population of a halo,
described by the ``halo occupation distribution'' (HOD), is assumed to
be independent of environment and assembly history of the
halo. Although some authors have shown that this ``halo assembly
bias'' can have an observable effect on galaxy clustering
measurements, the magnitude of this effect is, for the moment, much
smaller than the size of systematic errors
\citep{2007MNRAS.374.1303C}. This will be an important effect to test
in future surveys.

In this paper we will use this model to investigate the changing
relationship of dark matter and luminous matter from $z\sim1.2$ to the
present day. Until now, the majority of papers which have used HOD
modelling to interpret galaxy clustering have analysed either large,
low redshift surveys such as the SDSS \citep[see, for e.g.,
][]{2010arXiv1005.2413Z,2008MNRAS.385.1257B,2010arXiv1005.2413Z}, or
smaller ($\sim1$~deg$^2$) deep fields such as the COMBO-17 and COSMOS
surveys~\citep{2006A&A...457..145P,2009MNRAS.398..807S,2011arXiv1104.0928L}.
For the moment, spectroscopic surveys at higher redshifts are also
limited to small fields of
view~\citep{2007ApJ...667..760Z,2010MNRAS.406.1306A}.
\cite{2008ApJ...682..937B} has interpreted clustering measurements in
the framework of the halo model of slightly larger redshift baseline,
although their survey only covered $\sim7$ deg$^2$ and used five-band
photometric redshifts. None of these surveys had the required
combination of depth and areal coverage to make reliable measurements
from low redshift to $z\sim1$ of large samples of galaxies selected by
type and intrinsic luminosity. In contrast, this is an ideal task for
the Canada-France-Hawaii Legacy Survey (CFHTLS) with its unique
combination of depth, area, and image quality.

The five-band $u^\ast,g',r',i',z'$ CFHTLS photometry allows us to
measure photometric redshifts out to $z\sim1.2$ with $\sim 3 \%$
accuracy and small systematic errors \citep{2009A&A...500..981C}. The
large sky coverage, $\sim133$~deg$^2$ after masking, enables us to
probe a large range of halo masses. In addition, from the four
independent fields of the CFHTLS we can obtain a reliable estimate of
cosmic variance from the data itself. In this paper we construct a
series of volume-limited samples selected by type, luminosity, and
redshift, and use the halo model combined with the HOD formalism to
infer how each galaxy sample populates their hosting dark matter
haloes, and how this changes with look-back time.

The paper is organized as follows. In
Section~\ref{sec:observ-reduct-catal}, we present our data set and
catalogue production, including data reduction and photometric
redshift measurement. In Section~\ref{sec:clustering} we outline the
techniques we use to estimate galaxy clustering, and in
Section~\ref{sec:model} we present our model. Results are
described in Section~\ref{sec:results} and discussed in
Section~\ref{sec:discussion}. Conclusions are presented in
Section~\ref{sec:conclusion}.

Throughout the paper we use a flat $\Lambda$CDM cosmology ($\Omega_m~=~0.25$,
$\Omega_\Lambda~=~0.75$, $H_0 = 70$~km~s$^{-1}$ ~Mpc$^{-1}$ and
$\sigma_8=0.8$) with $h~=~H_{\rm0}/100$~km~s$^{-1}$~Mpc$^{-1}$. All
magnitudes are given in the AB system.

\section{Observations, reductions and catalogue production}
\label{sec:observ-reduct-catal}

\subsection{The Canada-France-Hawaii Legacy Survey}

\begin{figure*}
  \begin{center}
    \begin{tabular}{c@{}c@{}}
      \includegraphics[width=0.4\textwidth]{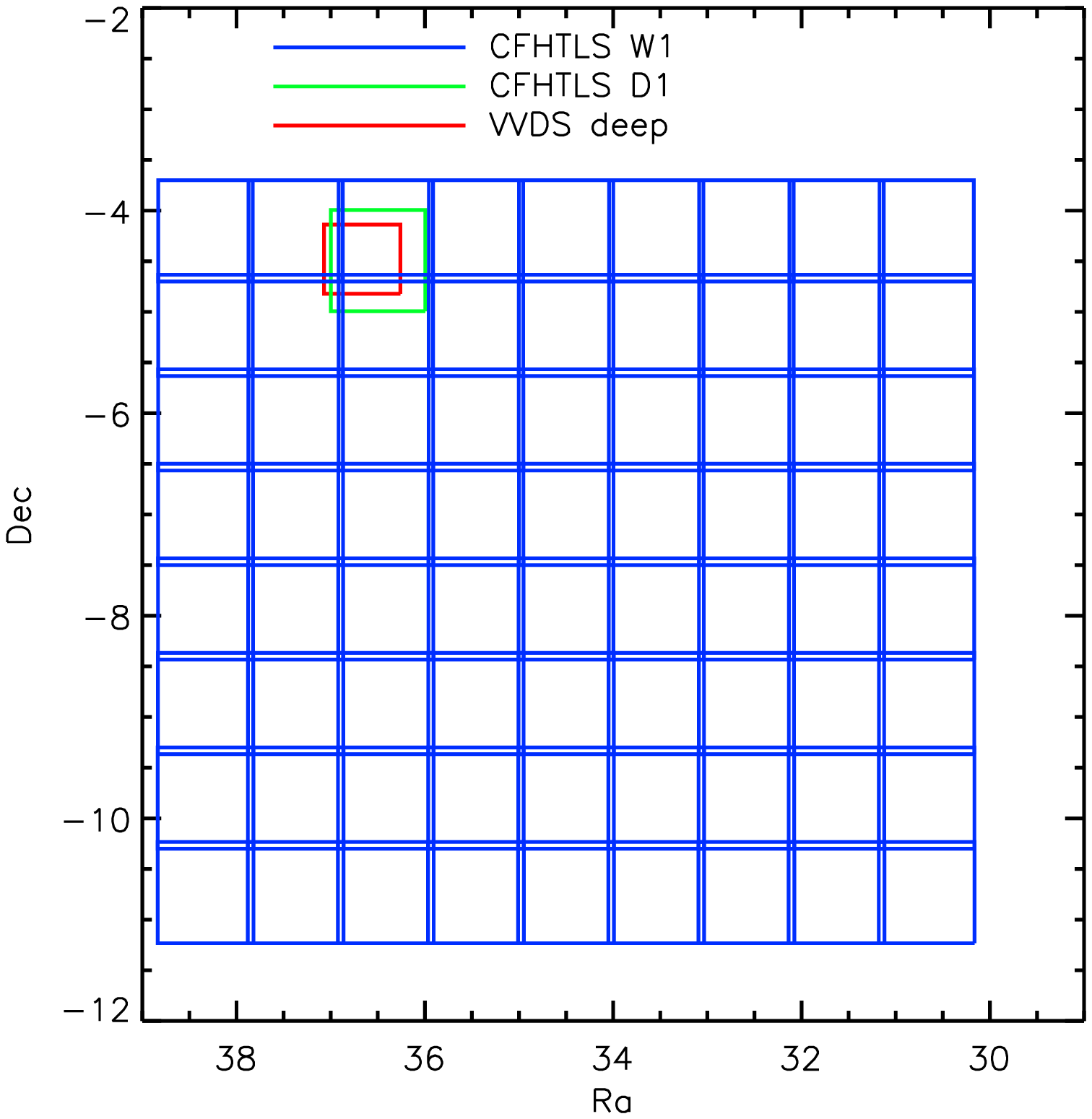}&
      \includegraphics[width=0.4\textwidth]{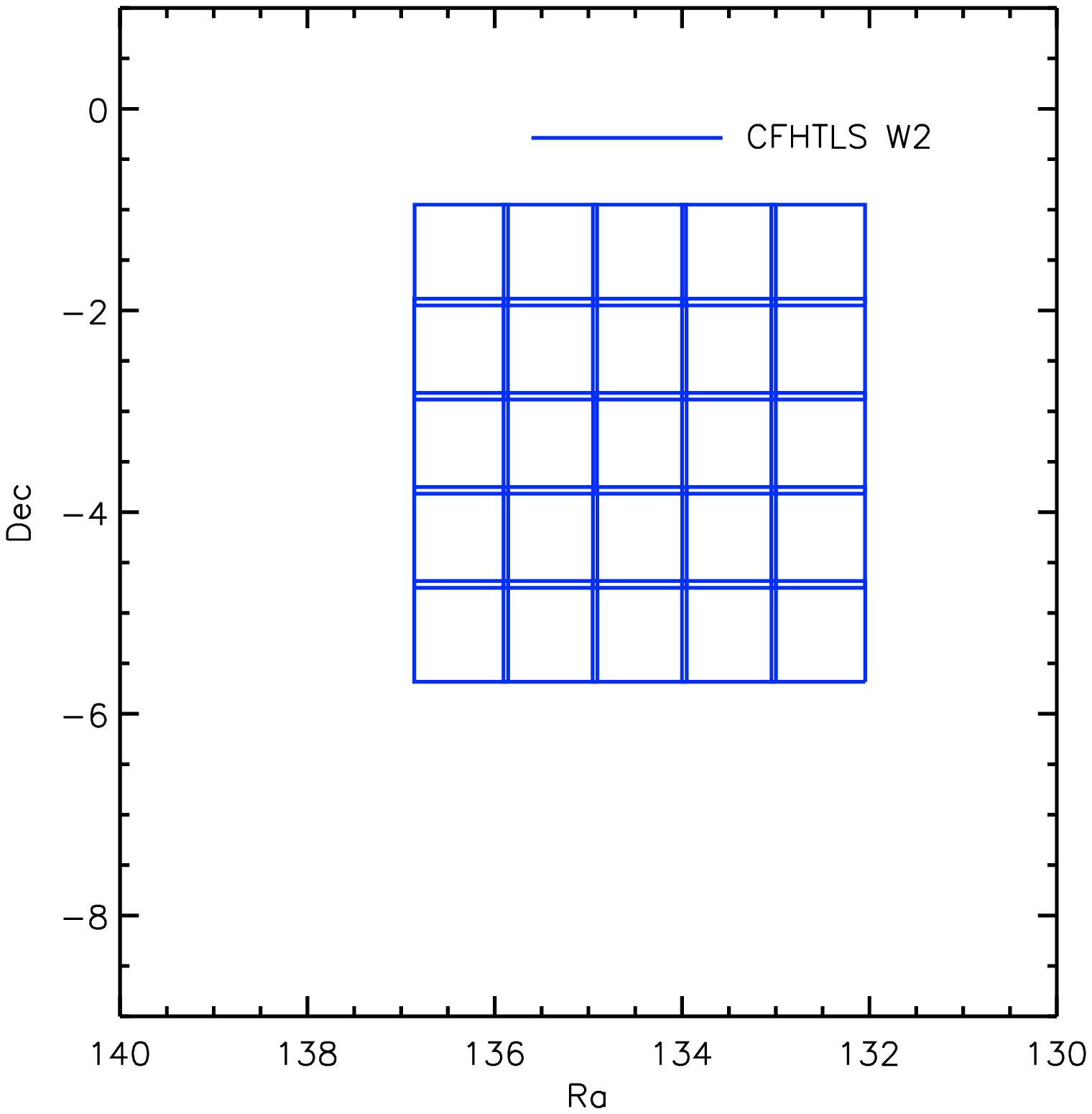}\\
      \includegraphics[width=0.4\textwidth]{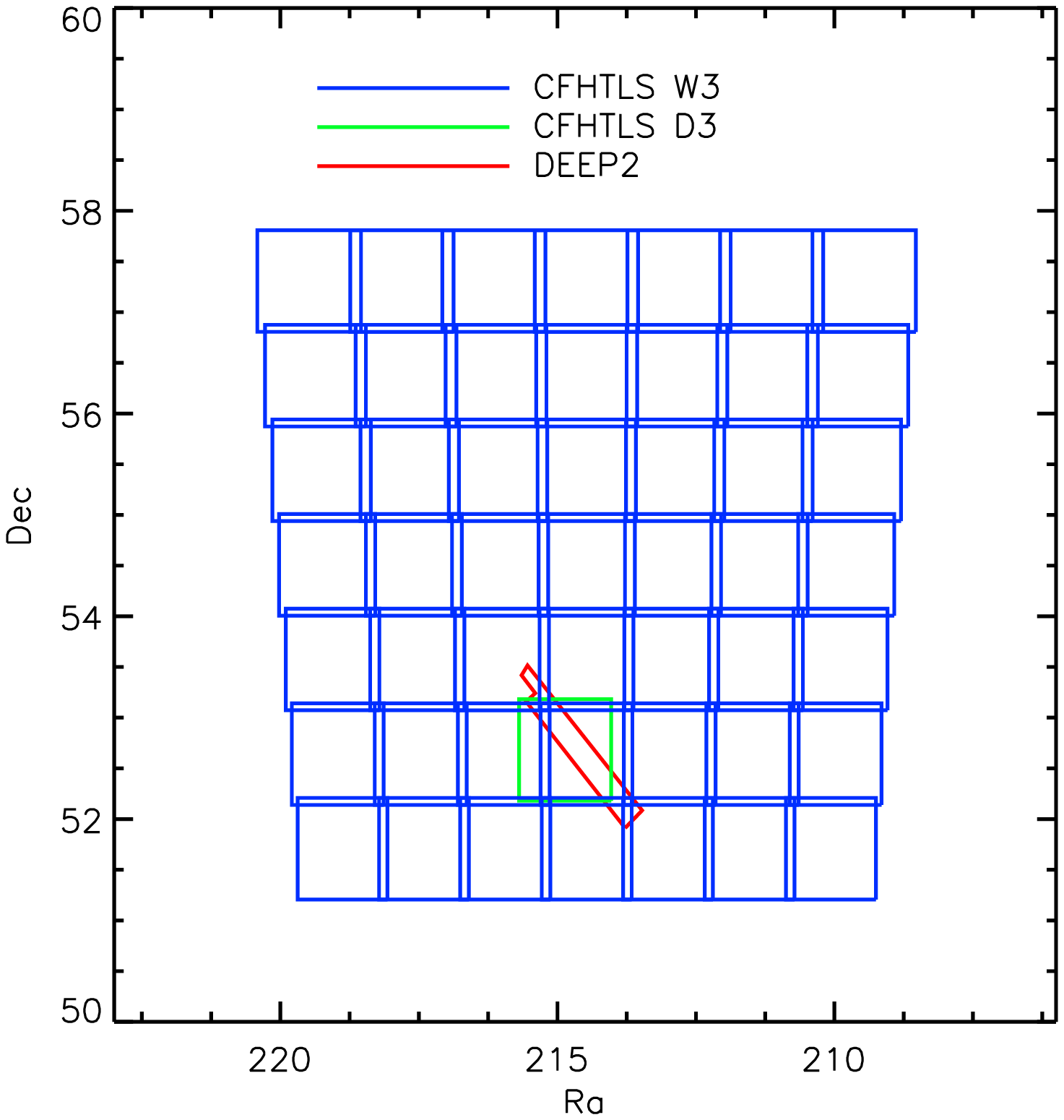}&
      \includegraphics[width=0.4\textwidth]{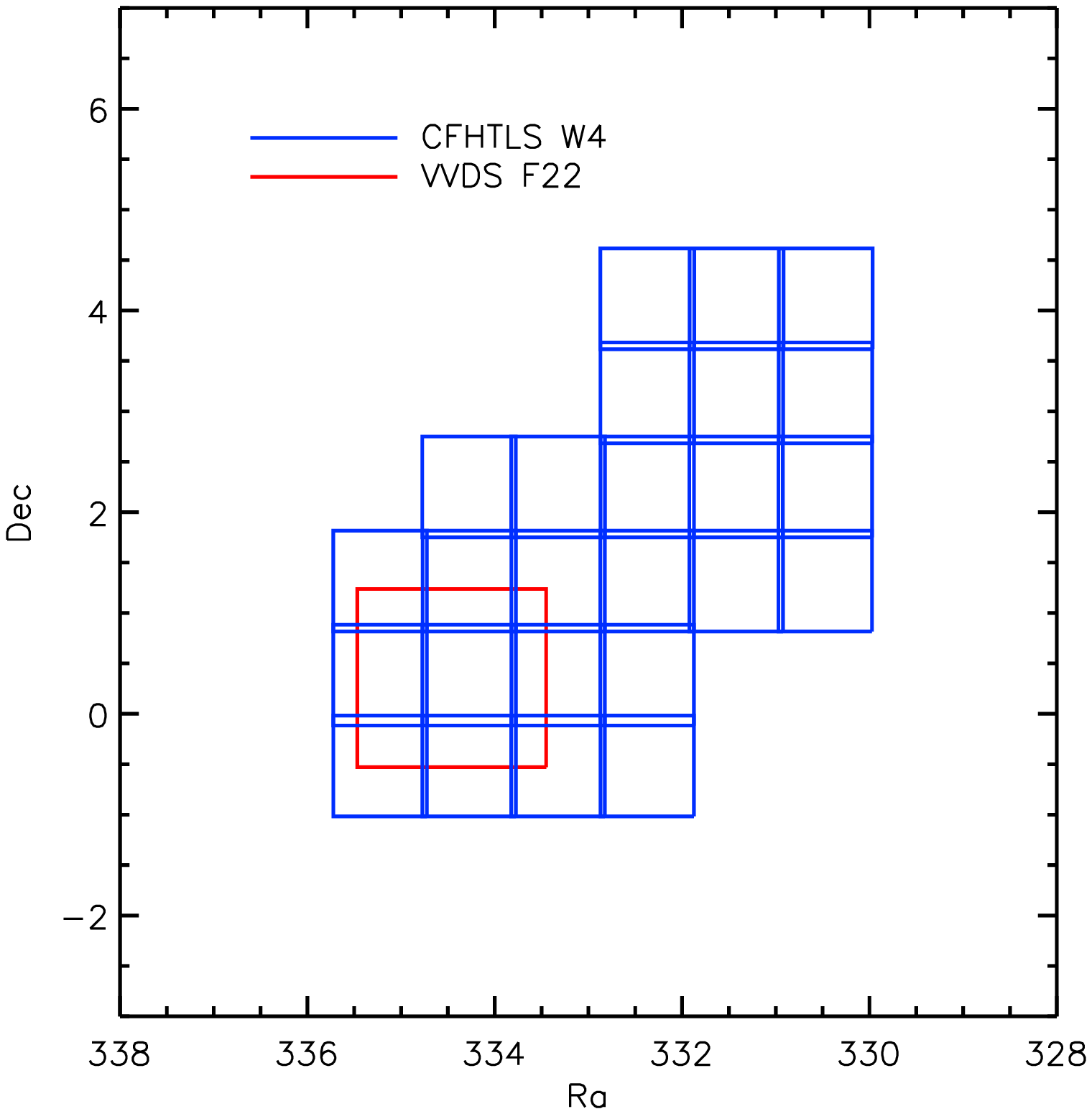}
    \end{tabular}
  \end{center}
  \caption{Sky coverage of W1,W2,W3 and W4 fields. Each blue square
    represents one MegaCam pointing. Spectroscopic data (VVDS F02,
    F22, and DEEP2) are in red. CFHTLS Deep fields overlapping with
    the wide survey (D1 and D3) are shown in green.}
  \label{CFHTLS_maps}
\end{figure*}

We use the ``T0006'' release of the CFHTLS
Wide\footnote{http://terapix.iap.fr/cplt/T0006-doc.pdf}
\citep{2009GaranovaT06}. The CFHTLS Wide has the mean limiting AB
magnitudes (measured as the 50\% completeness for point sources) $\sim
25.3,25.5,24.8,24.48,23.60$ in $u^\ast,g',r',i',z'$, respectively.  An
important feature in T0006 compared to previous releases is that for
each MegaCam pointing (``tile''), the TERAPIX group provides a
magnitude offset in each filter to account for tile-to-tile variations
in magnitude zero points. These offsets are calculated using a stellar
locus regression technique \citep{2009AJ....138..110H}. In each tile,
objects are detected on a $gri$-$\chi^2$ image
\citep{1999AJ....117...68S}. Galaxies are selected using
\texttt{SEXtractor} ``\texttt{mag\_auto}'' magnitudes
\citep{1996A&AS..117..393B}, and colours are measured in $3\arcsec$
apertures which are all well matched to the median size of galaxies at
our $i'<22.5$ magnitude limit. The sky coverage of the four CFHTLS
Wide fields is shown in Figure~\ref{CFHTLS_maps}.

The MegaCam $i$-band filter broke in 2006, and subsequent observations
were made with a new $i$-band filter denoted by ``$y$''. Data from
these two filters were treated separately with separate filter
curves. Using a small data set observed with both the new $y$- and the
old $i$-band, we detected no significant difference in photometric
redshift accuracy. For the rest of the paper, we use $i'$ to represent
both filters.

\subsection{Photometric redshift estimation}
\label{sec:photoz}

Our photometric redshifts were measured using LePhare
\citep{2002MNRAS.329..355A} following the procedures outlined in
\cite{2006A&A...457..841I} and \cite{2009A&A...500..981C}.  Our
primary template set is composed of the four
\cite{1980ApJS...43..393C} (CWW) observed spectra (Ell, Sbc, Scd, Irr)
complemented with two observed starburst spectral energy distributions
(SED) from \cite{1996ApJ...467...38K}.  This primary set of templates
is optimized using the VVDS Deep spectroscopic sample. We performed an
automatic calibration of the zero-points using spectroscopic redshifts
in the W1, W3 and W4 fields. The calibration is obtained by comparing
the observed and modelled fluxes, and done iteratively until the
zero-points converge.  For spectral types later than Sbc, we
introduced a reddening $E(B-V ) = 0$ to $0.35$ using the
\cite{2000ApJ...533..682C} extinction law.

Without near-infrared photometry, Lyman- and Balmer- breaks create
degeneracies between high and low photometric redshift
solutions. Following a Bayesian approach similar to
\cite{2000ApJ...536..571B} and \cite{2006A&A...457..841I}, we adopt
a ``prior'' based on the observed spectroscopic redshift
distribution to reduce the number of catastrophic failures. No
redshift solution is allowed which will produce a galaxy brighter
than $M_g = -24$.

Our photometric redshifts were calibrated with spectroscopic
samples. The VVDS Deep survey \citep{2005A&A...439..845L} is available
in the W1 field. The VVDS Deep is a pure magnitude-limited sample at
$I_{\rm AB} < 24$ and we used $5\,926$ secure redshifts (flags 4 and 5)
for photometric calibration, and for building the redshift
distribution prior.  From the VVDS F22 \cite[``VVDS WIDE'',
][]{2008A&A...486..683G} in W4 we used $4\,514$ galaxies limited to
$I_{\rm AB} < 22.5$. This sample has a success rate of 92\%.  Finally,
in the W3 field, $1\,267$ redshifts from the DEEP2 survey
\citep{2003SPIE.4834..161D} selected at $R_{\rm AB} < 24.1$ were used
for photometric calibration. No spectroscopic redshifts were available
in W2.

To estimate photometric redshifts, we adopted a slightly different
method than for previous studies to measure redshifts from the
probability distribution function (PDF).  Rather than directly using
the redshift that minimizes the $\chi^2$ distribution function, we
chose the median of the PDF.  It has been found that the maximum
likelihood estimate can lead to a systematic concentration of
photometric redshifts around discrete values (``redshift focusing''),
although we checked that the effect is very small for redshift bins of
size $\Delta z > 0.1$.

The redshift accuracy (``$\sigma$") of our samples with spectroscopic
redshift $z_{\rm s}$ and photometric redshift $z_{\rm p}$ is defined
as $\sigma_{\Delta_z}/(1+z_{\rm s})$, where $\Delta z = |z_{\rm p} -
z_{\rm s}|$ represents the difference between spectroscopic and
photometric redshift.  We use the normalised median absolute deviation
as accuracy estimator \citep[NMAD, ][]{1983ured.book.....H}, expressed
as $\sigma_{\Delta_z} = 1.48\times {\rm median}(|z_{\rm p} - z_{\rm
  s}|)$. ``Catastrophic'' redshifts are defined as objects with
$|z_{\rm p} - z_{\rm s}|/(1 + z_{\rm s}) > 0.15$. The percentage of
catastrophic redshifts is denoted by $\eta$. The dispersion and the
fraction of catastrophic redshifts increase from bright to faint
samples. The accuracy is $\sigma_{\Delta_z}/(1+z_{\rm s})=0.034$ (W1)
to $0.037$ (W4) at $i'< 21.5$ and reaches $0.039$ (W1) to $0.051$ (W3)
at $21.5 < i' < 22.5$. The failure rate increases from $\eta \sim
1.4\%$ at $i' < 21.5$ to $\eta \sim 5.5\%$ at $21.5 < i'<
22.5$. Galaxies with redder colours (El, Sbc) have more accurate
photometric redshift estimates ($\sigma_{\Delta_z}/(1+z_{\rm
  s})=0.031$, $\eta = 1.64\%$) than bluer ones (Scd, Irr and
starbursts, $\sigma_{\Delta_z}/(1+z_{\rm s})=0.038$, $\eta = 3.69\%$).

\begin{figure}
  \begin{center}
    \includegraphics[width=0.4\textwidth]{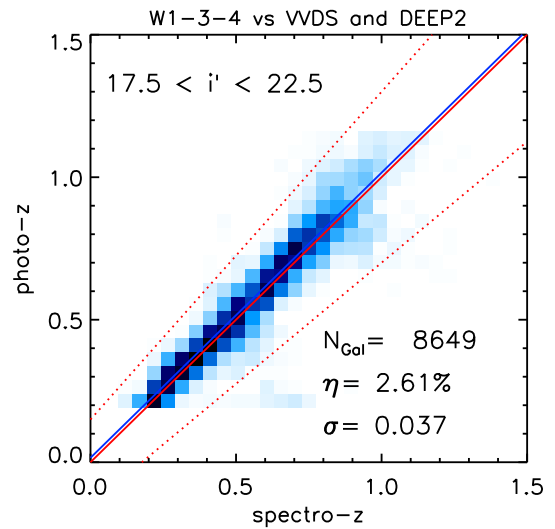}
  \end{center}
  \caption{Photometric redshift accuracy in the $0.2 < {\rm photo-}z <
    1.2$ interval. The figure shows the number density of photometric
    redshifts in W1, W3 and W4, versus spectroscopic redshifts from
    VVDS and DEEP2 surveys. The blue line is a linear fit to photo-z
    versus spectro-z. }
  \label{fig:zp_zs}
\end{figure}

We constructed a merged catalogue with unique objects from individual
tile catalogues provided by TERAPIX. We identified duplicate sources
positionally and chose the object closest to the centre of its
original tile. Although photometric redshifts were estimated to $i'<
24$, we selected objects brighter than $i' = 22.5$ to maintain a low
outlier rate \citep[see Table~3 of ][]{2009A&A...500..981C} and to
ensure catalogue completeness.  This sample covers a total effective
area of 133~deg$^2$.  We show the photometric redshift accuracy in the
secure redshift range covered in this study, $0.2 < z < 1.2$, compared
to spectroscopic redshifts in Fig.~\ref{fig:zp_zs}.

Stars appear as point-like sources on the image and can easily be
identified using their half-light radius at bright magnitudes;
however, at fainter magnitudes this estimator can be confused by
unresolved galaxies. We refined our star-galaxy separation by
combining this profile-magnitude criteria with a colour-based
criteria. We computed the half-light radius limit for each tile from
the median and dispersion of the brightest objects and classified as
stars all objects brighter than $i' = 20$ with flux radii limits
smaller than this amount. In the magnitude range $20 < i'< 22.5$, we
classified stars as objects smaller than the half-light radius {\it
  and} with $\chi^2({\rm star}) < \chi^2({\rm galaxy})$, where
$\chi^2({\rm object})$ is estimated from the stellar and galaxy
template libraries.

We eliminated objects observed with less than three filters and with
$\chi^2({\rm galaxy}) >100$. We suspect that these objects are false
detections or rare objects with wrong photometric redshift
estimates. Most of these objects reside in specific areas, such as
around bright stars and field borders where the masking procedure has
failed.

Using the photometric redshift, the associated best-fitting template
and the observed apparent magnitude in a given band, we can directly
measure the $k-$correction and the absolute magnitude in any
rest-frame band. Since at high redshifts the $k-$correction depends
strongly on the galaxy spectral energy distribution, it is the main
source of systematic error in determining absolute magnitudes. To
minimise $k-$correction uncertainties, we derive the rest-frame
luminosity at $\lambda$ using the object's apparent magnitude closer
to $\lambda \times (1 + z)$. We use either the $u*$, $g'$, $r'$, $i'$,
or $z'$ observed apparent magnitudes according to the redshift of the
galaxy \citep[the procedure is described in Appendix A of
][]{2005A&A...439..863I}.  For this reason the bluest absolute
magnitude estimate makes full advantage of the complete observed
magnitude set. However, as the $u-$band flux has larger photometric
errors, we decided to use $M_g$-band magnitudes.

\section{Clustering measurements}
\label{sec:clustering}

\subsection{Sample selection and galaxy number density estimation}
\label{sec:samples}

\begin{figure}
  \begin{center}
    \begin{tabular}{c@{}}
      \includegraphics[width=0.49\textwidth]{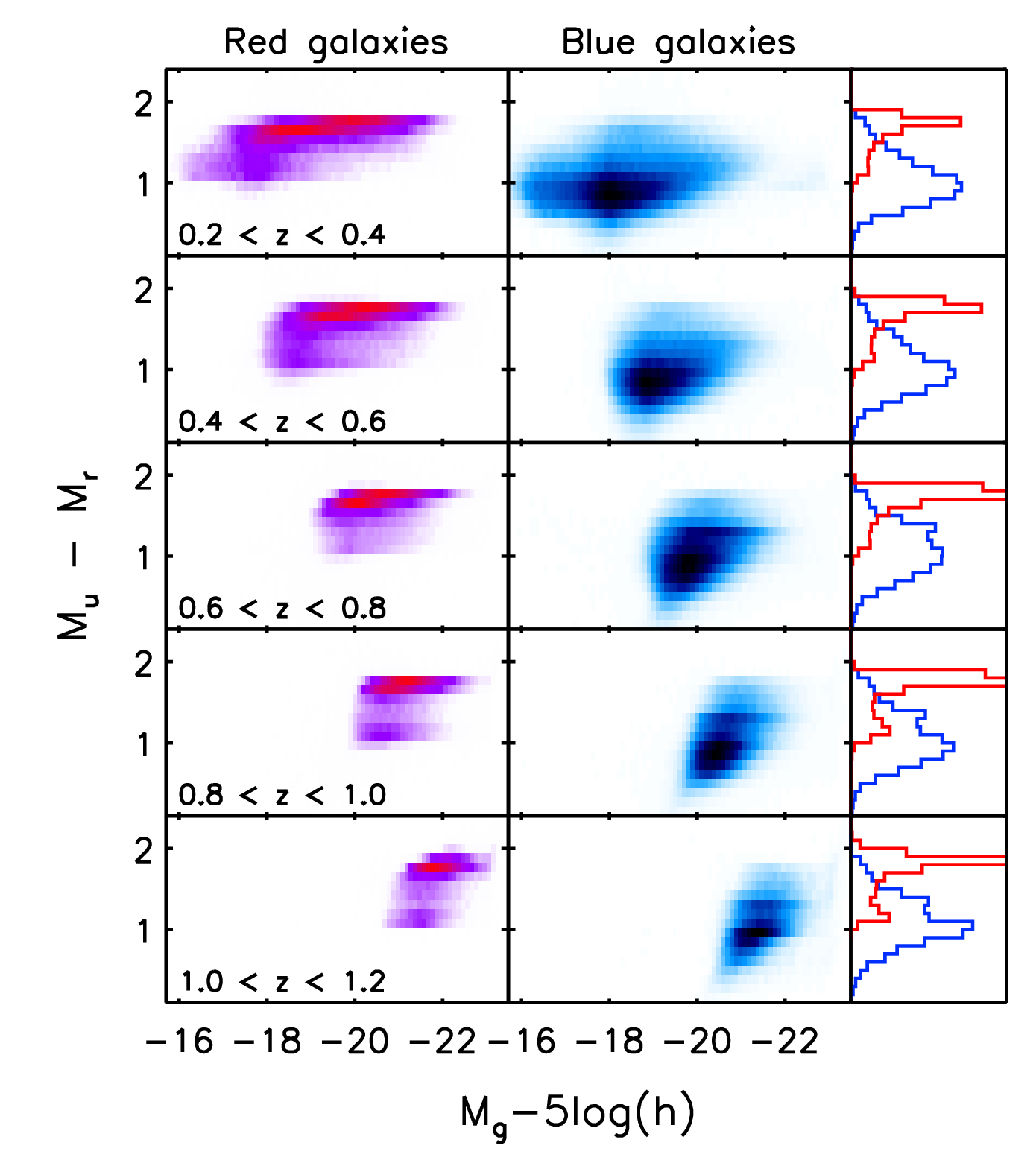}\\
      \includegraphics[width=0.49\textwidth]{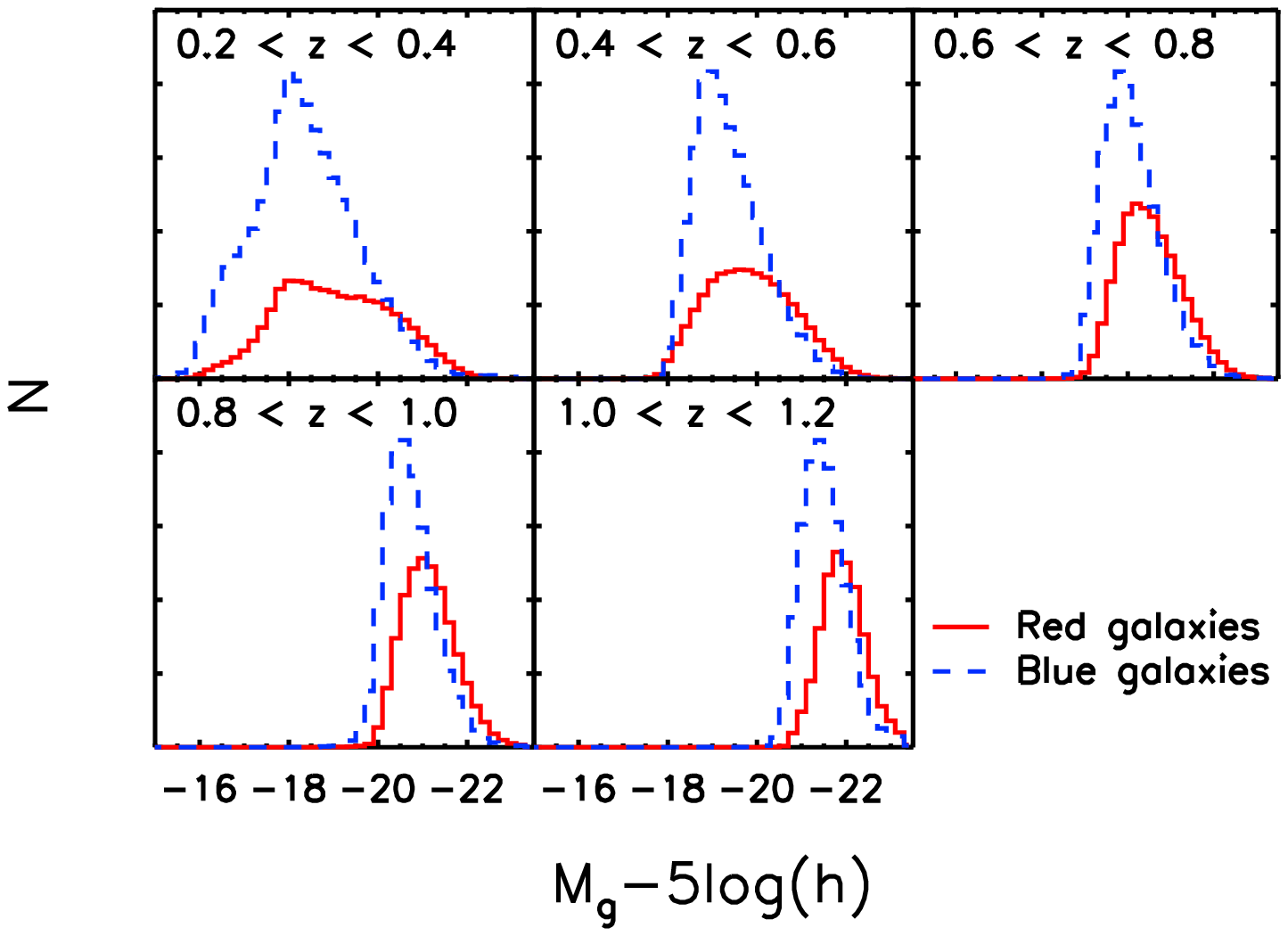}
    \end{tabular}
  \end{center}
  \caption{Top: type selection based on best fitting templates for the ``red''
    sample (left) and the ``blue'' sample (right) in the CFHTLS Wide. We
    show the colour distribution ($M_{u} - M_{r}$) as
    function of absolute magnitude ($M_{g}$) 
    and redshift (top to bottom) and number counts for ``red'' and ``blue''
    objects on the right panels. Bottom: number of red and blue
    galaxies as function of magnitude and redshift.}
  \label{fig:colors}
\end{figure}

Objects are selected with $i' < 22.5$ and divided into five redshift
bins: $0.2 < z < 0.4$, $0.4 < z < 0.6$, $0.6 < z < 0.8$, $0.8 < z <
1.0$, and $1.0 < z < 1.2$. A large bin width ($\Delta_z = 0.2$)
ensures a low bin-to-bin contamination from random errors ($\sigma_z <
0.1$). We investigate effects from systematic errors in
Sect.~\ref{sec:photozerr}.  The complete sample contains $2\,924\,730$
galaxies. To separate galaxies into ``red sequence'' and ``blue
cloud'' types, we used a criterion based on best-fitting galaxy
templates. We define ``red'' galaxies as objects with best-fitting CWW
templates (see Sect.~\ref{sec:photoz}) estimated as El and Sbc
(early-type), and ``blue'' galaxies estimated as Sbc, Scd, Im, SB1 and
SB2 (late-type).  Selecting galaxies by best-fitting template is a
more robust way of selecting galaxies by type, as compared to a simple
colour selection, and it is less sensitive to the effects of dust
extinction. Our resulting colour selection in the absolute
magnitude-redshift plane is shown in Fig.~\ref{fig:colors}.  We note
that a small amount of red-classified galaxies at high redshift $z >
0.8$ have slightly bluer colours. Their distribution peaks near the
blue galaxy distribution, suggesting that these objects could be blue
galaxies erroneously identified as red galaxies due to photometric
redshift errors.  We also note that a simple colour cut would not
exactly reproduce our selection. However, in the interests of
simplicity and clarity we keep the ``red''/''blue'' labels for the
rest of the paper.

\begin{figure*}
  \begin{center}
    \begin{tabular}{c@{}c@{}c@{}}
      \includegraphics[width=0.33\textwidth]{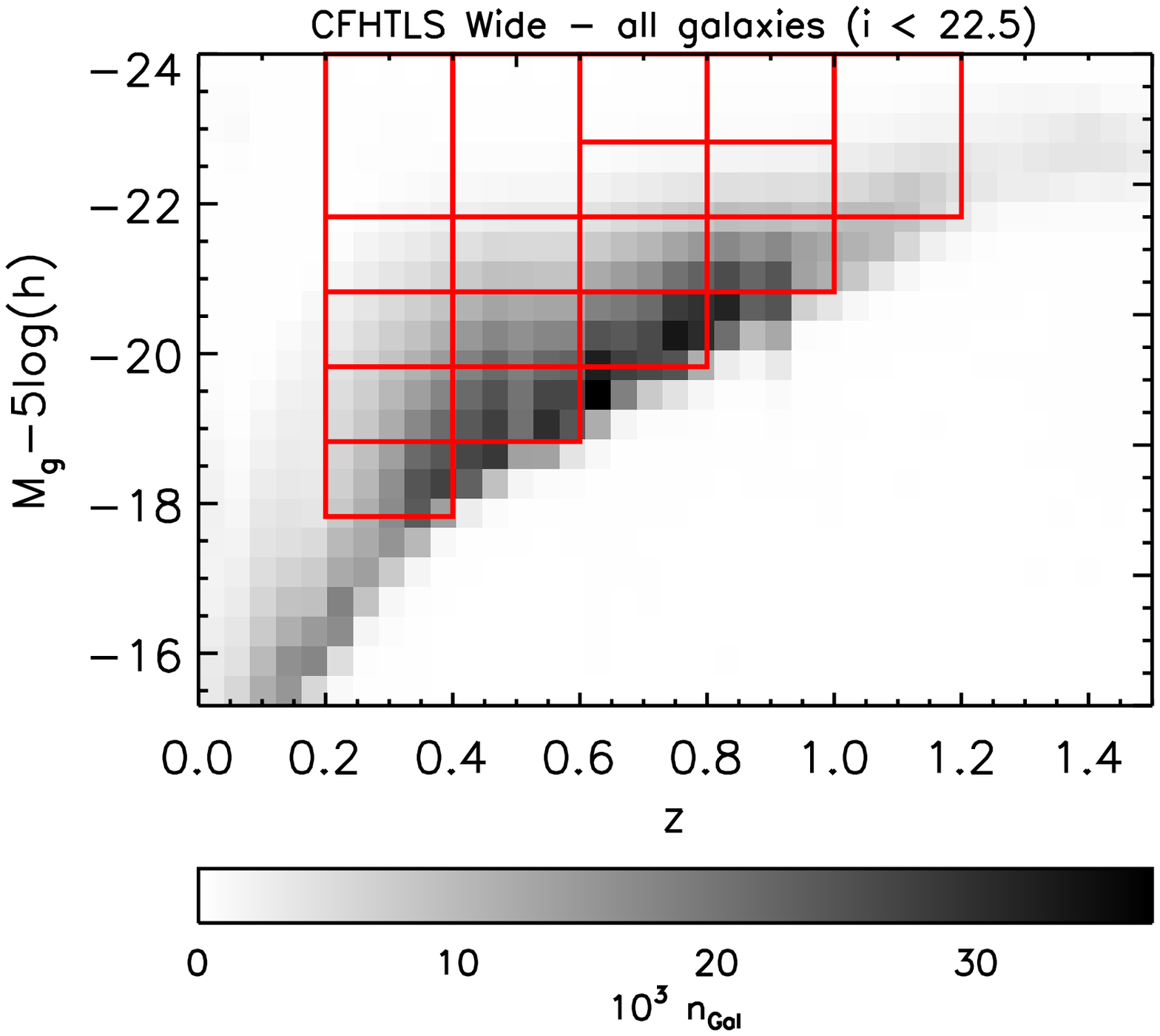} &
      \includegraphics[width=0.33\textwidth]{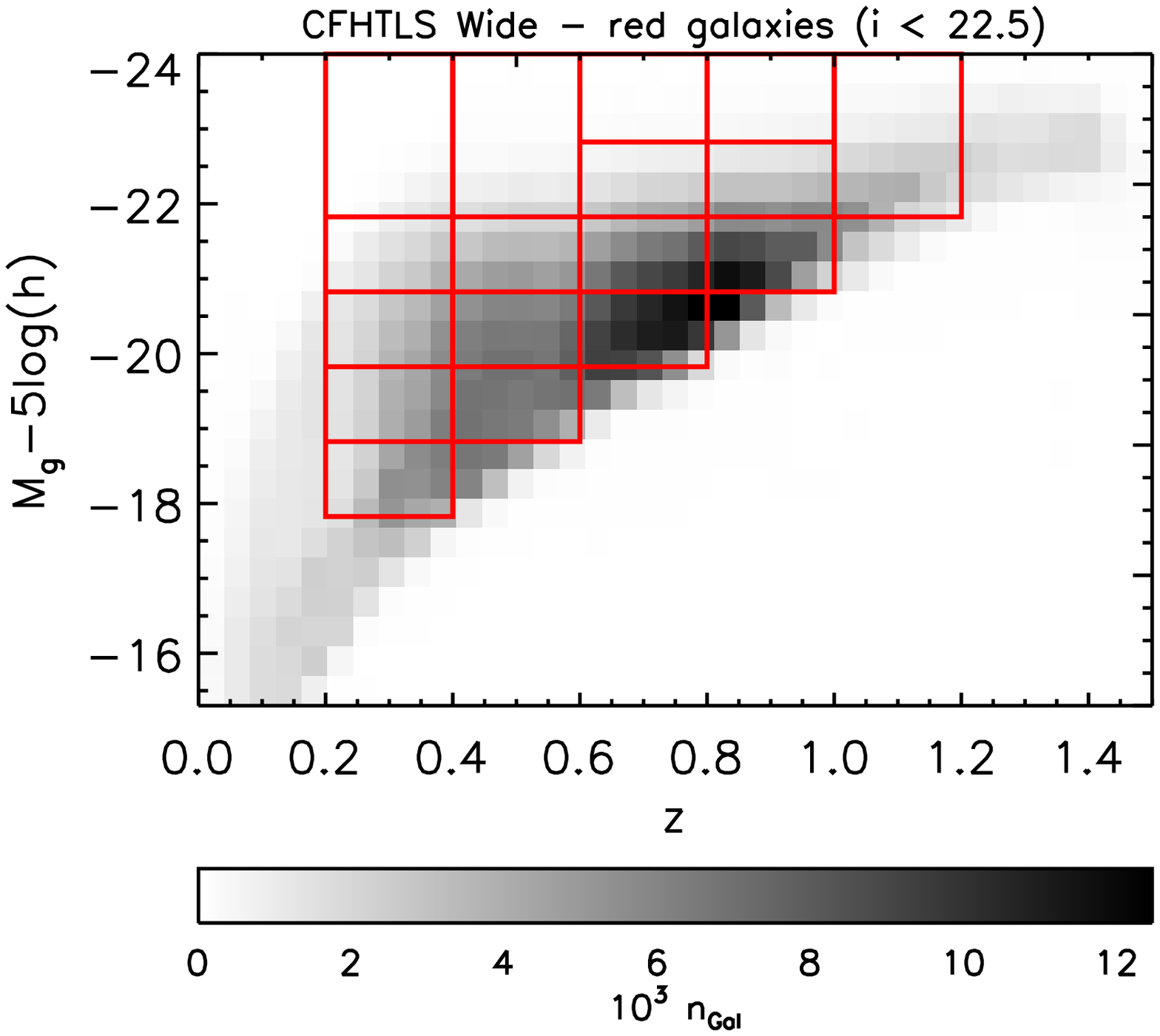}&
      \includegraphics[width=0.33\textwidth]{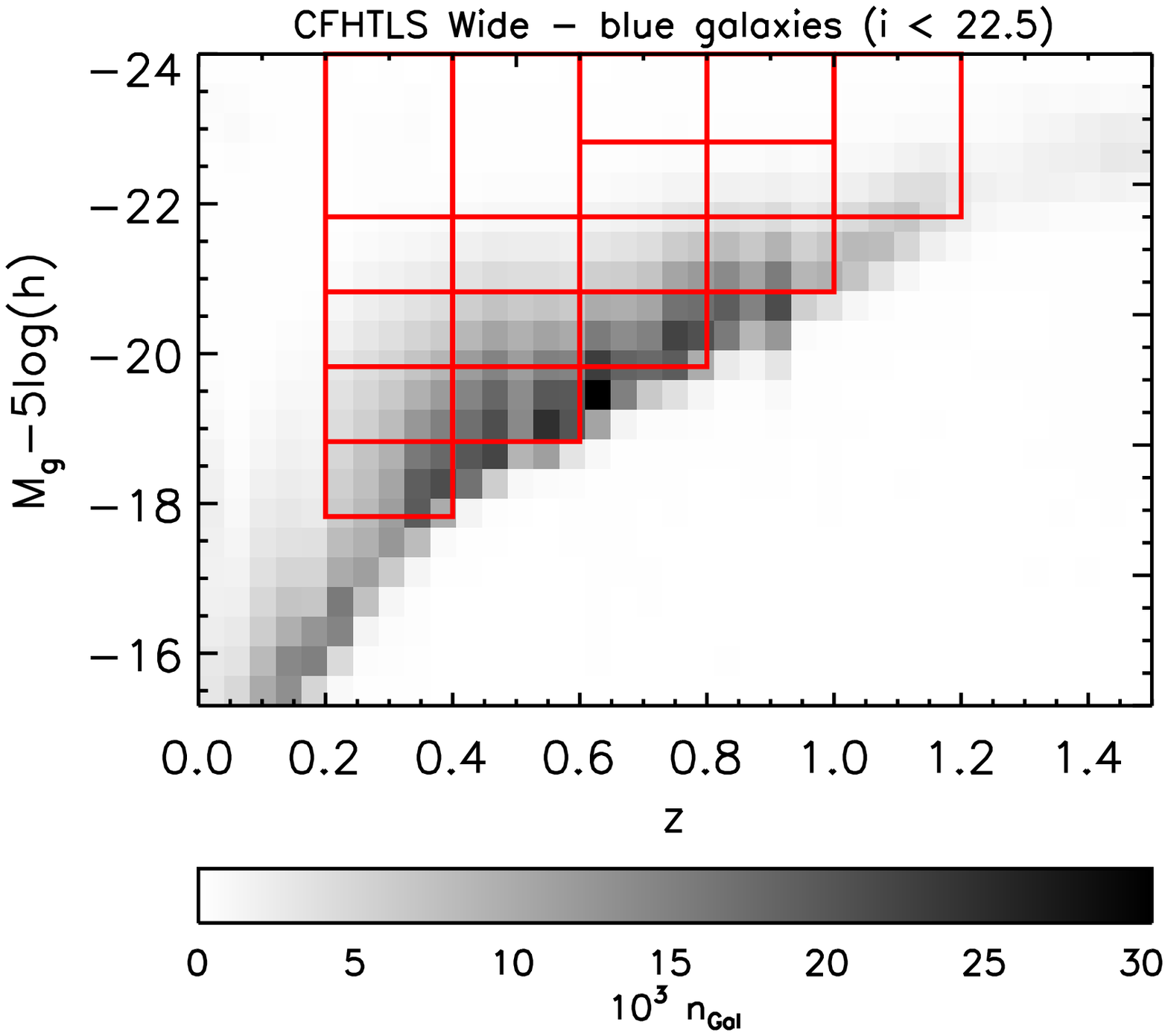}
    \end{tabular}
  \end{center}
  \caption{Sample selection in the full (left), red (center) and blue
    (right) galaxy samples. In each panel, the galaxy number density
    in the plane $M_g/z$ is shown; red rectangles represent the
    luminosity threshold samples.}
  \label{fig:samples}
\end{figure*}

We extract volume-limited luminosity-selected samples for each of the
``full'' (or ``all galaxies''), ``red'' and ``blue'' samples, using
$M_g$ absolute magnitude thresholds (hereafter denoted as ``luminosity
threshold samples''), from $M_g - 5 \log h= -17.8$ (fainter threshold
in the range $0.2 < z < 0.4$) to $M_g - 5 \log h = -22.8$ (brighter
threshold in the range $1.0 < z < 1.2$). The sample selection is
illustrated in Fig.~\ref{fig:samples}. For the rest of this paper we
will refer to these samples as simply full, red and blue.
Due to low numbers of pairs at small scales, luminous blue samples
were discarded.  We are left with 45 samples, each comprising on
average $\sim 153\,000$, $\sim 70\,000$ and $\sim 129\,000$ galaxies
for a typical full, red and blue sample, respectively.  The sample
properties are displayed in Tables~\ref{tab:HODall}, \ref{tab:HODred}
and \ref{tab:HODblue}.

Finally, in each redshift interval $[z_{\rm min}; z_{\rm max}]$ we
compute the galaxy number density:
\begin{equation}
  n_{\rm gal}^{\rm obs} = {N_{\rm
      total}}/\left[ {\Omega \int_{z_{\rm min}}^{z_{\rm max}} \frac{\ud V}{\ud
        z} \, \ud z} \right],
\end{equation}
where $\Omega$ represents the solid angle subtended by the survey, and
$\ud V/\ud z$ the volume element. Errors are estimated from the
weighted galaxy number density field-to-field variance.

\subsection{Photometric redshift uncertainties}
\label{sec:photozerr}

\begin{figure*}
  \begin{center}
    \begin{tabular}{c@{}c@{}c@{}}
      \includegraphics[width=0.33\textwidth]{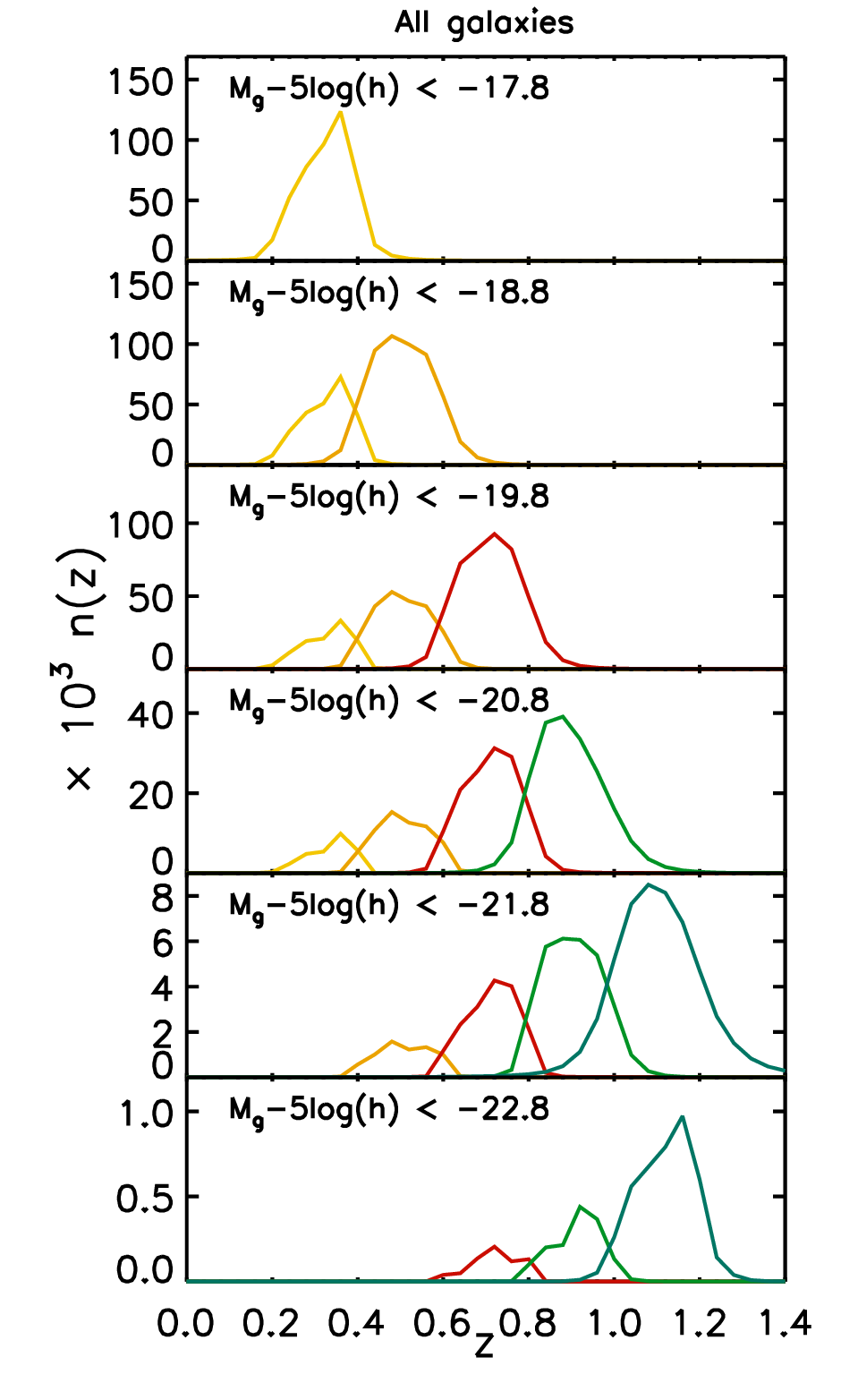} &
      \includegraphics[width=0.33\textwidth]{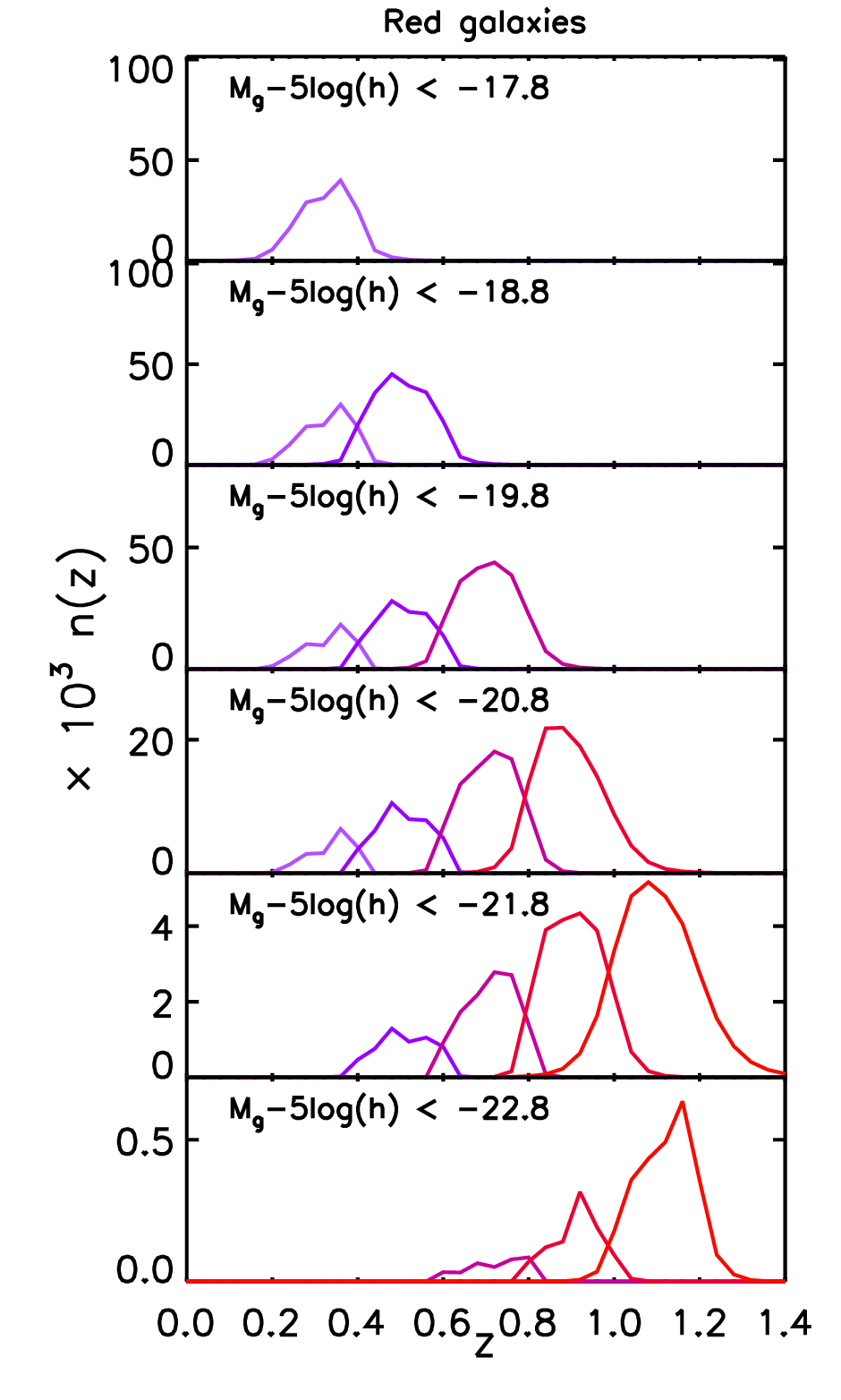}&
      \includegraphics[width=0.33\textwidth]{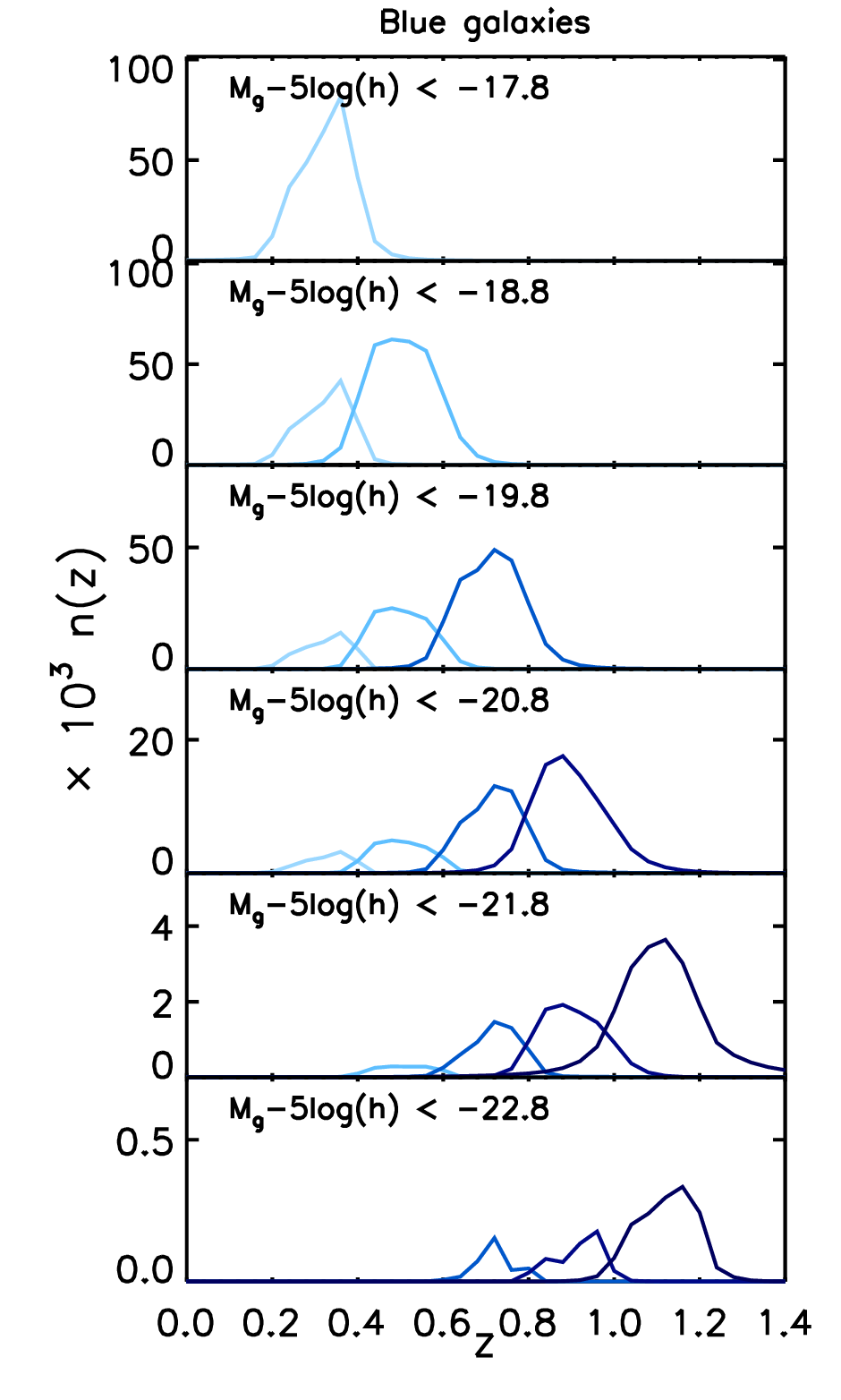}
  \end{tabular}
\end{center}
  \caption{Redshift distributions for the full (left), red (center) and blue
    (right) galaxy samples as function of absolute magnitude threshold.}
  \label{fig:nz}
\end{figure*}

Our modelled two-point correlation function is projected using the
measured redshift distributions. In order to take into account
statistical errors on redshifts, we select galaxies in the redshift
range considered and convolve the observed redshift distributions with
the estimated photometric redshift errors, derived from the
probability distribution functions. We construct a Gaussian error
distribution for each galaxy centred on the median redshift of the
PDF, with a width corresponding to the 68\% confidence limits of the
PDF, and normalised to unity. We then sum these Gaussians to construct the
redshift distribution resampled to a redshift bin width of 0.04.
Redshift distributions for each sample are illustrated in
Fig.~\ref{fig:nz}.

To further assess the quality of photometric redshifts, we perform
the cross-correlation analysis introduced in
\citet{2010MNRAS.408.1168B}. The measurement of the angular
correlation functions for galaxies in different photo-$z$ bins is used
to constrain the fraction of galaxies that are scattered into
``wrong'' redshift bins due to photo-$z$ errors.
We measure the bin-to-bin cross correlation function for the full, red
and blue samples, respectively. A non-zero correlation between
adjacent redshift bins is present in all cases. This may be due to the
presence of large-scale structures extending over several redshift
bins. More importantly, this is also due to photometric redshift
scatter, which results in the leakage of galaxies into neighbouring
bins. This error contribution to the redshifts is taken into account
by the convolution of the redshift distribution with the errors,
therefore, we do not consider adjacent bins further in the analysis of
photometric redshifts uncertainties.

The angular cross-correlation of galaxies in non-adjacent redshift
bins is much lower, indicating a small fraction of catastrophic
outliers. We use the ``global pairwise analysis'' method to measure
the contamination between two redshift bins $i$ and $j$. In this
approximation, the following linear combinations of the angular
cross-correlation function $w_{ij}$ and the two auto-correlation
functions, $w_{ii}$ and $w_{jj}$, respectively, are expected to cancel
for all angular scales $\theta_t$,
\begin{eqnarray}
  d_t & = & w_{ij}(\theta_t) \left( f_{ii} f_{jj} + f_{ij} f_{ji} \right)
  - w_{ii}(\theta_t) \frac{N_i}{N_j} f_{ij} f_{jj} 
  - w_{jj}(\theta_t) \frac{N_j}{N_i} f_{ji} f_{ii}  \nonumber \\
  & = & 0.
  \label{d_t}
\end{eqnarray}
Here, $N_i \; (N_j)$ is the observed number of galaxies in bin $i \;
(j)$. The contamination $f_{ij}$ is the number of galaxies with true
redshift in bin $i$, but misidentified into bin $j$, as a fraction of
the true number of galaxies in bin $i$.  For each bin pair ($i, j$),
the leakage of the other redshift bins is neglected. This
approximation is valid for contamination fractions of up to 10\%
\citep{2010MNRAS.408.1168B}. With this, the fraction of galaxies which
stay in their bin is $f_{ii} = 1 - f_{ij}$. We fit the two parameters
$f_{ij}$ and $f_{ji}$ in Eq.~\ref{d_t} by performing a $\chi^2$
null-test on $d_t$. For the covariance $\langle d_t d_s \rangle$, we
take into account the correlation between angular scales for each of
the three correlation functions using a Jackknife estimate (see next
section). We neglect the sub-dominant covariance between different
correlation functions. This corresponds to using the first three terms
in Eq.~A4 of \cite{2010MNRAS.408.1168B}.

\begin{figure}
  \begin{center}
    \begin{tabular}{c@{}c@{}}
      \multicolumn{2}{c}{\includegraphics[width=0.49\textwidth]{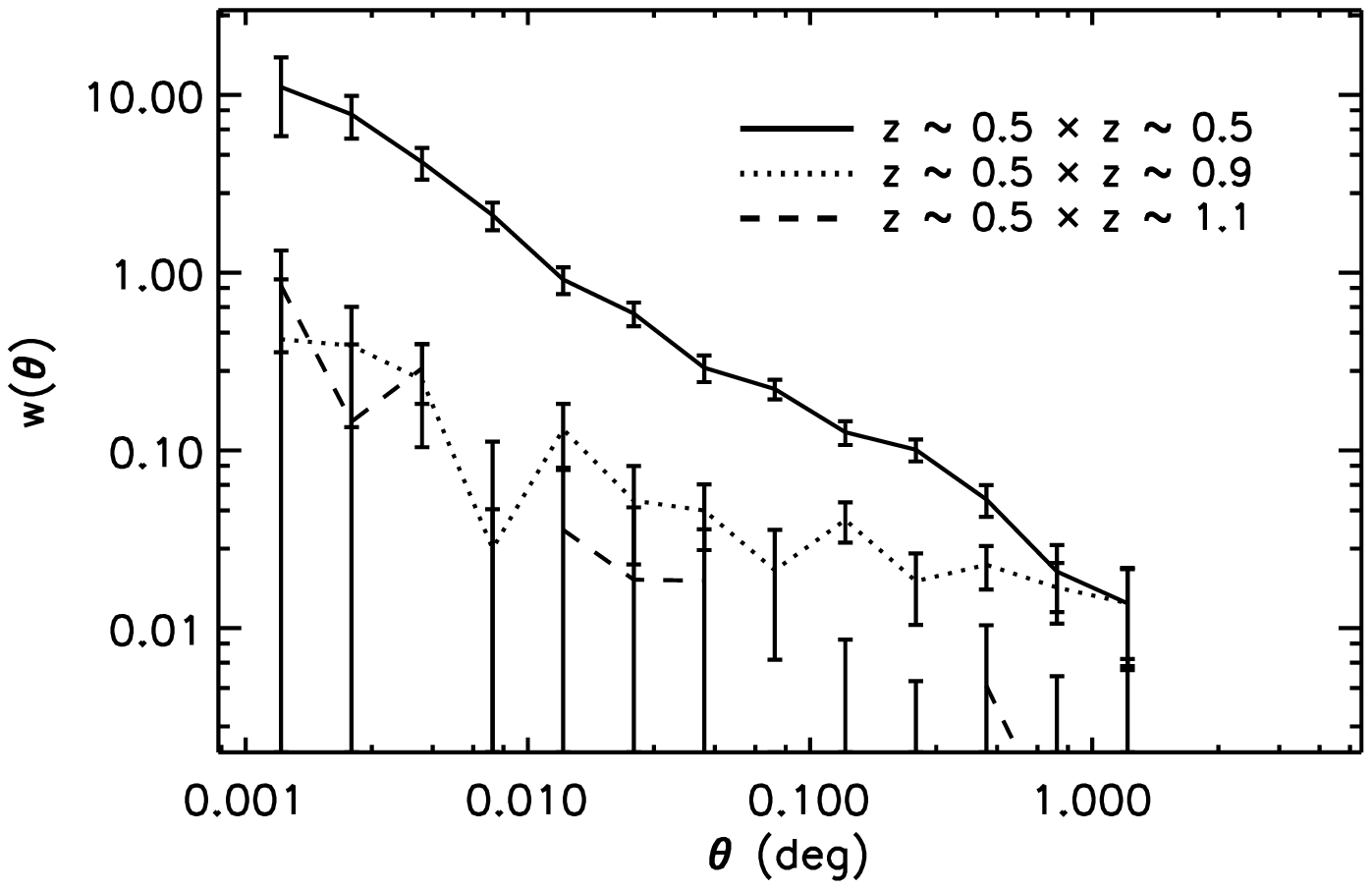}}\\
      \includegraphics[width=0.24\textwidth]{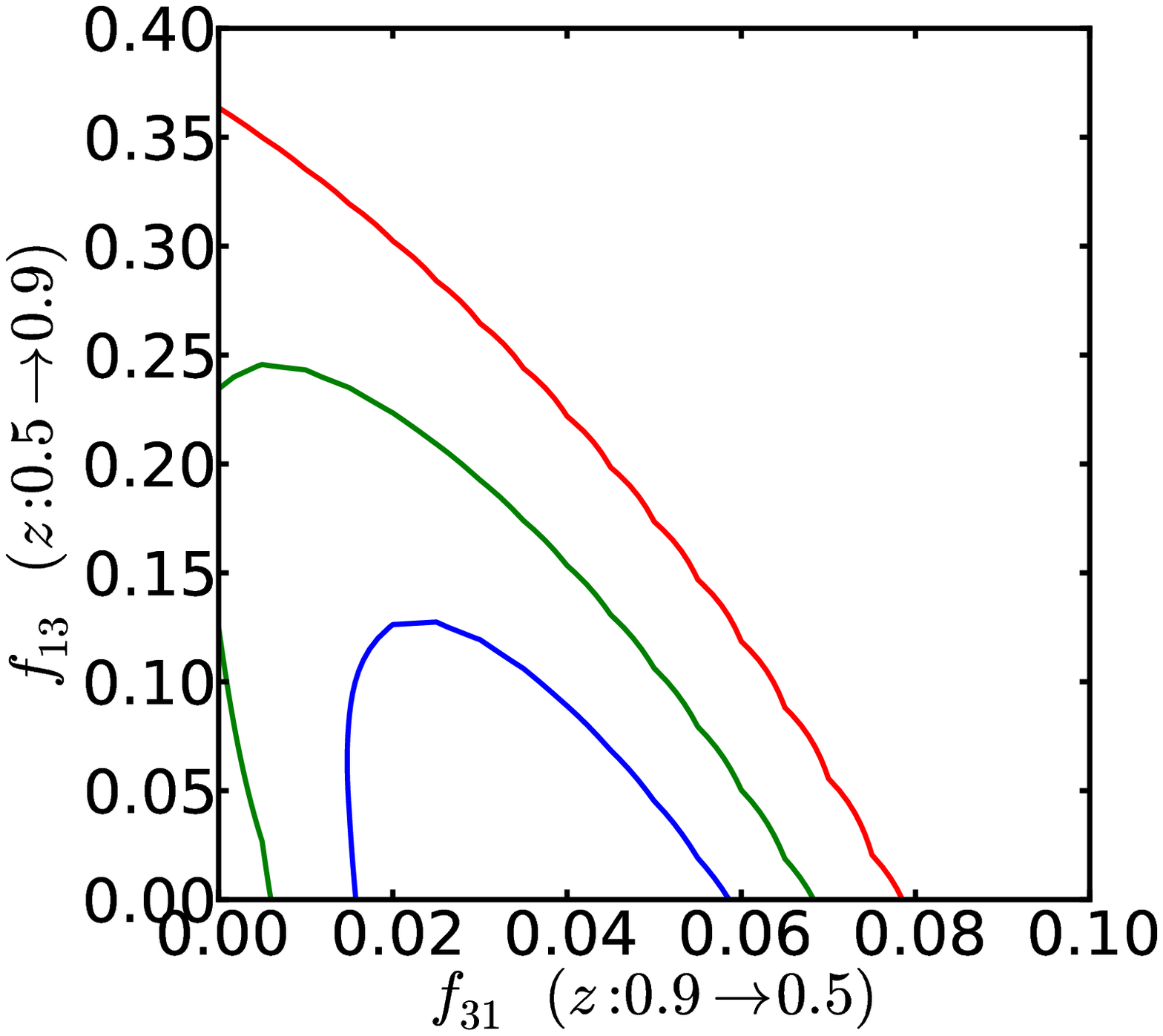}&
      \includegraphics[width=0.24\textwidth]{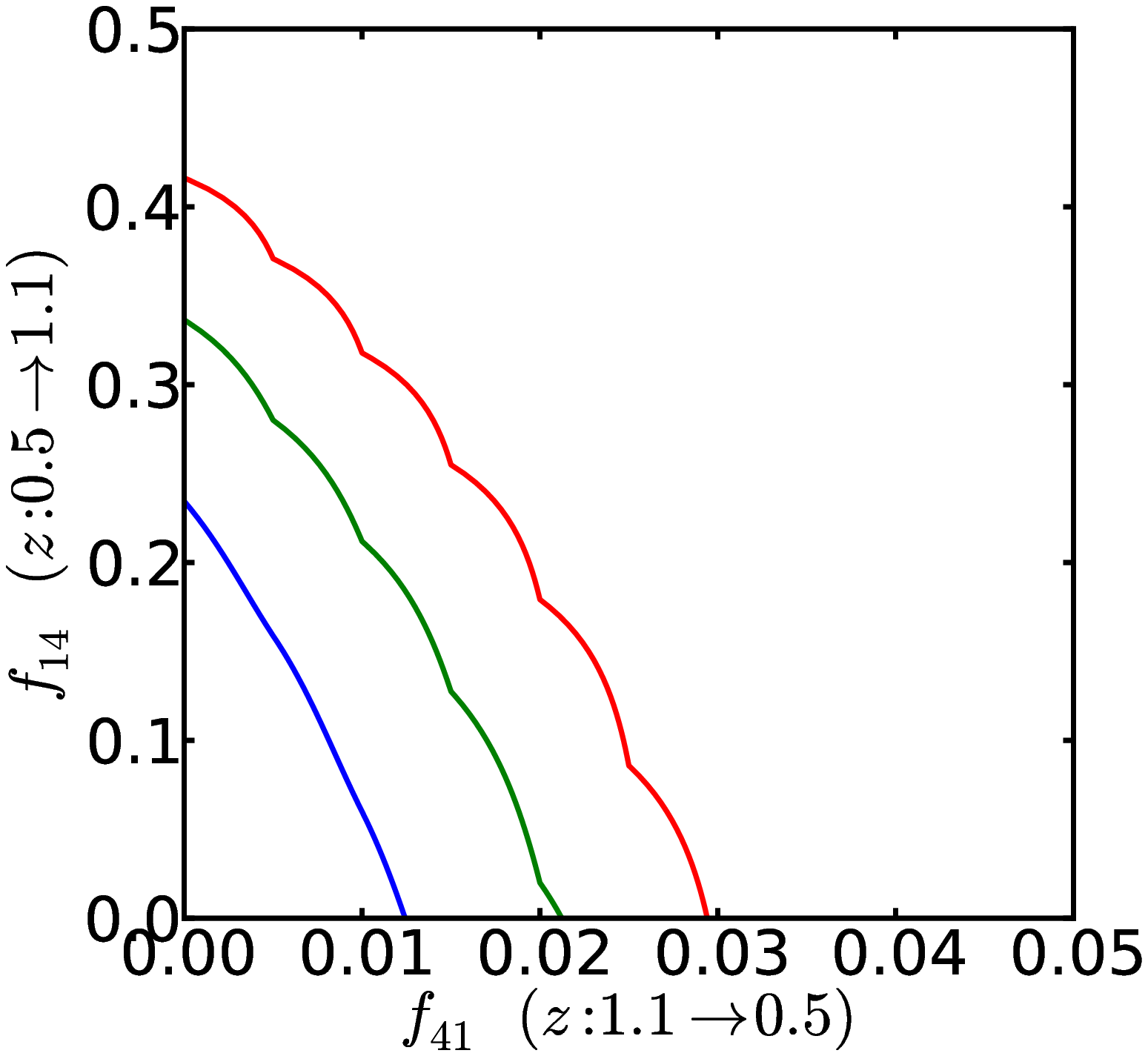}
    \end{tabular}
  \end{center}
  \caption{Cross correlation analysis between redshift bins for the full
    sample. Top: auto-correlation in the redshift bin $0.4 < z < 0.6$
    (straight line) and cross-correlation between the bins  $0.4 < z
    < 0.6$ and $0.8 < z < 1.0$  (dotted line) and between $0.4 < z <
    0.6$ and $1.0  < z < 1.2$ (dashed line). 
    Bottom: quantitative estimates of the contamination (percentage
    of galaxies scattered) from a pairwise
    analysis between redshift bins $0.4 < z < 0.6$ and $0.8 < z < 1.0$
    (left) and  between $0.4 < z < 0.6$ and $1.0 < z < 1.2$ (right). 
    The contours show the 68.3 (blue), 95.5 (green) and 99.7 (red) confidence regions.}
  \label{fig:cross}
\end{figure}

Due to degeneracies between the parameters $f_{ij}$ and $f_{ji}$,
large values for $f_{ij} \; (i>j)$ cannot be ruled out in
principle. However, for the full and red galaxy samples, the
contamination fractions are consistent with zero in most cases. The
blue galaxy samples are slightly worse, but contaminations are
consistent with values between 2\% and 10\%. Together with the very
low outlier rate for the sub-sample with spectroscopic redshifts (see
Fig.~\ref{fig:zp_zs}), these results further strengthen our confidence
in the photometric redshift estimates in this work and to use them to
measure angular correlation functions. Our measurements for the full
sample are shown in Fig.~\ref{fig:cross}.

Bin-to-bin mixing caused by photometric redshift errors will reduce
our measured clustering signal, given that objects in separate bins
are uncorrelated. This would lead to an underestimation of halo
masses, as low mass haloes tend to be less clustered than more massive
ones.  However, it is not trivial to predict how galaxy number density
estimates would be affected by such errors, which may result in
incomplete or contaminated samples. Since the halo number density
decreases monotonically with halo mass, galaxy number density
decreases with increasing host halo mass. Therefore, a contaminated
sample will underestimate halo mass fitting estimates. Conversely, an
incomplete sample will tend to overestimate halo masses.

To understand how photometric redshift errors could affect our results
as function of redshift and luminosity threshold, we compared our
measurements in the CFHTLS Wide with those from the ``Deep'' component
of the CFHTLS \citep{2009GaranovaT06}, where more precise photometric
redshifts are available \citep[see Table~3
in][]{2009A&A...500..981C}. Photometric redshifts in the CFHTLS Deep
were computed using the method described in Sect.~\ref{sec:photoz} and
to avoid cosmic variance uncertainties we performed our tests in
overlapping areas between the Wide and the Deep, D1/W1 and D3/W3, over
a total area of 2~deg$^2$.

These tests show that faint luminosity threshold samples are slightly
incomplete at low redshift, probably as a consequence of catastrophic
errors moving low redshift galaxies to higher redshift. We report that
the difference in galaxy density between the Wide and the Deep
estimates rarely exceeds the field-to-field variance error
estimate. However, brighter and brighter samples become more and more
contaminated by spurious objects, and in the worst case (the brightest
sample in the range $0.8 < z < 1.0$), the number of objects in the
Wide is more than three times higher than in the Deep, much larger
than the 40\% field-to-field variance. A higher galaxy number density
estimate will enhance the effect of a reduced clustering signal, which
may result in underestimated halo masses for brighter samples up to a
few sigmas. In the worst case, this represents a bias of $\sim0.5$ in
$\log M_{\rm min}$.

\subsection{Angular correlation function}

We measure the two-point angular correlation function $w$ using
the \cite{1993ApJ...412...64L} estimator

\begin{equation}
  w (\theta) = \frac{N_{\rm r}(N_{\rm r}-1)}{N_{\rm d}(N_{\rm d}-1)} 
  \frac{\langle DD \rangle}{\langle RR \rangle} - \frac{N_{\rm
      r}-1}{N_{\rm d}} \frac{\langle DR \rangle}{\langle RR \rangle} + 1 \, ,
\end{equation}
where $\langle DD \rangle$ is the number of galaxy pairs, $\langle RR
\rangle$ the number of random pairs, $\langle DR \rangle$ the number
of galaxy-random pairs, all in a bin around the angular separation
$\theta$. $N_{\rm d}$ and $N_{\rm r}$ are the number of galaxies and
random objects, respectively.  A random catalogue is generated for
each sample. In order to find an acceptable compromise between small
Poisson errors and computational requirements we scale $N_{\rm r}$
depending on the number of galaxies in our input catalogues, from $50
\times N_{\rm d}$ in the case of low density data catalogues to $2
\times N_{\rm d}$ for high density catalogues. We measure $w$ using a
fast two-dimensional tree code.  At large separations $\theta$,
instead of counting individual galaxy pairs, we correlate boxes by
multiplying the number of objects in the two boxes. For  that, we
define a threshold angle $\alpha_w = \theta / d_{\rm b}$, where
$d_{\rm b}$ is the box size. Below this threshold, the number of
objects inside the box is taken into account instead of individual
objects. We found that $\alpha_w=0.05$ gives accurate results at low
computational cost.

We correct our galaxy correlation measurements for the
``integral constraint'' \citep{1977ApJ...217..385G}, a correction factor
which arises from the finite area of our survey $\Omega^2$:
\begin{equation}
  w(\theta) = w_{\rm mes}(\theta) + w_C \, .
\end{equation}

The correction factor $w_C$ can be estimated as follows:
\begin{equation}
  w_C = \frac{1}{\Omega^2} \int\!\!\!\!\int w(\theta) \, \ud \Omega_1 \ud \Omega_2 \, ,
\end{equation}
assuming a simple power law fitted on the data 
with slope $\gamma$ and amplitude $A_w$: 
\begin{equation}
  w_{\rm mes}(\theta) = A_w \theta^{1-\gamma} - w_C = A_w (\theta^{1-\gamma} - C)\, . 
\end{equation}
As $A_w$ varies with each sample, we first compute C using Monte Carlo
integration over random pairs
\begin{equation}
  C =  \frac{\sum \theta^{1-\gamma} RR (\theta)}{\sum RR(\theta)}\, ,
\end{equation}
and then correct $w_{\rm mes}(\theta)$ using
\begin{equation}
  w(\theta) = w_{\rm mes}(\theta)
  \frac{\theta^{1-\gamma}}{\theta^{1-\gamma} - C}\, . 
\end{equation}
The integral constraint for the four Wide fields is
$C\sim0.5$. Assuming $\gamma = 1.8$, this leads to a correction 
of $\sim10\%$ at $\theta=0.1$ deg and up to a factor of two at $\theta=1.0$
deg. We note that assuming a power-law for $w(\theta)$ to estimate $C$
could be a source of error; it is neglected in the current analysis as 
the correction is smaller than our Jackknife error estimates.

We combine our galaxy samples from the four CFHTLS fields into a
single catalogue for the correlation function measurement. Computing
$w(\theta)$ on the four fields independently, and using a
pair-weighted average would not lead to the same clustering estimate
at large scales. Relative photometric offsets are expected to vary
more from one field to another than within the same field, as the four
fields are non-overlapping.  Since our parent catalogue is selected by
apparent magnitude cut ($i'< 22.5$), even a small difference in
photometric offsets will lead to a spurious field-to-field variation
of the galaxy number density. In addition, different photometric
offsets could bias the photometric redshift selection and also
increase the field-to-field density fluctuations.  A 0.02 magnitude
offset variation would result in a $\sim2\%$ variation in galaxy
number density, which may bias the two-point correlation at very large
scales. Because we expect these effects to occur at the scale of
individual field (a few degrees), we adopt a conservative cut and
limit our measurements to $\theta < 1.5 \deg$. We do not measure $w$
at separations below $3\farcs6$ to avoid blended objects. Two-point
correlation function measurements and errors are provided in the
appendix~\ref{sec:wtheta}.

\subsection{$w(\theta)$ error estimates}

\begin{figure}
  \begin{center}
    \includegraphics[width=0.24\textwidth]{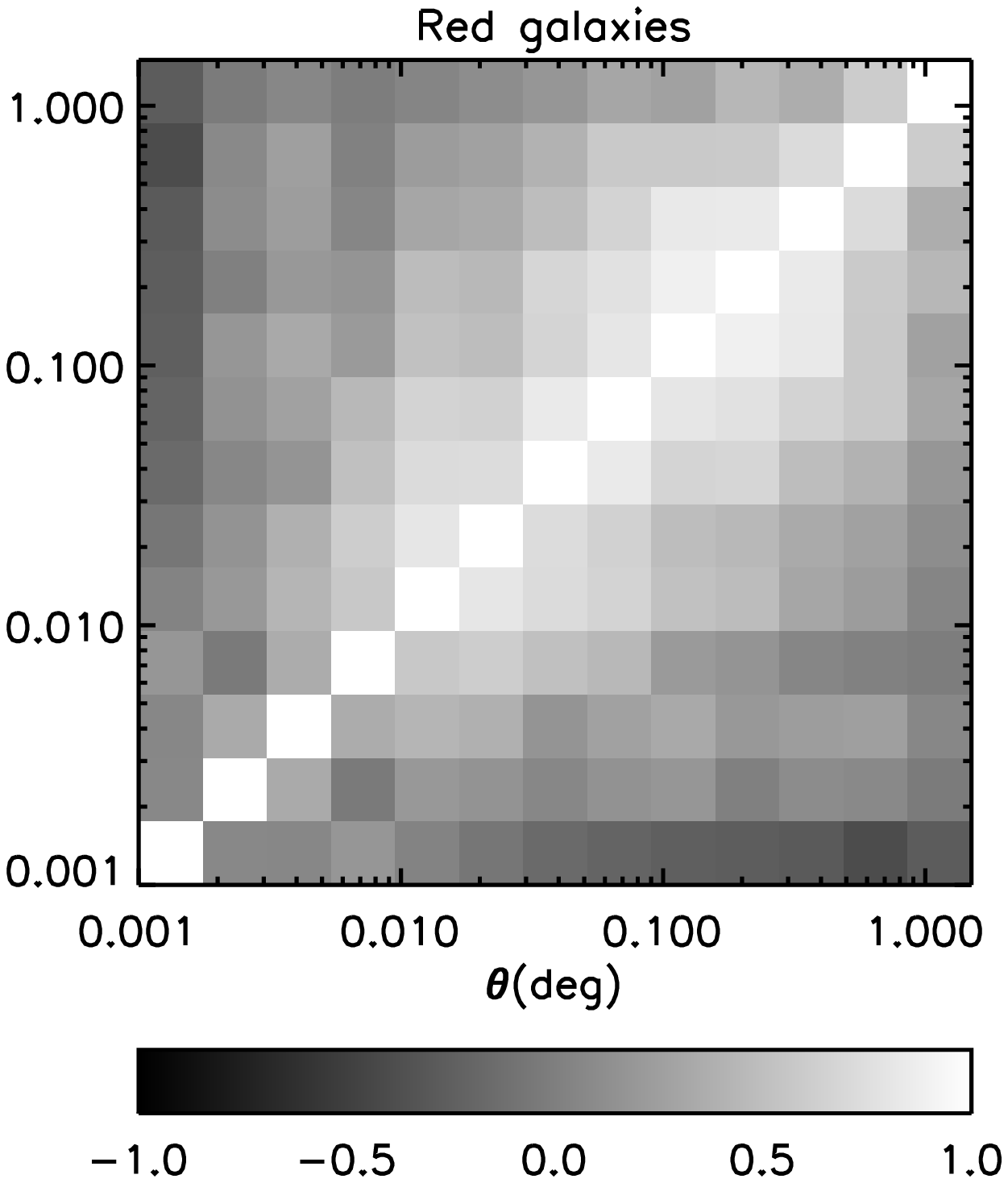}
    \includegraphics[width=0.24\textwidth]{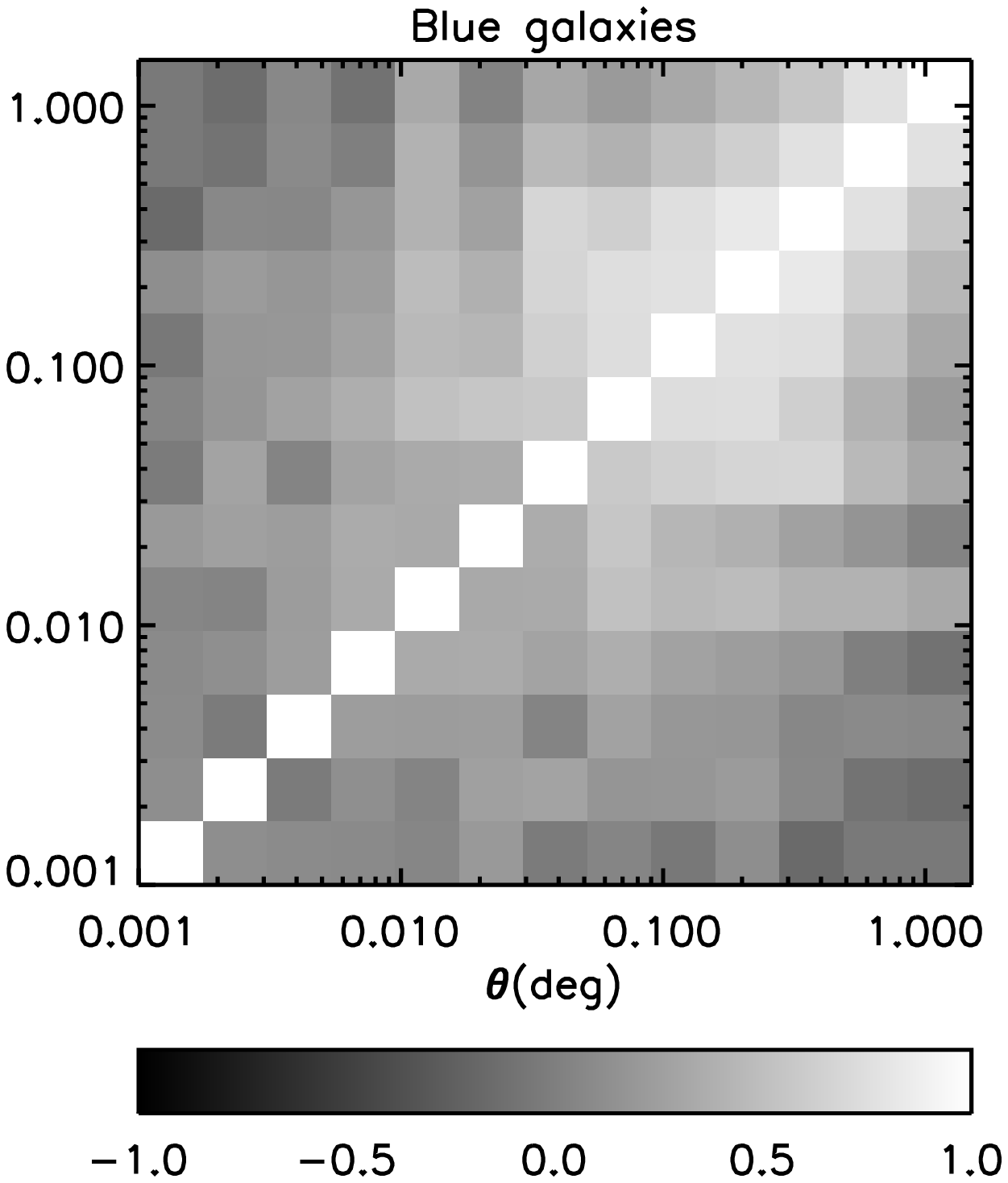}
  \end{center}
  \caption{Covariance matrix correlation coefficients of $w(\theta)$
    from Jackknife estimates, for galaxy samples with $0.4<z<0.6$ and
    $M_g - 5\log h < -19.8$. Left panel: red sample. Right panel: blue sample.}
  \label{fig:cov_mat}
\end{figure}

We estimate statistical errors on the two-point correlation function
using the Jackknife internal estimator. We divide all samples into 
$N = 68$ sub-samples of about $2\deg^2$. Removing one sub-sample at a time 
for each Jackknife realisation, we compute the covariance matrix as
\begin{equation}
  \mathrm{C}(w_i,w_j) = \frac{N-1}{N} \sum_{l=1}^{N} (w_i^l -
  \overline{w}_i) (w_j^l - \overline{w}_j) \, ,
  \label{covariance}
\end{equation}
where $N$ is the total number of subsamples, $\overline{w}$ the mean
correlation function, and $w^l$ the estimate from the $l^{\rm th}$
Jackknife sample. Since $w(\theta)$ is computed from a combined
catalogue including all four fields, our Jackknife estimate leads to a
noisy but fair estimate of cosmic variance. We apply the
correction factor given by \cite{2007A&A...464..399H}
to the inverse covariance matrix to compensate for
the bias introduced by the noise. We show the
correlation coefficient of the covariance matrix, $r_{ij} =
C_{ij}/\sqrt{C_{ii} C_{jj}}$ for two red and blue samples,
respectively, in Fig.~\ref{fig:cov_mat}.

\section{Filling haloes with galaxies: the halo occupation model}
\label{sec:model}
\subsection{The halo occupation distribution}
\label{sec:halo-occup-distr}

In order to relate galaxies to the dark matter haloes which host them
we have implemented an analytic model of galaxy clustering, the halo
model \citep[for a review see][]{2002PhR...372....1C}, 
which contains at its core a prescription for how galaxies
populate haloes, namely the halo occupation distribution. Our model follows closely the approaches used
in recent works; further details and references can be found in
Appendix \ref{sec:model_app}.  The key assumption underlying our halo
occupation distribution function is that the number of galaxies $N$ in
a given dark matter halo depends only on the halo mass $M$; it does
not depend on environment of formation history of the
haloes. Furthermore, following \citet{2005ApJ...633..791Z}, we express
$N(M)$ as the sum of two terms, corresponding to the contribution from
the central galaxy $N_{\rm c}$ and the satellite galaxies $N_{\rm s}$:
\begin{equation}
N(M) = N_{\rm c}(M)\times \left [ 1 + N_{\rm s}(M) \right] \, ,
\end{equation}
where
\begin{eqnarray}
\label{eq:nc}
N_{\rm c}(M) & = &\frac{1}{2} \left [ 1 + \mathrm{erf} \left (\frac{\log M - \log
      M_{\rm min}}{\sigma_{\log M}}  \right ) \right ] \, , \\
N_{\rm s}(M) & = & \left (  \frac{M - M_0}{M_1}
\right )^\alpha \, .
\end{eqnarray}
The smooth transition for central galaxies expresses the uncertainties
in the galaxy formation process \citep{2007ApJ...667..760Z}. The factor
$N_{\rm c}(M)$ for the satellite number in $N(M)$ accounts for the fact that a
halo cannot be populated by satellite galaxies without the presence of
a central galaxy.

For a given cosmology and dark matter halo profile, our model has five
adjustable parameters. $M_{\rm min}$ is the mass scale for which 50\%
of haloes host a galaxy. To reflect the scatter in the luminosity-halo
mass relation, a smooth transition of width $\sigma_{\log M}$ is
used. \cite{2007ApJ...667..760Z} show that if this scatter is small,
the above expression takes the identical form as the distribution of
central galaxies given a halo mass $M$, integrated over the entire
luminosity range above the luminosity threshold. Thus $M_{\rm min}$
also represents the halo mass scale for central galaxies, whose mean
luminosity $\langle L_{\rm c} \rangle$ is equal to the luminosity
threshold $L_{\rm min}$. 

This simple relation between $M_{\rm min}$ and $L_{\rm c}$ is based on
the hypothesis that stellar mass (or luminosity) has a power law
dependence on halo mass, which may not be exact over the entire mass
range.  \cite{2011arXiv1103.2077L} recently proposed a model in which
this relation assumes a more realistic form. The authors showed that
in their model $M_{\rm min}$ and $\sigma_{\log M}$ take different values
than those computed with other models. However, in the mass range 
over which we compare our results with theirs in
Sect.~\ref{sec:discussion} ($M_{\rm h}\sim10^{12} h^{-1} M_{\odot}$), $M_{\rm min}$ values do not
differ by more than 10\% \citep[see Fig.~3 in][]{2011arXiv1103.2077L}.

The number of satellite galaxies as function of halo mass follows a
power law with slope $\alpha$ and amplitude $M_1$. $M_1$ then
represents the characteristic scale for haloes hosting one satellite
galaxy. At lower masses, the dependence becomes steeper and the
transition mass scale occurs at $M \sim M_0$.

\begin{figure*}
  \begin{center}
    \includegraphics[width=\textwidth]{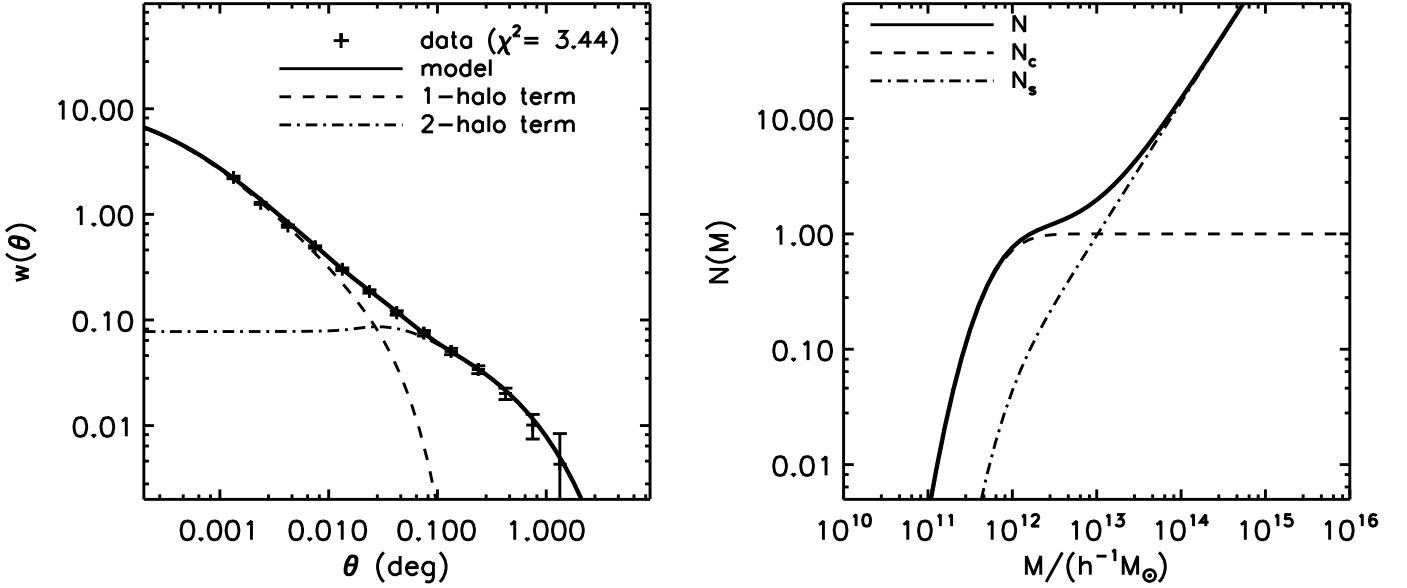}
  \end{center}
  \caption{Example of a measured $w(\theta)$ (for all galaxies in the
    redshift range $0.4 < z < 0.6$ and for $Mg - 5\log h < -19.8$),
    as well as the best-fitting model, as described in
    sec. \ref{sec:model}. Left: $w(\theta)$ measurement and
    model. Right: $N(M)$, showing the central term $N_{cent}$ and the
    satellite term $N_{\rm s}$.}
  \label{fig:example}
\end{figure*}

We show in Fig.~\ref{fig:example} an example of measured $w(\theta)$
and its best-fitting model, together with the best-fitting HOD function
$N(M)$. The total galaxy correlation function is
the sum of two terms. At distances much smaller than the virial
radius, the one-halo term contains contributions from galaxy pairs
within a single halo, whereas at large distances the two-halo term
contains contributions from pairs in separate haloes. 

\subsection{Deduced parameters}

From the HOD model we obtain deduced parameters describing galaxy
properties. The mean galaxy bias $b_{\rm g}$ at redshift $z$ is the
mass-integral over the halo bias $b_{\rm h}$ (see
Sect.~\ref{sec:model_app}) weighted by the number of galaxies,
\begin{equation}
  \label{eq:bias}
  b_{\rm g}(z) = \int \ud M \, b_{\rm
    h}(M, z)
  \, n(M, z) \frac{N(M)}{n_{\rm gal}(z)} \, .
\end{equation}
The dark-matter mass function $n$ is given in Eq.~\ref{eq:nufnu},
and the halo bias $b_{\rm h}$ in Eq.~\ref{eq:halo_bias}. The total
number of galaxies is
\begin{equation}
  \label{eq:ngal}
  n_{\rm gal}(z) = \int N(M) \, n(M,z) \, \ud M \, .
\end{equation}
Similarly to the galaxy bias, the mean halo mass for a galaxy
population is
\begin{equation}
   \label{eq:Mhalo}
  \langle M_{\rm halo} \rangle (z) = \int \ud M \, M \, n(M, z)
  \frac{N(M)}{n_{\rm gal}(z)} \, .
\end{equation}
The fraction of central galaxies per halo is
\begin{equation}
 f_{\rm c}(z) = \int \ud M \, n(M, z)
 \frac{N_{\rm c} (M)}{n_{\rm gal}(z)} \, .
\end{equation}
Consequently, the fraction of satellite galaxies is
\begin{equation}
   \label{eq:fsat}
  f_{\rm s}(z) = 1 - f_{\rm c}(z) \, .
\end{equation}

\subsection{Population Monte Carlo sampling}
\label{sec:pmc}

We use the ``Population Monte Carlo'' (PMC) technique to sample
likelihood space \citep{2009PhRvD..80b3507W, KWR10,2011arXiv1101.0950K} and have implemented
it in a publicly-available code,
 \textsc{CosmoPMC}\footnote{\texttt{http://cosmopmc.info}}. Contrary to
the widely-used Monte Carlo Markov Chain method (MCMC), PMC is an
adaptive importance-sampling technique \citep{CGMR03,
  cappe:douc:guillin:marin:robert:2007}, which has two principal advantages: first, the
parallel sampling algorithm combined with our fast HOD code means that
each run can be quickly computed. Secondly, PMC does not have issues
with chain convergence. Instead, the perplexity (defined below) is a
diagnostic for the reliability of a given sampling run.

Points are sampled from a
simple, so-called importance sampling function $q$. Each sample point
$\vec \theta_n$ is attributed a weight $\bar w_n$ which is the ratio
of the posterior $\pi$ to the importance function,
\begin{equation}
  \bar w_n \propto \frac{\pi(\vec \theta_n)}{q(\vec \theta_n)}; \;\;\;
    \sum_{n=1}^N \bar w_n = 1.
\end{equation}
The initial importance sampling function $q$ is a mixture of 
seven multi-variate Gaussians. A
PMC run consists of a number of iterations, in the course of which the
importance function is adapted to better match the posterior
distribution. We run PMC for typically 10 iterations using $10\,000$
sample points per iteration.

As a stopping criterion for the iterations we take the
perplexity $p$ which is defined as
\begin{equation}
  p =  \frac 1 N \exp\left( - \sum_{n=1}^N \bar w_n \log \bar w_n \right).
\end{equation}

The perplexity $p \in [0;1]$ is a measure of the distance between the
posterior and the importance function, and it approaches unity if the
two distributions are identical. The perplexity is a measure of the
adequacy and efficiency of the sampling.  Values of $p$ above about
0.6 indicate a good agreement between the posterior and the importance
function.

The evaluation of the likelihood function, which is the most
time-consuming process for most sampling tasks, is easily
parallelisable in importance sampling with little overhead.
Furthermore, we have optimized several aspects of the numerical
computations of the HOD model. For example, we employ the
\texttt{FFTLog} \citep{2000MNRAS.312..257H} algorithm to perform
Fourier transforms and make use of tabulated values to improve
efficiency.  The computation of the angular correlation function $w$
on a range of scales is performed in under a second on a standard
desktop. This module is part of the latest public version (v1.1)
of \textsc{CosmoPMC}. Typically, on an eight-core desktop (16 threads)
a single PMC run takes around 90 minutes for a total of $100,000$
sample points. Our PMC technique allows us to efficiently sample the
parameter space for the large number of galaxy samples used in this
work.

\subsection{Likelihood function}

For each galaxy sample we simultaneously fit both the projected angular correlation
function $w$ and the number density of galaxies $n_{\rm gal}$, by
summing both contributions in log-likelihood:
\begin{eqnarray}
  \chi^2 &=& \sum_{i,j}\left[w^{\rm obs}(\theta_{i}) -
    w^{\rm model}(\theta_{i})\right]\left(C^{-1}\right)_{ij}\left[w^{\rm obs}(\theta_{j}) -
    w^{\rm model}(\theta_{j})\right]
  \nonumber\\
  & & + \frac{\left[n_{\rm gal}^{\rm obs} - n_{\rm gal}^{\rm
      model}\right]^2}{\sigma^2_{n_{\rm gal}}} \, ,
\end{eqnarray}
where $n_{\rm gal}^{\rm model}$ is given by Eq.~\ref{eq:ngal}, at the
mean redshift of the sample.  The data covariance matrix $\mathrm{C}$
is approximated by Eq.~\ref{covariance}. The error on the galaxy
number density $\sigma_{n_{\rm gal}}$ contains Poisson noise and
cosmic variance. The latter is estimated from the field-to-field
variance between the four Wide patches.

The likelihood function is
\begin{equation}
  \mathcal{L} = \exp(-\chi^2/2) .
\end{equation}
The product of prior $P$ and likelihood defines the posterior
distribution, $\pi = P \mathcal{L}$. We sample a five-dimensional
  space where $\{\log_{10} M_{\rm min}, \log_{10} M_1, \log_{10} M_0,
  \alpha, \sigma_{\log M} \}$ is the parameter vector. We use flat
  priors for all parameters; the ranges are $\log_{10} M_{\rm min} \in [11; 15],
  \log_{10} M_1 \in [12; 17], \log_{10} M_0 \in [8; 15], \alpha \in [0.6; 2]$ and
  $\sigma_{\log M} \in [0.1; 0.6]$.

\section{Results}
\label{sec:results}

\subsection{Clustering measurements}
\label{sec:results-wtheta}

\begin{figure*}
  \begin{center}
    \includegraphics[width=\textwidth]{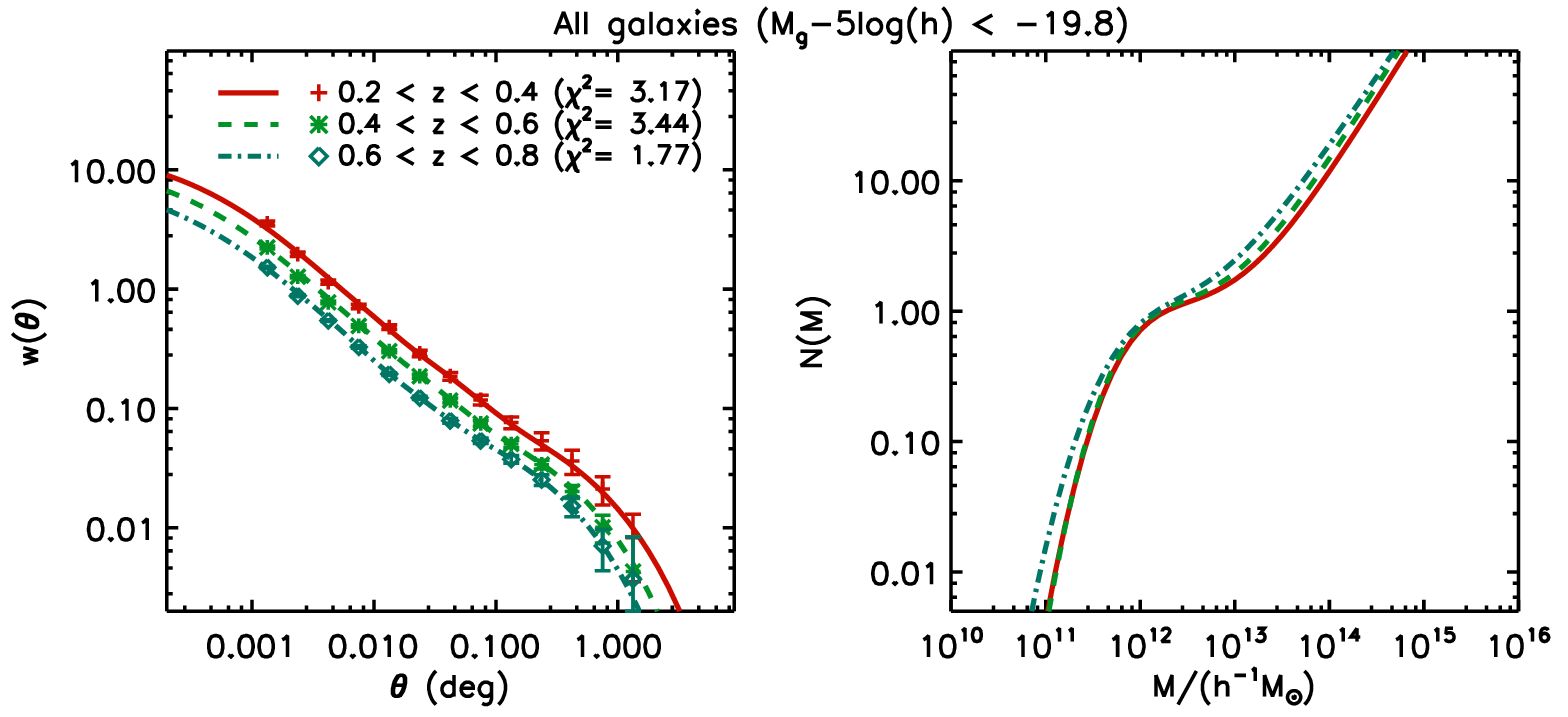}  
    \includegraphics[width=\textwidth]{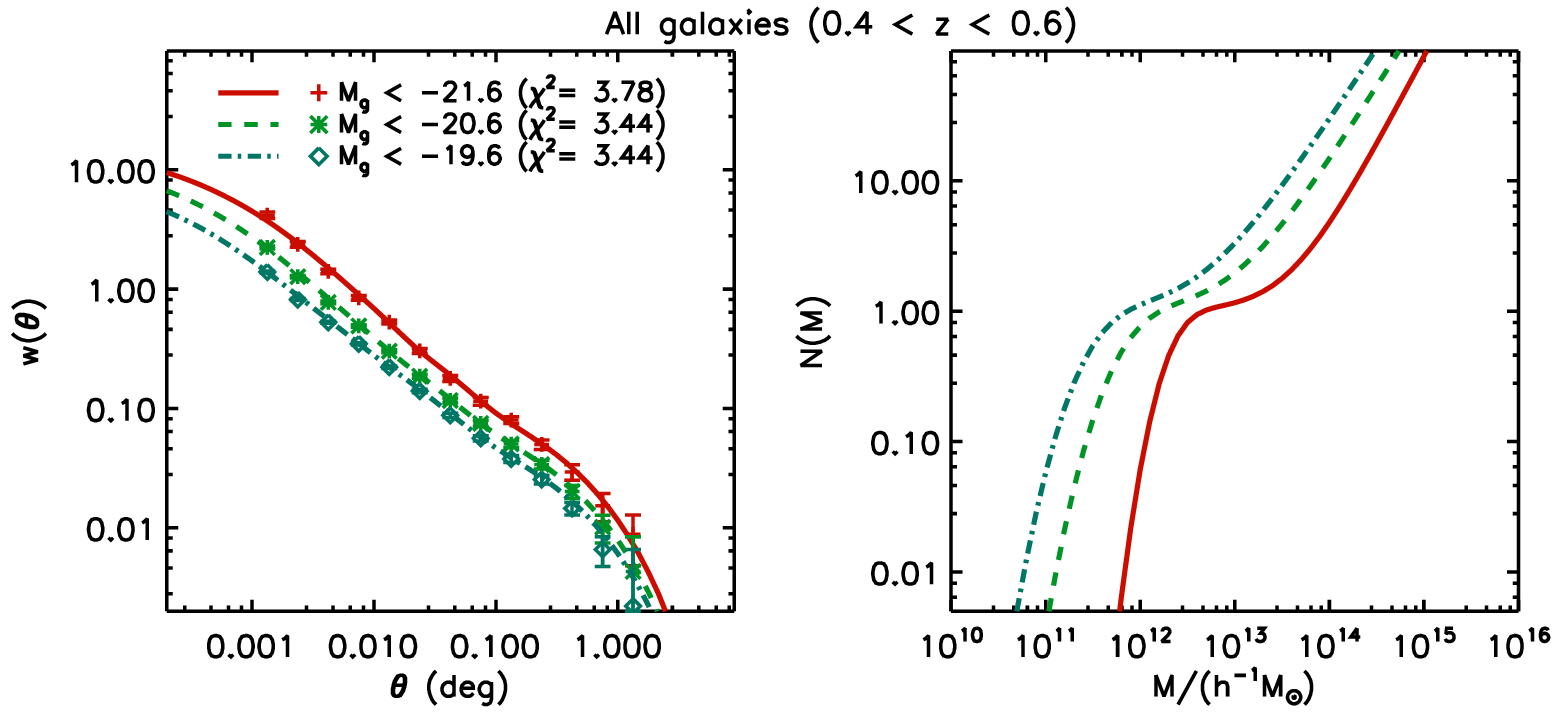}
  \end{center}
  \caption{$w(\theta)$ measurements (left) with their best fit models for all
    galaxies, and the resulting galaxy occupation function (right) with
    respect to halo mass. We illustrate the redshift 
    evolution (top) in the luminosity threshold $M_g -5\log h<
    -19.8$, and the luminosity dependence (bottom) in the
    redshift range $0.4 < z < 0.6$.}
  \label{fig:w_res_all}
\end{figure*}

\begin{figure*}
  \begin{center}
    \includegraphics[width=\textwidth]{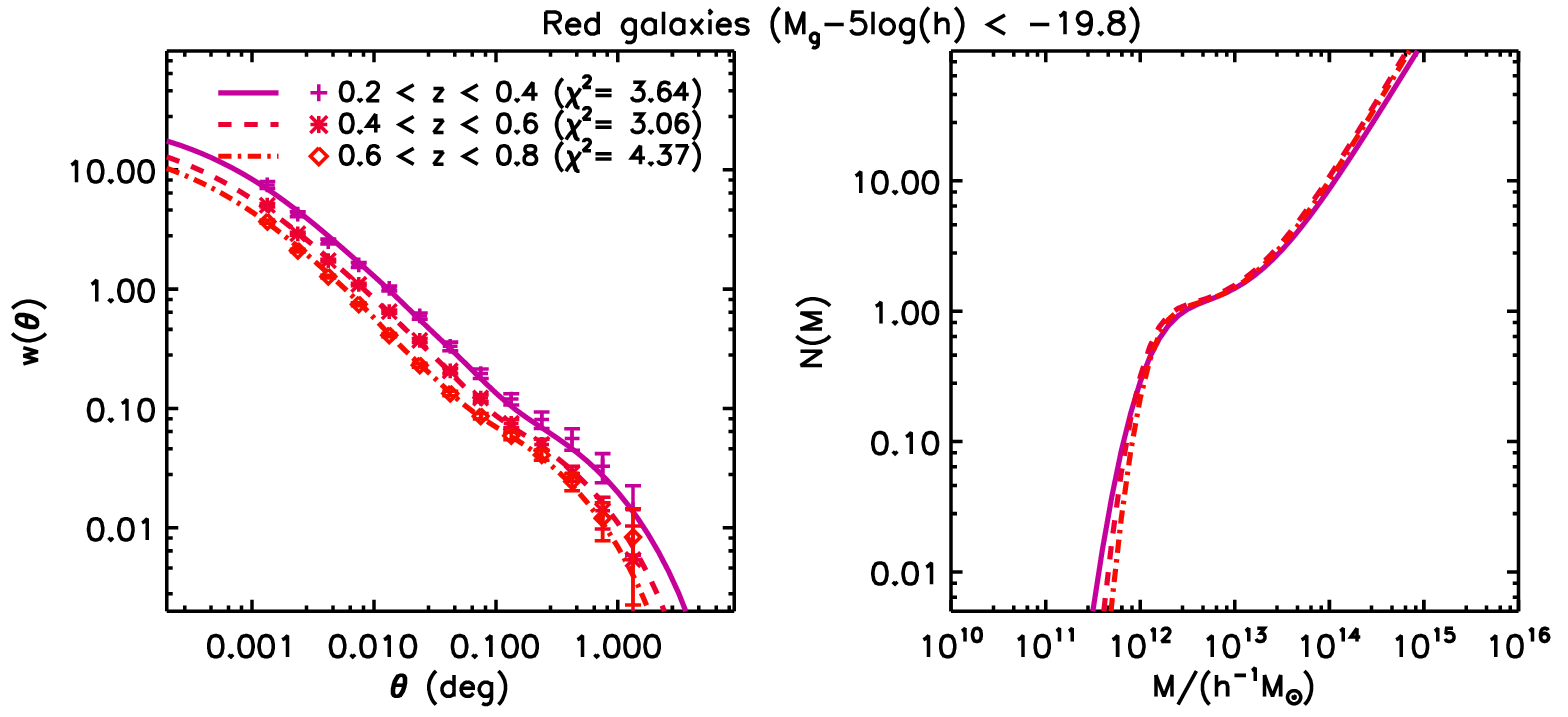}
    \includegraphics[width=\textwidth]{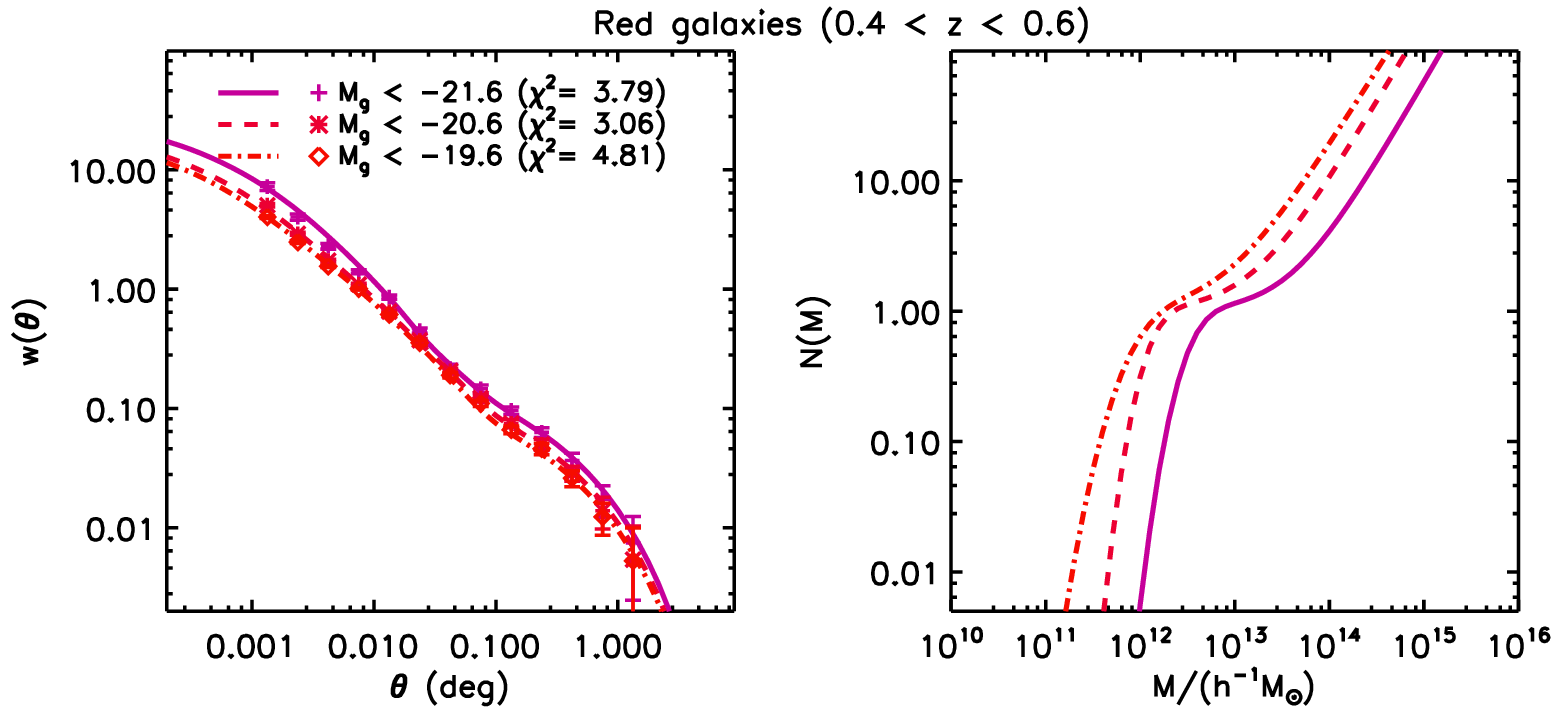}
  \end{center}
  \caption{$w(\theta)$ measurements (left) with their best fit models for red
    galaxies, and the resulting galaxy occupation function (right) with
    respect to halo mass. We illustrate the redshift 
    evolution (top) in the luminosity threshold $M_g -5\log h<
    -19.8$, and the luminosity dependence (bottom) in the
    redshift range $0.4 < z < 0.6$.}
  \label{fig:w_res_red}
\end{figure*}

\begin{figure*}
  \begin{center}
    \includegraphics[width=0.49\textwidth]{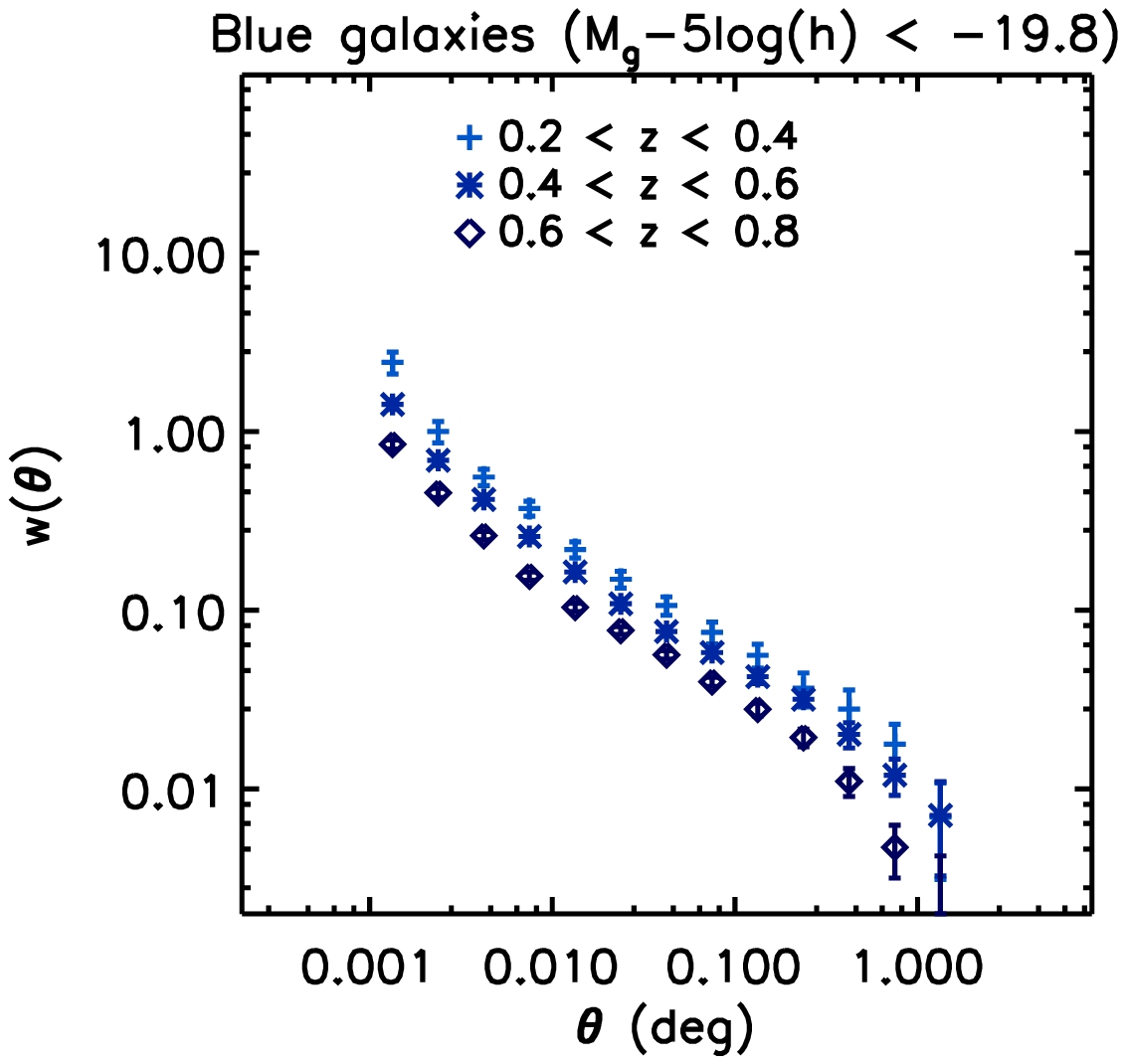} 
    \includegraphics[width=0.49\textwidth]{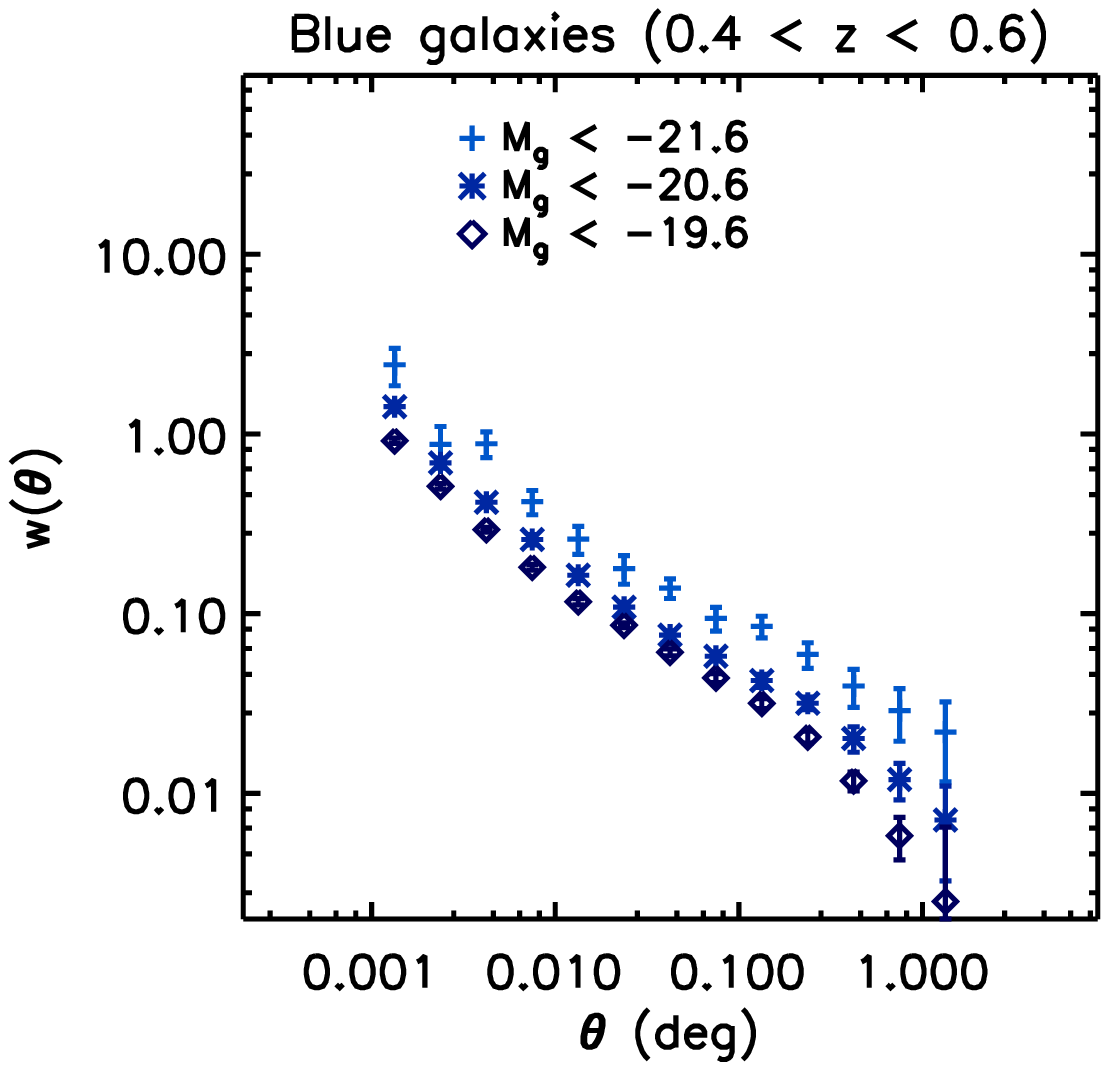}
  \end{center}
  \caption{$w(\theta)$ measurements for blue galaxies. We illustrate
    the redshift evolution (left) at the luminosity threshold $M_g
    -5\log h < -19.8$, and the luminosity dependence (right) in the
    redshift range $0.4 < z < 0.6$.}
  \label{fig:w_res_blue}
\end{figure*}

Our measurements are presented in Figure~\ref{fig:w_res_all},
\ref{fig:w_res_red} and \ref{fig:w_res_blue} for the full, red and
blue samples, respectively. For clarity, we show three samples
selected with the same luminosity threshold ($M_g - 5\log h < -19.8$)
and three samples covering the same redshift range ($0.4 < z < 0.6$).
At the same luminosity and redshift, red samples are more clustered on
small scales than the full sample, and blue galaxies are the least
clustered.  At a constant luminosity threshold, $w(\theta)$ increases
with redshift for the full, red and blue samples.

On the left panels of Figs.~\ref{fig:w_res_all} and
\ref{fig:w_res_red}, we plot our best fit $w(\theta)$ models.  Model
parameters and deduced quantities are shown in Tables~\ref{tab:HODall}
(full) and \ref{tab:HODred} (red).  An example of sample distributions
of best-fitting HOD parameters is given in
Fig.~\ref{fig:contours_HOD}.  Since the halo model aims to reproduce
the clustering in threshold samples as a sum of contributions from
central and satellite galaxies, both types of galaxy must be present.
We presume that in general, central galaxies are brighter than their
satellites, so that the latter assumption is valid for full samples
selected by luminosity threshold. Also, we assume that red satellites
are systematically associated with a (brighter) red central galaxy. On
the contrary blue satellites may belong to a red central galaxy.  For
this reason, we defer fitting our model to our pure blue samples. We
plan to address the blue population in a future work, perhaps
considering separate halo occupation functions for both red and blue
populations \citep{2009MNRAS.398..807S,2009MNRAS.392.1080S}.

In most cases, we find a good agreement between the data and our
model, with $\chi^2/\rm dof$ from 1.77 to 10.21 for full samples and
from 1.20 to 6.21 for red samples.  We report a few large $\chi^2$
values in some full samples, mostly at high redshift and for bright
luminosity thresholds, where the galaxy number density is small. 
We suspect systematic photometric redshift errors to be responsible for
an underestimated clustering signal and systematic errors in 
the galaxy number density estimates (see Sec~\ref{sec:photozerr}).
Perplexity (see Sect.~\ref{sec:pmc}) is larger than 0.6 for most samples,
indicating that our model accurately describes the observations.

\subsection{Halo model fitting}
\label{sec:results-HOD}

In the right panels of Figs.~\ref{fig:w_res_all} and
\ref{fig:w_res_red} we show our best fit $N(M)$, the halo
occupation function. First, we note that at a given redshift interval,
$M_{\rm min}$, the average mass for which 50\% of haloes contain one
galaxy increases with luminosity. Secondly, we note that the evolution
in $N(M)$ with redshift at a fixed luminosity is less pronounced than
the evolution with luminosity at a fixed redshift, whereas the
amplitude of $w(\theta)$ significantly decreases with redshift.

\begin{figure*}
  \begin{center}
    \includegraphics[width=0.49\textwidth]{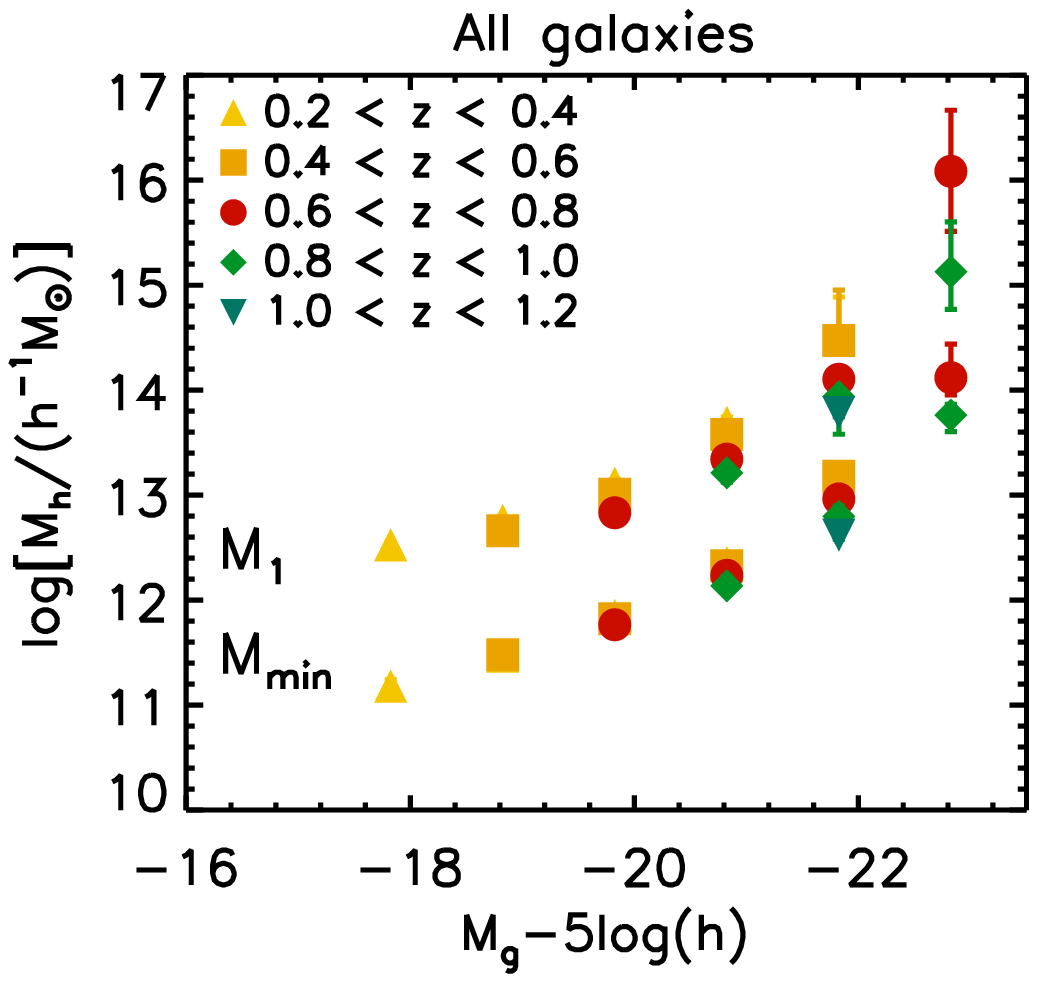}
    \includegraphics[width=0.49\textwidth]{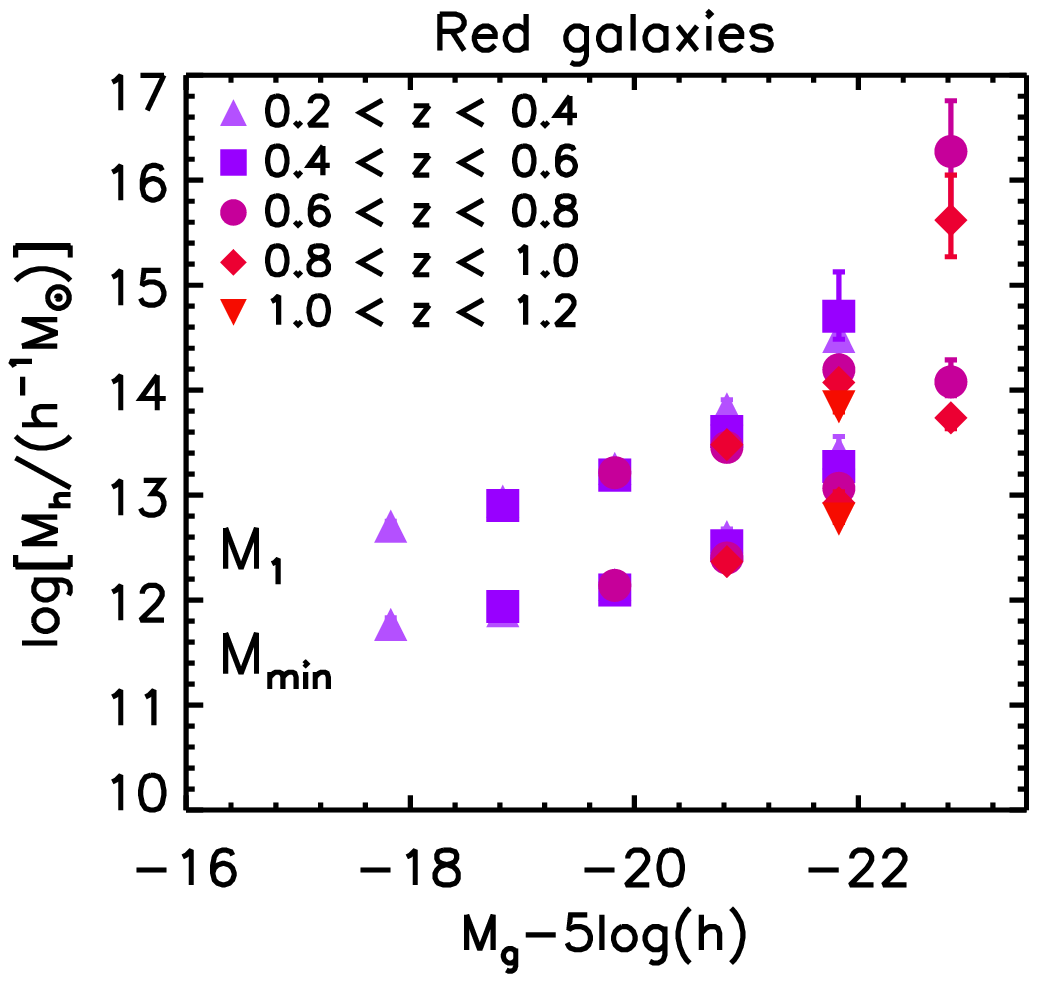}
    \includegraphics[width=0.49\textwidth]{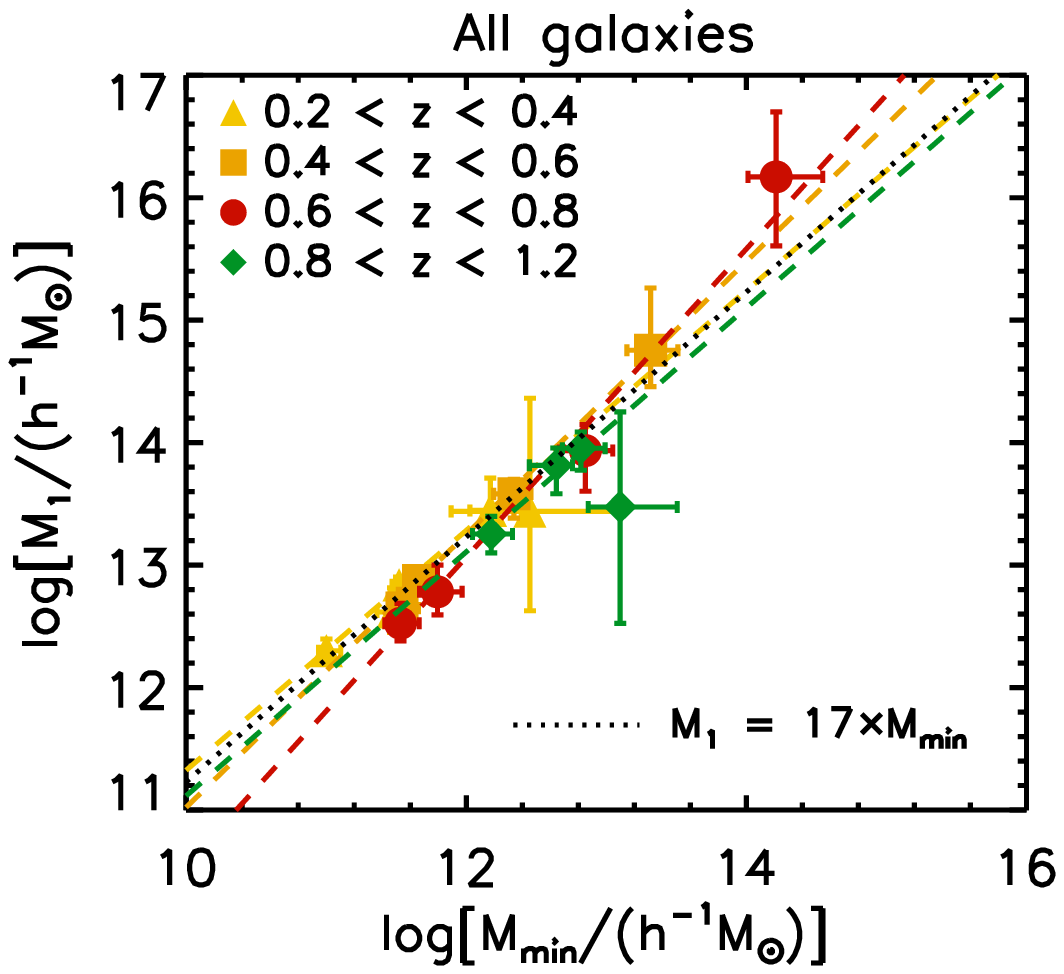}
    \includegraphics[width=0.49\textwidth]{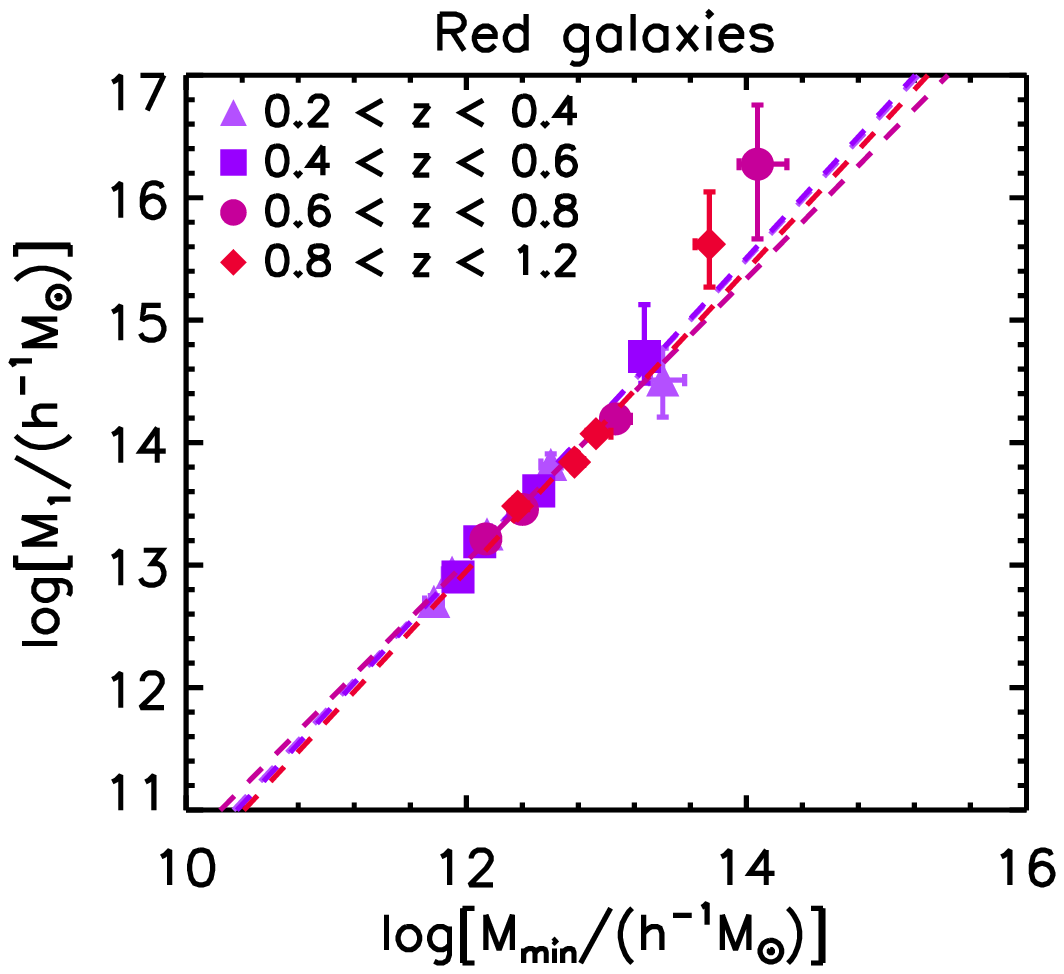}
  \end{center}
  \caption{Top panels: halo mass estimates $M_{\rm min}$ and $M_{1}$
    for all (left) and red (right) galaxy samples as function of
    luminosity threshold ($M_g$). Bottom panels: $M_{\rm 1}$ versus
    $M_{\rm min}$ in different redshift bins. The dotted lines
    represent a linear relation and the dashed lines are a power-law
    fit to $M_1$ versus $M_{\rm min}$ in each redshift bin.}
  \label{fig:M1_Mmin_no_evol}
\end{figure*}

Halo masses $M_1$ and $M_{\rm min}$ are tightly constrained, whereas
$M_0$, $\sigma_{{\rm log}M}$ and $\alpha$ are poorly constrained.
$M_{\rm min}$ and $M_1$ measurements are displayed in the top panels
of Fig.~\ref{fig:M1_Mmin_no_evol} with respect to luminosity threshold
($M_g$). For every sample, halo masses increase with luminosity.
$M_{\rm min}$ covers the range $10^{11}$ to $10^{14} h^{-1} M_{\sun}$.
$M_1$ is significantly larger than $M_{\rm min}$, ranging from
$10^{12.5}$ to $10^{16} h^{-1} M_{\sun}$. These trends are consistent
with those found in both the local
\citep{2010arXiv1005.2413Z,2005ApJ...630....1Z} and distant Universe
studies \citep{2007ApJ...667..760Z,2010MNRAS.406.1306A}.  At high
luminosity thresholds, where the number of objects is small, error
bars on our best-fitting parameters are correspondingly larger. This
trend is especially pronounced for $\alpha$, where the small number of
satellites in high mass haloes makes it difficult to constrain this
parameter. However, we note $\alpha$ increases with luminosity
threshold, suggesting that for massive haloes, bright galaxies will be
more numerous than faint ones. $\sigma_{{\rm log}M}$ remains
relatively stable for red galaxies ($\sim 0.3$), as also seen in other
data sets and numerical simulations.  For the full sample, values are
slightly larger (up to $\sim 0.5$ in some cases). Our fitted
$\sigma_{\log M}$ values are slightly larger than previous works. We also
made fits where values $\sigma_{\log M}>0.6$ were allowed, and in some
cases large values ($>1.0$) were found. We therefore decided
to impose the restriction that $\sigma_{\log M}<0.6$. Sample
incompleteness (due to errors in the photometric redshifts), leading
to missing central galaxies, could explain such high $\sigma_{\log M}$
values.

At constant luminosity threshold, in both red and full samples, $M_1$
and $M_{\rm min}$ decrease with increasing redshift. This effect is
partially due to a selection effect caused by the dimming and
reddening of stellar populations with time; for a constant luminosity
cut one selects less and less massive galaxies at higher redshifts. We
will return to this measurement in Sect.~\ref{sec:discussion} where we
attempt to separate this passive stellar evolution effect from the
intrinsic stellar-to-halo mass evolution.

At faint luminosities, $M_{\rm min}$ values are higher for red
galaxies than for full samples. As we explained previously, one
assumption of our model is that our red threshold samples have
luminous red central galaxies. On the other hand, it is also possible,
especially for fainter samples, that blue satellite galaxies inhabit
haloes with red central galaxies. In such a case for the faint red
galaxy sample, a red central galaxy will be seen ``alone'' in a more
massive halo.  This would also increase the halo mass estimate found
for red central galaxies as compared to those in the full sample.

The bottom panels of Fig.~\ref{fig:M1_Mmin_no_evol} show $M_1$ with
respect to $M_{\rm min}$ for full and red samples. The parameters are
strongly correlated and we fitted a power law (dashed line) in each
redshift bin.  For the full sample, we compare our results with a
linear relation, $M_1 = 17\times M_{\rm min}$ (dotted line), based on
the coefficient observed in the SDSS ($z\sim0.1$) by
\cite{2010arXiv1005.2413Z}. Our results agree with them in the lowest
redshift bin, $z\sim0.3$, but depart from this value at higher
redshift.

We find that although $M_1/M_{\rm min}$ relation does not vary
significantly with luminosity threshold it does show some indications
of evolution with redshift. We note that $M_1/M_{\rm min}$ decreases
with redshift for $M_{\rm min} \lesssim 10^{13} h^{-1} M_{\odot}$ and
increases with redshift at higher mass. For red galaxies, $M_1$ is no
longer proportional to $M_{\rm min}$ but follows a power law. For
$M_{\rm min} \sim 10^{12} h^{-1} M_{\odot}$, the ratio $M_1/M_{\rm
  min}$ is $\sim12$ \citep[as compared to $\sim 10$ for luminous
red galaxies found by][]{2009ApJ...707..554Z} and shows a much smaller redshift evolution
over the entire mass range.

\begin{figure*}
  \begin{center}
    \includegraphics[width=0.49\textwidth]{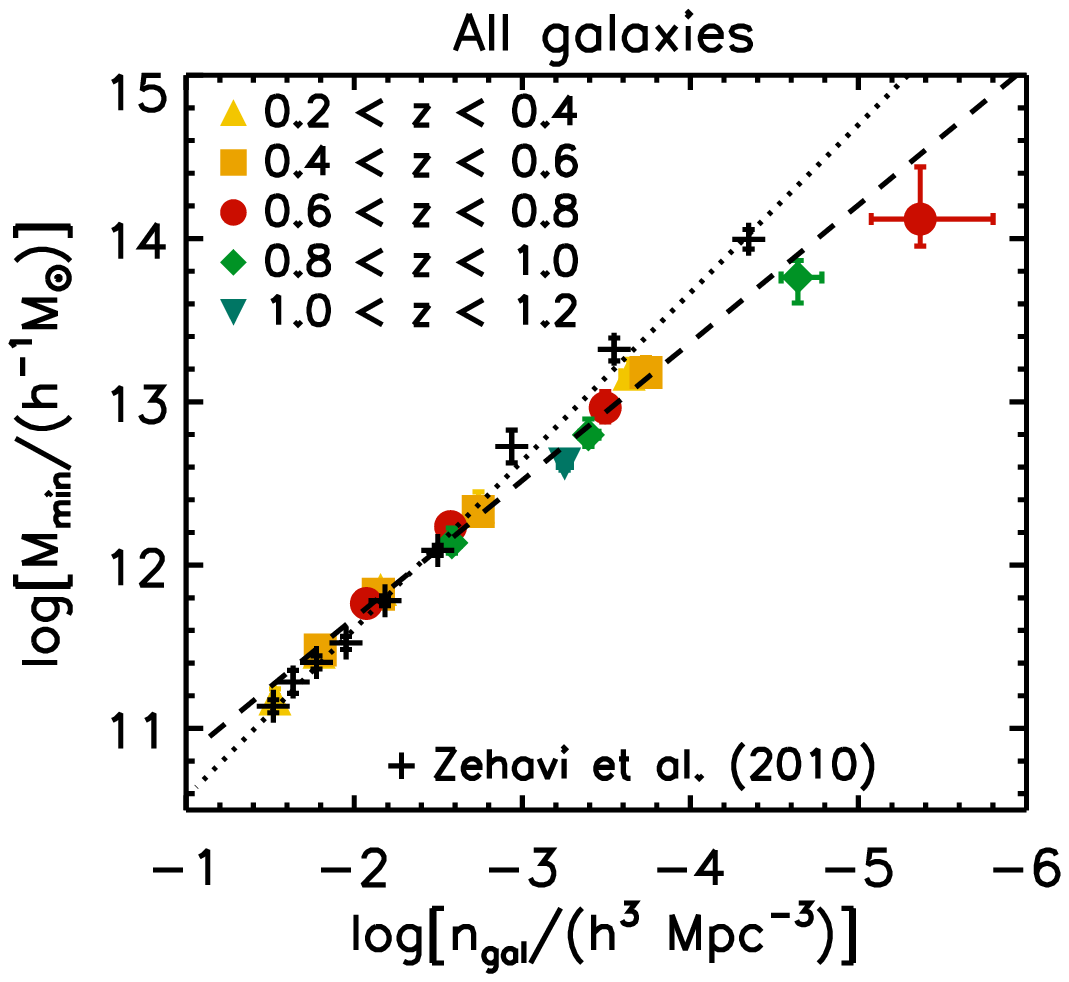}
    \includegraphics[width=0.49\textwidth]{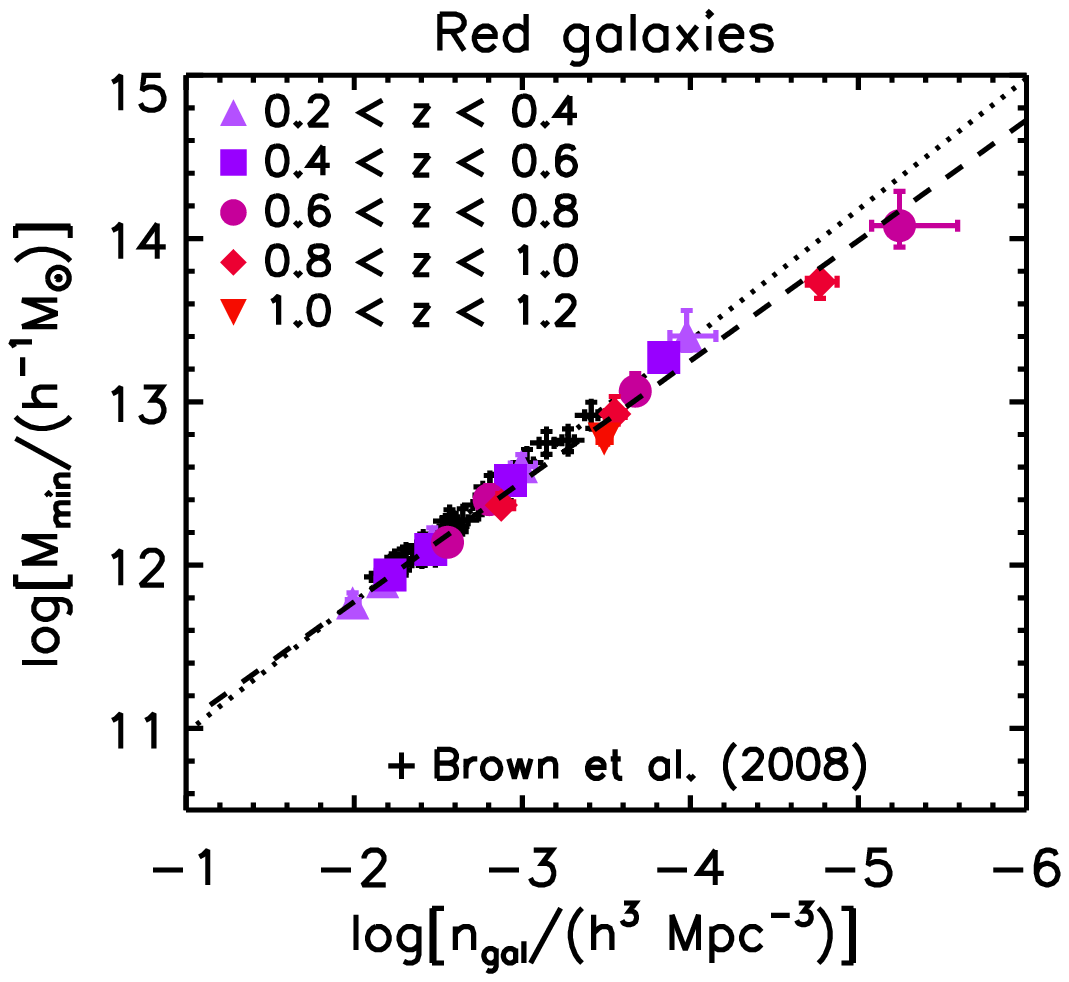}
  \end{center}
  \caption{$M_{\rm min}$ as a function of galaxy number density for all
    (left) and red (right) galaxies.  The dashed line in each panel
    represents a power law fit to the data, and the dotted line
    represents a power law fit to the results given in
    \cite{2008ApJ...682..937B} (red galaxies) and in the SDSS
    by \cite{2010arXiv1005.2413Z} (all galaxies). We converted
      $M_{200}$ halo masses to virial masses $M_{vir} $ when necessary.}
  \label{fig:Ngal}
\end{figure*}

Figure~\ref{fig:Ngal} shows the relationship between $M_{\rm min}$ and
the galaxy number density $n_{\rm gal}$ for red and full samples. As
one would expect, progressively rarer objects are found in more
massive haloes.  We do not detect a significant evolution of the
slope of our $M_{\rm min}$ versus $n_{\rm gal}$ relationship with
redshift. For red galaxies, we overplot the measurements from
\cite{2008ApJ...682..937B} with which we are in excellent agreement,
even though it covers a much smaller area, For the full galaxy
sample we compare with SDSS
measurements \citep{2010arXiv1005.2413Z}. The agreement is good at
high surface densities; however we note a departure for low density
samples. This difference suggests that at constant density, galaxies
tend to reside in more massive haloes at low redshift ($z\sim0$)
than at higher redshift ($z>0.6$). Part of this difference could be
caused by systematic photometric redshift errors.  As we discussed
in Sec~\ref{sec:photozerr}, bright samples may be contaminated by
fainter objects, and the combined effect of higher density and lower
clustering signal will result in reducing $M_{\rm min}$. In the most
pessimistic case outlined in Sec~\ref{sec:photozerr}, where the
density estimate would be overestimated by a factor 3, $\log M_{\rm
  min}$ would be biased low by a few sigmas (up to $\sim0.5$).

The relationship between $M_{\rm min}$ and
$n_{\rm gal}$ can be well approximated by a power law and we 
fitted galaxy samples at all redshifts simultaneously.
We found:
\begin{equation}
  M_{\rm min}^{\rm all} = 10^{10.0} \times [n_{\rm gal}^{\rm all}/(h^3\,{\rm Mpc^{-3}})]^{-0.84} \, h^{-1} \, M_{\sun} \, ,
\end{equation}
for all galaxies and
\begin{equation}
  M_{\rm min}^{\rm red} = 10^{10.3} \times [n_{\rm gal}^{\rm red}/(h^3\,{\rm Mpc^{-3}})]^{-0.74} \, h^{-1} \, M_{\sun} \, ,
\end{equation}
for red galaxies.

\section{Discussion}
\label{sec:discussion}

The key issue we would like to address is to understand how galaxy
formation and evolution depend on the properties of the underlying
dark matter haloes.  Ideally, to relate halo masses to their stellar
content, one would select our samples by stellar mass. However,
stellar mass estimates computed from our five-band optical data
alone would suffer from large uncertainties. Keeping this limitation in
mind, we adopt an intermediate approach where we determine an empirical
relation to convert the observed luminosity threshold of each
sample to an approximately mass-selected sample using a reference
sample with accurate stellar masses. This method allows to transform
luminosities into masses and provides an estimate of the likely
errors in such a transformation.

\subsection{Transforming to stellar mass threshold samples}
\label{sec:masstol}

As a consequence of passive stellar evolution, the characteristic
rest-frame $M_g$-band luminosity of a given galaxy type evolves
significantly over the redshift range $0.2 < z < 1.2$
\citep{2005A&A...439..863I,2006A&A...455..879Z,2007ApJ...665..265F}.

\begin{figure*}
  \begin{center}
    \includegraphics[width=0.49\textwidth]{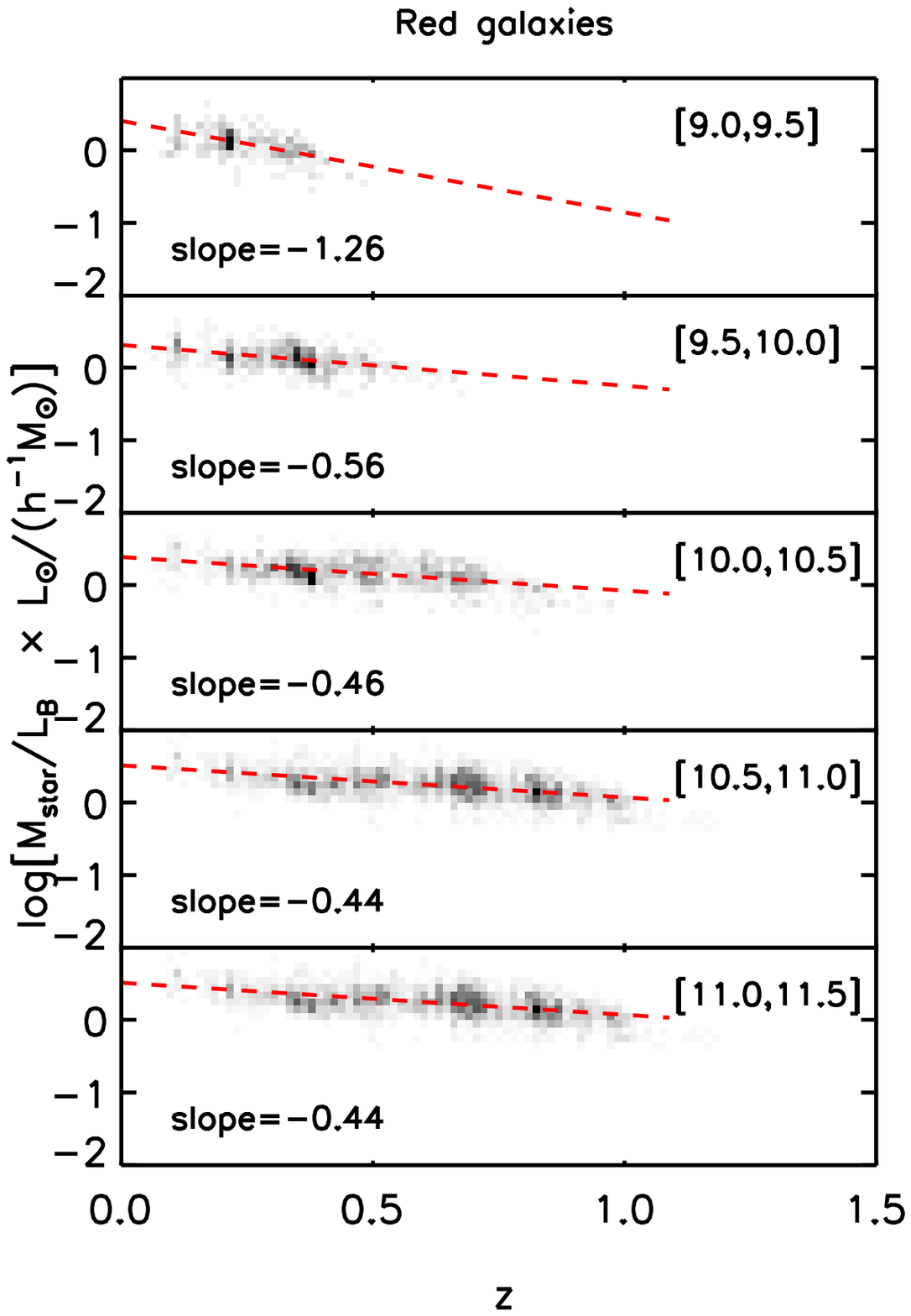}
    \includegraphics[width=0.49\textwidth]{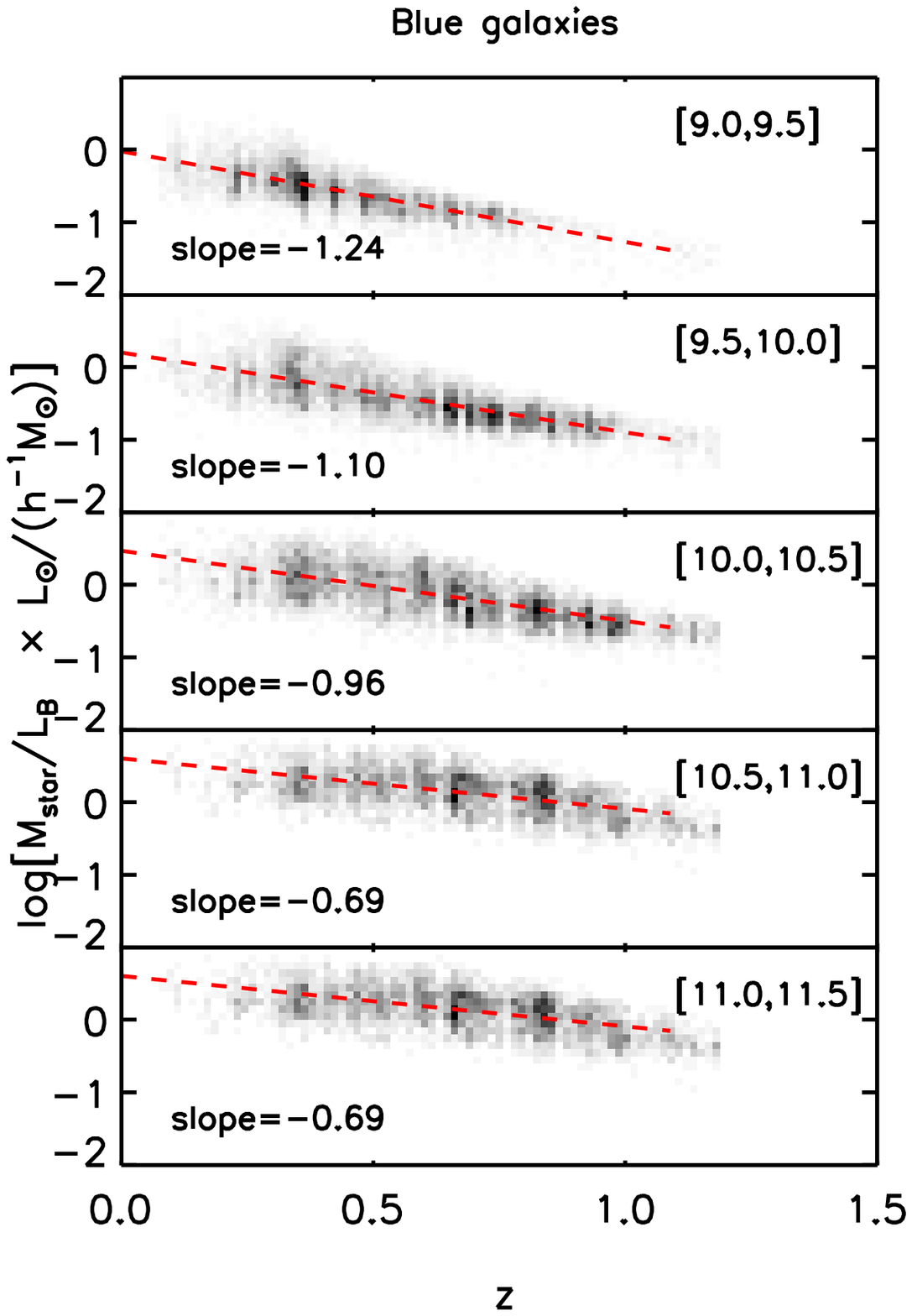}
  \end{center}
  \caption{Stellar mass-to-luminosity relations for red (left) and blue
    (right) galaxies, as a function of stellar mass and redshift as
    measured in the COSMOS survey \citep{2009ApJ...690.1236I}. Each
    panel shows a separate slice in stellar mass. Samples are fitted
    with a power law (dashed line).}
   \label{fig:ML}
 \end{figure*}

In order to account for these effects, we have established a simple
approximation to relate stellar mass and luminosity based on COSMOS
30-band photometry~\citep{2010ApJ...709..644I}. We used the COSMOS
$B-$band luminosities in order to be as close as possible to our
$M_g$-band selected samples.  Red galaxies are selected as $M(NUV)-M_R >
3.5$ and blue galaxies as $M(NUV)-M_R\le 3.5$. We then fit the stellar
mass-to-luminosity ratio with power-law functions 
as function of redshift between $z=0$ and
$z=1.5$ for five stellar mass bins from $10^{9.0}$ to
$10^{11.5}$. These results are displayed in Fig.~\ref{fig:ML}.

Due to the ageing of stellar populations, the average stellar
mass-to-light ratio increases with time. For red galaxies, the slope
depends weakly on stellar mass at intermediate masses ($10^{9.5} -
10^{11.5} h^{-1} M_{\sun}$) and ranges from $-0.56$ to $-0.44$. The
slope is steepest in the lowest mass bin, but the number of objects in
this bin is small. Based on these results we adopt the following
relationship for the red galaxy population:
\begin{equation}
  \label{eq:corrred}
  \log \left( \frac{M_{\rm star}}{L_B} \right )_{\rm red} =  \log
  \left ( \frac{M_{\rm star}}{L_B}
  \right ) _{{\rm red, }z=0} -(0.5\pm0.1)z \, ,
\end{equation}
where the $0.1$ error accounts for the scatter between mass bins.
This evolution is consistent with the luminosity function
evolution measured for red galaxies in the DEEP2 and COMBO17
surveys \citep{2007ApJ...665..265F,2006ApJ...647..853W}, where 
the characteristic absolute magnitude $M_B^{\ast}$ decreases by
about 1.2--1.3 mag per unit redshift ($M_B \propto -2.5\log L_B $).

For the blue sample, the evolution in redshift is more pronounced than
for the red sample, as also seen in DEEP2 \citep{2006ApJ...647..853W}
and VVDS \citep{2006A&A...455..879Z}.  Unfortunately, in this case,
the slope depends on stellar mass, ranging from $-0.69$ to $-1.10$ in
the mass range $[10^{9.5},10^{11.5}]$, and has a larger scatter.  The
full sample is dominated by blue galaxies at faint luminosities but by
red galaxies at bright luminosity (see Fig.~\ref{fig:colors}).
Therefore a simple relation between luminosity and stellar mass valid
for all luminosities cannot be obtained. For the full sample we choose
to apply the same correction as for the red sample. As blue galaxies
evolve more rapidly than red ones such a correction underestimates the
expected evolution of the full samples at faint luminosity and
therefore cannot be used to construct stellar-mass selected
samples. In the following sections, we discuss how our results depend
on this correction.

In the following sections we use these expressions to calculate
``corrected'' luminosities, denoted as $L'$.  Note that these
corrections are equivalent to shifting observed galaxy luminosities to
redshift $z=0$.  Our results are displayed as a function of relative
luminosity threshold $L'/L_{\ast}$, where $L_{\ast}$ corresponds to
$M^{\ast}_g - 5\log h = -19.81$ and $-19.75$ for the full and red
samples, respectively, measured in the local Universe \citep[and
references therein]{2007ApJ...665..265F}.  It is also worth mentioning
that if galaxies would experience a pure passive luminosity evolution
with cosmic time, this would result in constant HOD parameters for a
given $L'/L_{\ast}$ as function of redshift. However, deduced
parameters (galaxy bias, mean halo mass and satellite fraction) are
expected to vary due to the evolving halo mass function and halo bias
(see Eqs.~\ref{eq:bias}--\ref{eq:fsat}).

 \subsection{The stellar-to-halo mass relationship}
 \label{sec:lighttomass}

\begin{figure*}
  \begin{center}
    \includegraphics[width=0.49\textwidth]{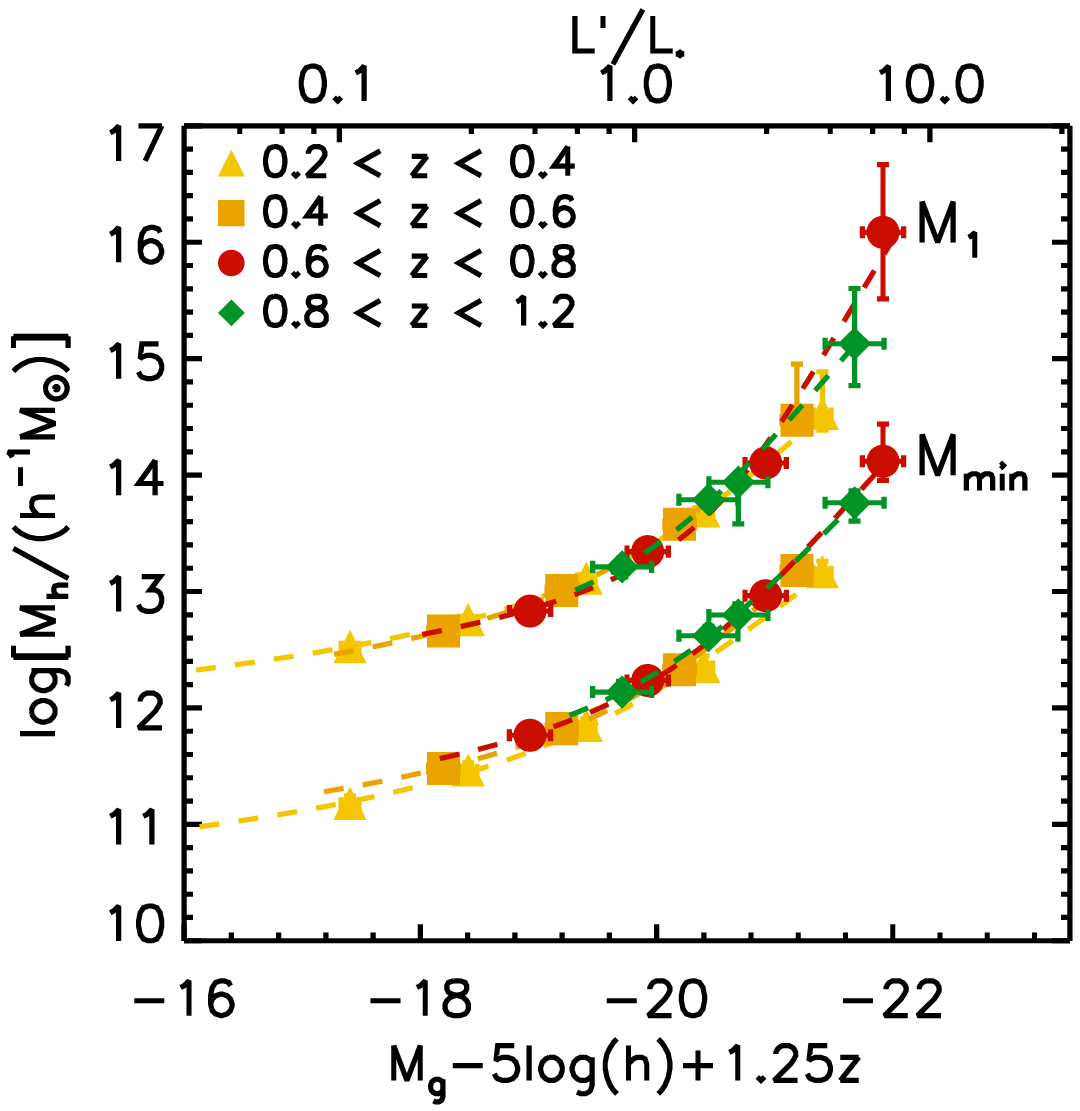}
    \includegraphics[width=0.49\textwidth]{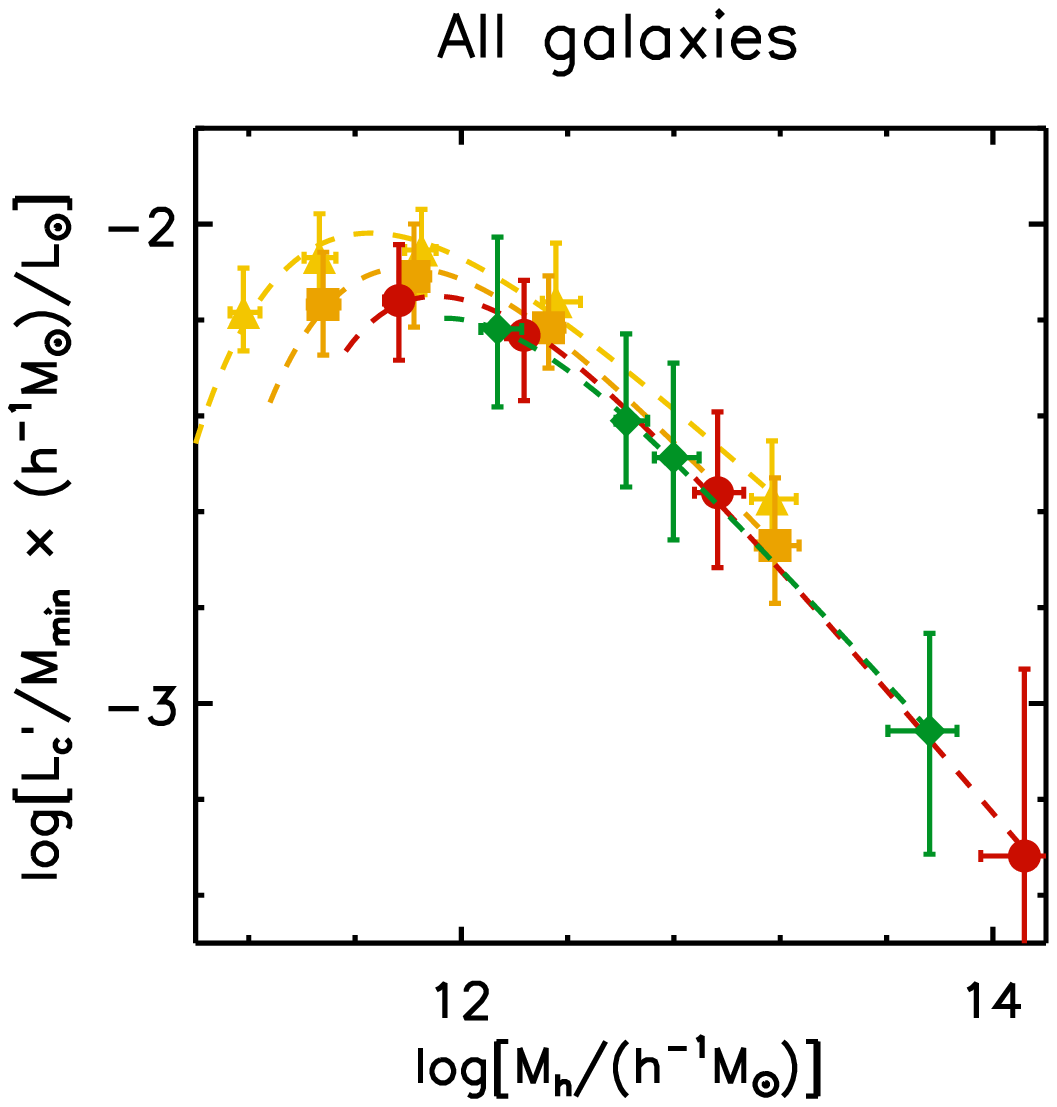}
    \includegraphics[width=0.49\textwidth]{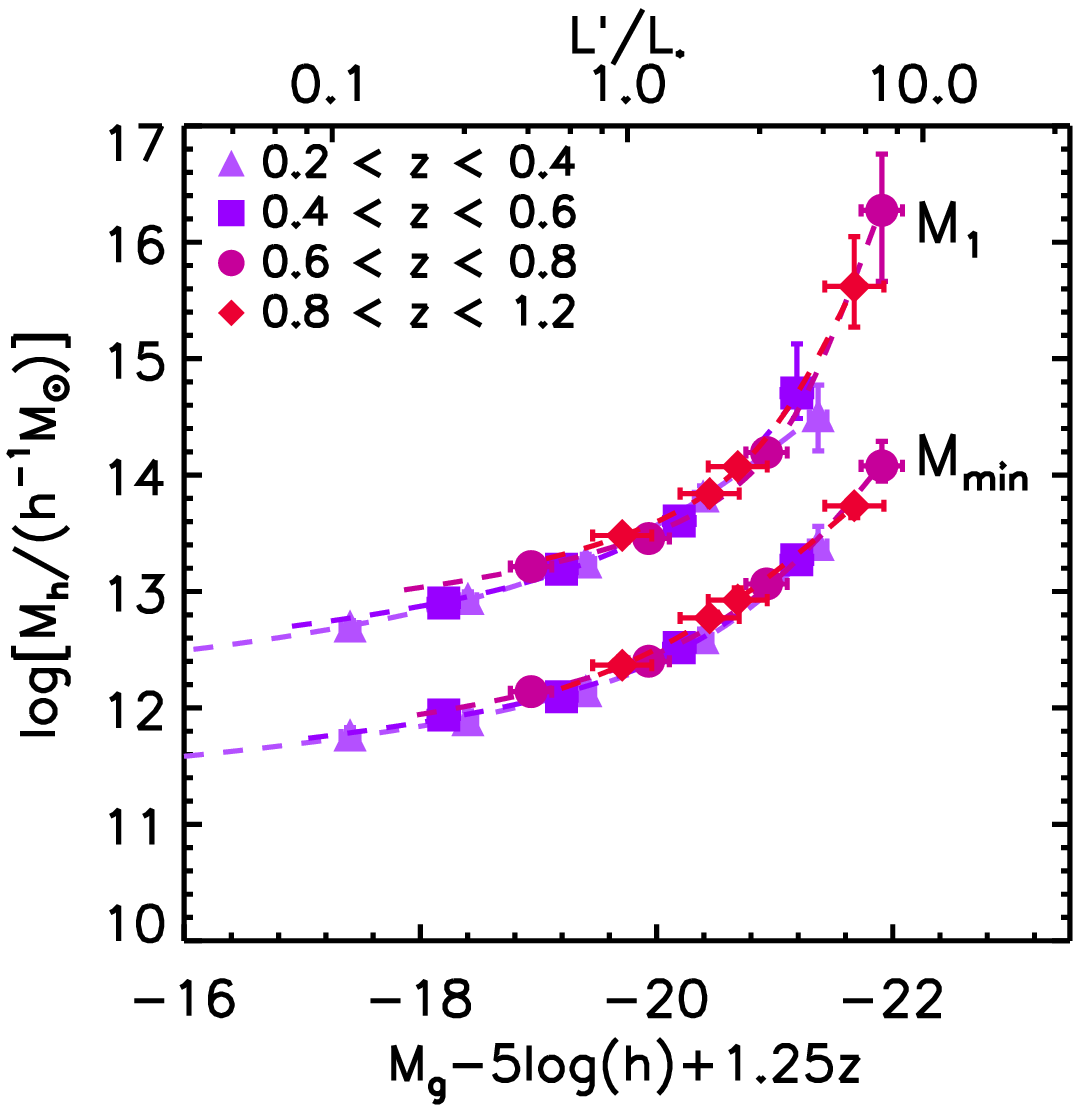}
    \includegraphics[width=0.49\textwidth]{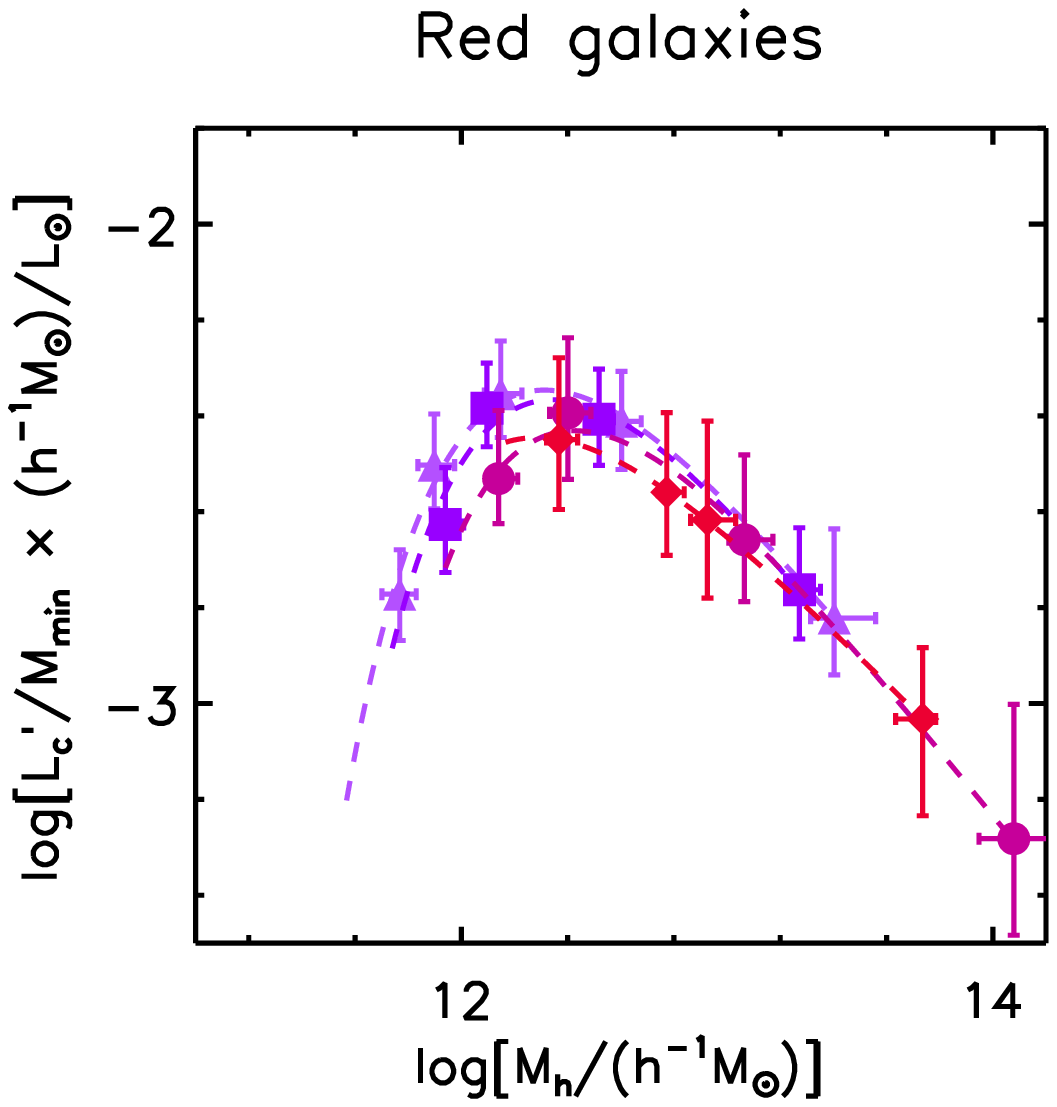}
  \end{center}
  \caption{Left panels: halo mass estimates $M_{\rm min}$ and $M_{1}$
    for all (top) and red (bottom) galaxy samples, as function of
    luminosity threshold, corrected for passive redshift evolution to
    approximate stellar mass selected samples.  The dashed lines
    correspond to Eq.~\ref{eq:zehavi}.  Right panels: light-to-halo
    mass ratios $L'_{\rm c}/M_{\rm min}$ (Eq.~\ref{eq:zehaviM_L} with 
    identical parameters as those fitted with Eq.~\ref{eq:zehavi}) as
    function of halo mass.}
  \label{fig:M1_Mmin}
\end{figure*}

In our model, $M_{\rm min}$ represents the hosting halo mass for
central galaxies whose median luminosity equals the luminosity
threshold of the sample (see Sect.~\ref{sec:halo-occup-distr}). The
ratio $L'/M_{\rm min}$ is therefore the ratio of luminosity to dark
matter mass for central galaxies. \cite{2010arXiv1005.2413Z} showed
that the relation between central galaxy luminosity $L'_{\rm c}$
(corresponding to our $L'$) and the halo mass could 
be approximated by:
\begin{equation}
  \label{eq:zehavi}
  \frac{L'_{\rm c}}{L_\ast} = A \left ( \frac{M_{\rm h}}{M_{\rm t}} \right )
  ^{\alpha_M} \exp \left ( -\frac{M_{\rm t}}{M_{\rm h}} + 1 \right ) \, ,
\end{equation}
where A, $\alpha_M$ and $M_{\rm t}$ are free parameters and $M_{\rm
  h}$ represents $M_{\rm min}$ or $M_1$. 

This expression encapsulates the idea that there exists a ``transition
halo mass'' ($\sim ~ M_t$) where the ratio of central galaxy
luminosity to dark matter halo mass reaches a maximum.  This
transition mass represents the halo mass for which baryons have been
most efficiently converted into stars until the time of observation,
resulting from past and on-going star formation over cosmic history.

We fit Eq.~\ref{eq:zehavi} for full and red samples for each redshift
slice. For our two highest redshift bins where the number of points are
less than the number of free parameters, we perform a simultaneous fit
over both redshift bins. Results are plotted in Fig.~\ref{fig:M1_Mmin}
and best-fitting parameters given in Table~\ref{tab:zehavi}.

In our lowest redshift bin ($0.2 < z < 0.4$), our full sample has
best-fitting parameters slightly different than in the SDSS
\citep{2007ApJ...667..760Z,2010arXiv1005.2413Z}, $A=0.20 \pm 0.01$
(compared to 0.32) and $M_{\rm t}=2.24 \pm 0.1\times 10^{11} h^{-1}
M_{\sun}$ (compared to $3.08\times 10^{11} $). We note that the
difference is likely due to the different selection ($M_r$
in the SDSS). At high luminosities, the difference with their relation 
increases. For a given luminosity, we find lower halo
masses, $L_{\rm c} \propto M^{0.5\pm0.01}$, compared to 
$L_{\rm c} \propto M^{0.26-0.28}$, which follows the trends observed 
in Fig.~\ref{fig:Ngal}, and might be due to contaminated samples at 
low density. In the red sample, we measure $L_{\rm c}\propto
M^{0.31\pm0.16}$, whereas \cite{2009ApJ...707..554Z} found $L_{\rm
  c}\propto M^{0.5}$ for luminous red galaxies in the SDSS.
Both $M_{\rm min}$ and $M_1$ increase with redshift for the full
sample. The dependence with redshift is 
weak for red galaxies, although we detect a
slight increase in the very highest redshift bins.

For the full sample we assumed a correction based on the evolution of
the stellar mass-to-luminosity ratio of red galaxies
(Eq.~\ref{eq:corrred}). As the characteristic luminosity of blue
galaxies evolves more rapidly than red ones and as the faint sample is
dominated by blue galaxies a legitimate concern is the dependence of
our results on our corrections to stellar mass.  Several corrections
assuming the blue galaxy luminosity evolution (using slopes given in
Fig.~\ref{fig:ML}) were tested. If a larger correction is applied,
$M_{1}$ and $M_{\rm min}$ show a similar but more pronounced
trend. For this reason we conclude that the increase of halo mass
parameters with redshift for the full sample is a robust result
(although the exact amount of evolution observed depends on the
correction we apply).

\subsection{Redshift evolution of the SHMR and comparison with literature values}
\label{sec:redsh-evol-shm}

\begin{figure*}
  \begin{center}
    \includegraphics[width=\textwidth]{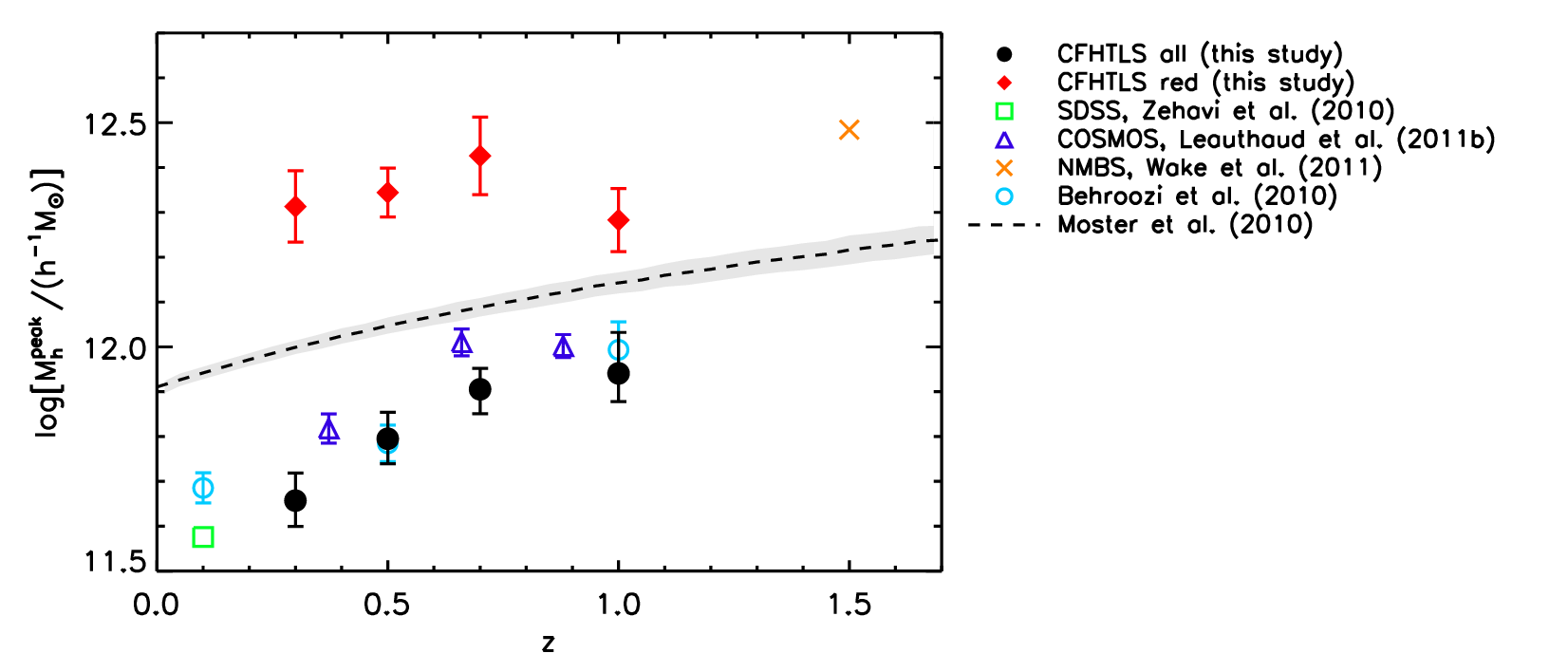}
  \end{center}
  \caption{Position of the peak $M^{\rm peak}_{\rm h}$ 
    corresponding to $\max(L'_{\rm c}/M_{\rm min})$ for the full
    (filled circles) and red (filled diamonds) samples as function of
   redshift. Error bars are derived from the nearest data point to the
   peak. We compare our results with measurements in SDSS
   \citep{2010arXiv1005.2413Z}, in
   COSMOS \citep{2011arXiv1104.0928L} and in the NEWFIRM Medium Band Survey
   \citep[NMBS,][]{2011ApJ...728...46W}. We convert
   $M_{200}$ halo masses (corresponding to a sphere 
   with overdensity 200 times the background density) 
   into virial masses $M_{\rm vir}$ (our definition) using the method 
   described in Appendix C of \cite{2003ApJ...584..702H}. We also
   display numerical predictions based on abundance-matching
   method from \cite{2010ApJ...717..379B} and
   \cite{2010ApJ...710..903M}, where the shaded region shows the
   1$-\sigma$ uncertainty.}
 \label{fig:Mmin_0}
\end{figure*}

To study the redshift evolution of the stellar-to-halo mass ratio, we
consider the relationship between corrected galaxy luminosity and the
host halo mass from Eq.~\ref{eq:zehavi}:
\begin{equation}
  \label{eq:zehaviM_L}
  \frac{L'_{\rm c}}{M_{\rm h}} = A \,\frac{L_{\ast}}{M_{\rm t}}  \left ( \frac{M_{\rm h}}{M_{\rm t}} \right )
  ^{\alpha_M-1} \exp \left ( -\frac{M_{\rm t}}{M_{\rm h}} + 1 \right ) \, ,
\end{equation}
where the maximum is located at:
\begin{equation}
  \label{eq:zehavi_peak}
  M^{\rm peak}_{\rm h} = \frac{M_{\rm t}}{1-\alpha_M} \, .
\end{equation}

The right panels of Fig.~\ref{fig:M1_Mmin} show the $L'_c/M_{\rm min}$
relation for full and red samples as function of halo mass fitted by
Eq.~\ref{eq:zehavi}. Best fitting parameters are listed in
Table~\ref{tab:zehavi}. For less massive haloes, the relative stellar
mass content increases with halo mass and reaches a maximum around
$M_{\rm min} \sim 10^{12} h^{-1} M_{\sun}$. In more massive haloes,
the stellar-to-halo mass ratio sharply decreases. Our measurements are
consistent with recent observations measuring the stellar-to-halo mass
ratio as function of halo mass
\citep{2006MNRAS.368..715M,2007ApJ...667..760Z,2010MNRAS.406..147F,
  2010arXiv1005.2413Z,2011MNRAS.410..210M,2011ApJ...728...46W,2011arXiv1104.0928L},
and ``abundance-matching'' techniques based on $N$-body simulations
\citep{2009ApJ...696..620C,2010ApJ...710..903M,2010ApJ...717..379B}. Similar
trends are found in semi-analytic simulations \citep[see, for
e.g.,][]{2008MNRAS.391..481S}, in which physical processes (such as
supernovae and AGN feedback) preventing gas from cooling and quenching
star formation must be taken into account to match observations.

Our measured peak values $M^{\rm peak}_{\rm h}$  (Eq.~\ref{eq:zehavi_peak},
with error bars from the nearest data points) are plotted as a function
of redshift in Fig.~\ref{fig:Mmin_0}.  Comparing to literature
measurements at lower redshifts, the peak stellar-to-halo mass
relation is $M^{\rm peak}_{\rm h}  = 4.54\times10^{11} h^{-1} M_{\odot}$,
similar to the local Universe value \citep[$\approx 4.2\times10^{11}
h^{-1} M_{\odot}$, ][]{2010arXiv1005.2413Z}.  For our full sample, the
peak location gradually increases with redshift.

At higher redshifts, we compare our measurements with those of
\cite{2010ApJ...710..903M} and \cite{2010ApJ...717..379B}. These
authors derived the stellar-to-halo mass relation using an
abundance-matching technique, linking dark matter halo mass functions
from $N$-body simulations with observed galaxy stellar mass
functions. \cite{2010ApJ...710..903M} estimate the redshift evolution
of the stellar-to-halo mass ratio (their Table~7) and from this we
derived analytically the position of the peak, as well as associated
errors. Values from \cite{2010ApJ...717..379B} at redshift 0.1, 0.5
and 1.0 were supplied to us (P. Behroozi, private communication). Both
\citeauthor{2010ApJ...710..903M} and
\citeauthor{2010ApJ...717..379B}'s measurements are based primarily on
galaxy stellar mass measurements covering relatively small areas (a
few hundred square arcminutes). Our full sample is in excellent
agreement with \citeauthor{2010ApJ...717..379B} but differs
significantly from \citeauthor{2010ApJ...710..903M}, although in both
cases the two data sets show the same general trend: $M^{\rm peak}_{\rm h}$ shifts
gradually towards higher halo masses at higher redshift. The
\citeauthor{2010ApJ...710..903M} points do not seem to agree very well
with most observations and moreover overestimate the position of the
peak at low redshift.

Measurements in the COSMOS field are also shown
\citep{2011arXiv1104.0928L}; their points are in general larger than
most measurements, with the exception of
\citeauthor{2010ApJ...710..903M}. It has already been noted that there
are several rich structures in the COSMOS field
\citep{2007ApJS..172..314M,2009A&A...505..463M,2010MNRAS.409..867D}
 which lead to higher correlation functions, especially at $z\sim0.8$; cosmic variance could
perhaps be the origin of this discrepancy.

For red galaxies, the peak is at larger halo masses,
$M^{\rm peak}_{\rm h} =20.6\times10^{11} h^{-1} M_{\odot}$. This is not surprising
as $M_{\rm min}$ is larger for red galaxies at faint luminosities. The
red galaxy peak position undergoes only little evolution to $z\sim1$.  Our
intermediate-redshift red galaxy points are consistent with higher
redshift measurements \citep{2011ApJ...728...46W}; these authors used
a near-infrared survey to select galaxies by stellar mass, and we
would expect their selection to be dominated by 
red galaxies.

The results presented here represent the first time a single data set
covering a statistically significant area has been used to derive the
evolution of the peak position as a function of redshift and galaxy
type in a self-consistent manner.  The peak position, $M^{\rm
  peak}_{\rm h}$, can be interpreted as representing the halo mass at
which the stellar mass content is most efficiently accumulated in
haloes, either by star formation or merger processes.  The movement of
this peak towards higher halo masses at higher redshifts is consistent
with a picture in which the sites of efficient stellar mass growth
migrate from low-mass to high-mass haloes at higher redshift.  This
``anti-hierarchical'' evolution, frequently referred to as ``halo
downsizing'', resembles closely the scenario first sketched by
\cite{1996AJ....112..839C}, in which the maximum luminosity of
galaxies undergoing star-formation declines steadily to the current
day. Conversely, the fact that we observe constant transition masses
for the red population suggests that the stellar-to-halo mass ratio
does not significantly evolve in haloes with mass $\sim10^{12} h^{-1}
M_{\odot}$ and supports observations that intermediate-mass red
galaxies have experienced no significant stellar mass growth by star
formation since $z\sim1.2$.

We tried several corrections to convert to stellar mass thresholds
based on the slope measured for blue galaxy evolution (see
Fig.~\ref{fig:ML}), and we found that the peak luminosity depends only
weakly on these corrections. Only the highest-redshift point changes
by around one sigma if we apply the extreme ``blue galaxy'' correction
to our full sample. For this reason we conclude that the increase of
$M^{\rm peak}_{\rm h}$ with redshift for the full sample is a robust
result.

\subsection{Redshift evolution of halo satellite fraction}
\label{sec:frsat}

The fraction of galaxies which are satellites in a given dark matter
halo is a sensitive probe of the past evolutionary history of the halo
which may be modified by processes such as major mergers (which can
decrease the satellite fraction with time) or environmental effects
which operate primarily to reduce the number of satellites, such as,
but not limited to, the restriction of gas supply or ``strangulation''
\citep{1980ApJ...237..692L} or ram-pressure stripping
\citep{1972ApJ...176....1G}. Satellite galaxies may also merge with
their central galaxy.  Observational evidence based on pair fraction
measurements and theoretical studies using numerical simulations now
shows that mergers seem to play a significant (although not dominant)
role in the evolution of massive galaxies since $z\sim1$
\citep{2007ApJ...655L..69W,2010MNRAS.402.1796W,2010ApJ...709.1018V,
  2010ApJ...719..844R,2011arXiv1104.0389Z}.  Our objective in this
section is to see if measurements of the satellite fraction can
provide new insights into the evolutionary history of haloes.

\begin{figure*}
   \begin{center}
     \begin{tabular}{c@{}c@{}c@{}}
       \includegraphics[width=0.33\textwidth]{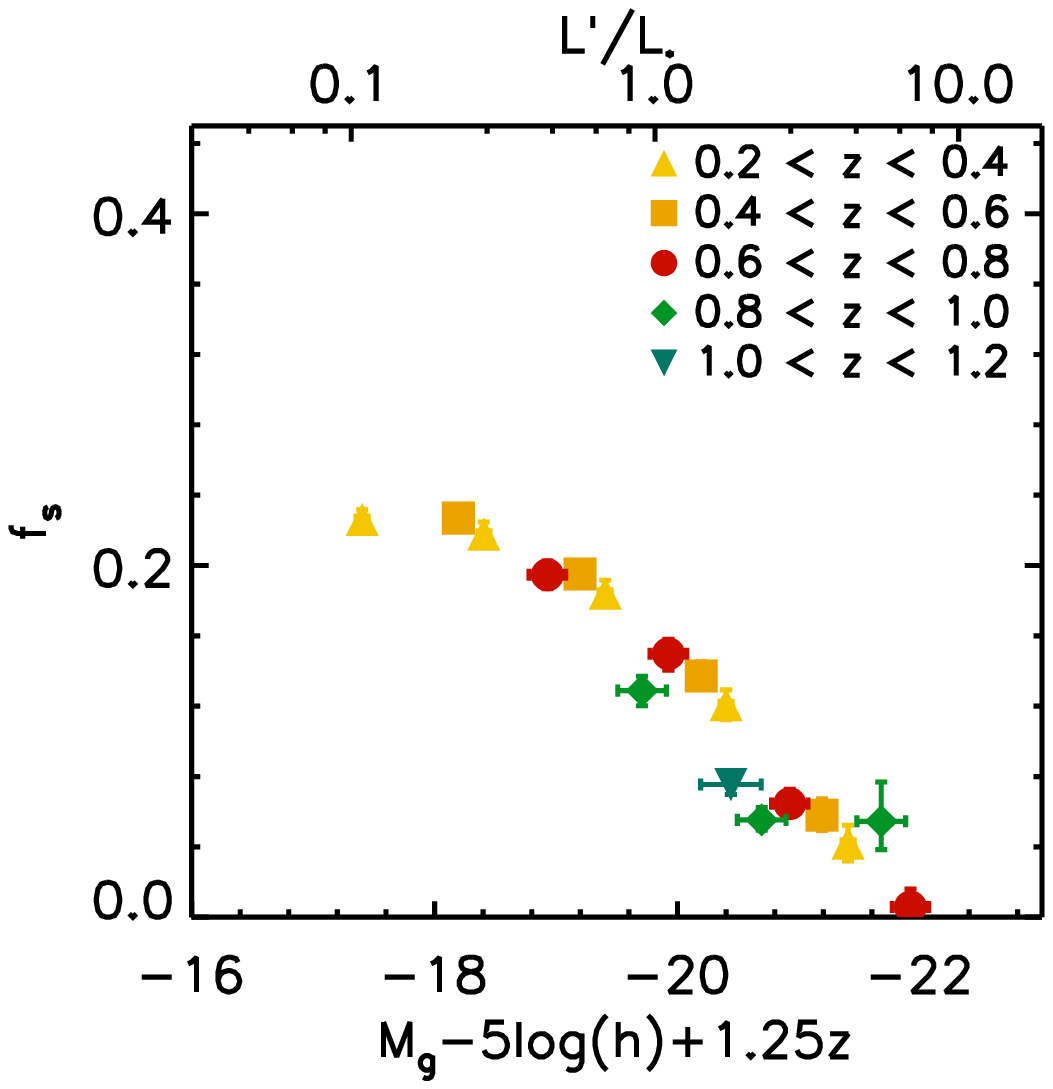}&
       \includegraphics[width=0.33\textwidth]{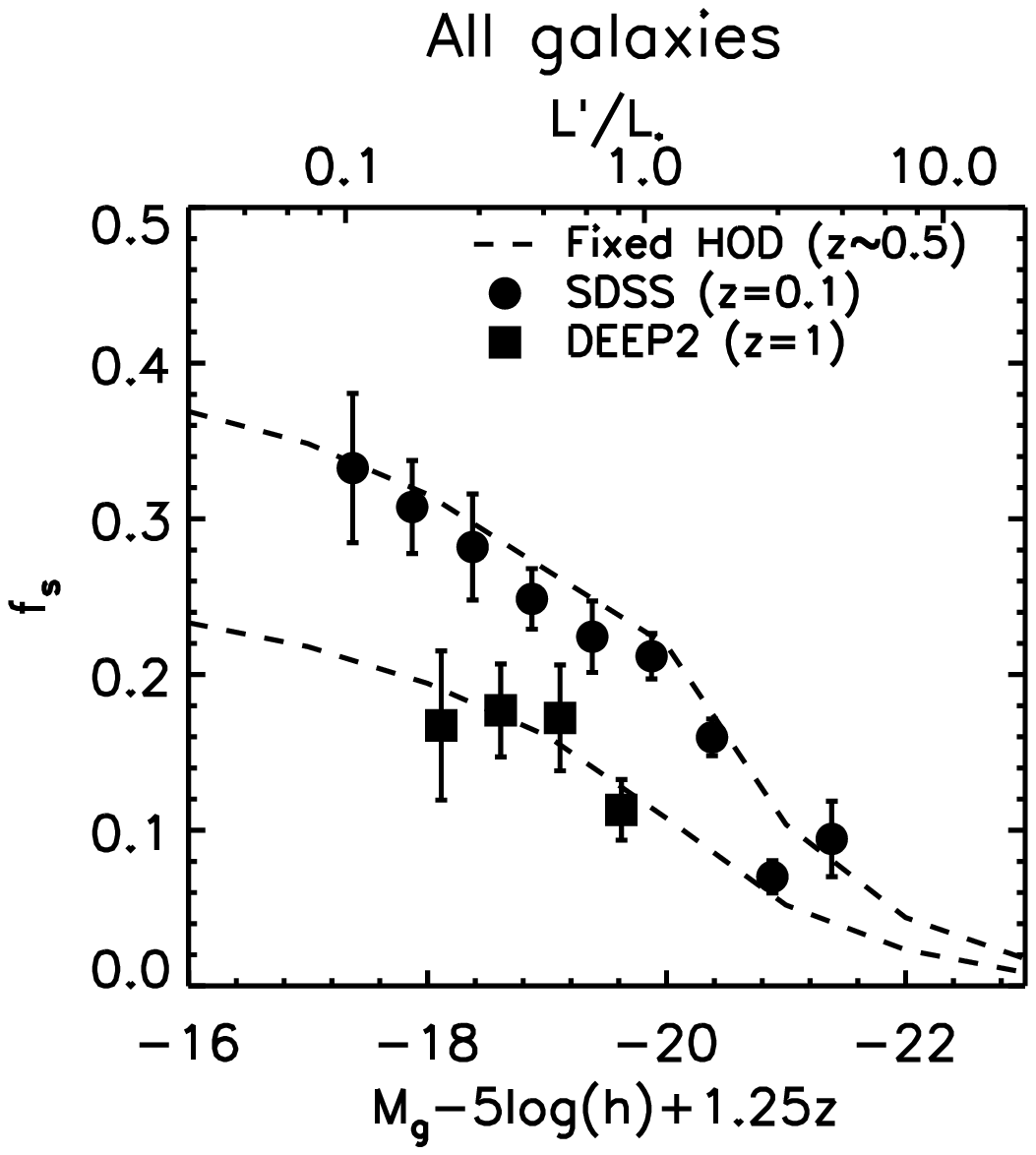}&
       \includegraphics[width=0.33\textwidth]{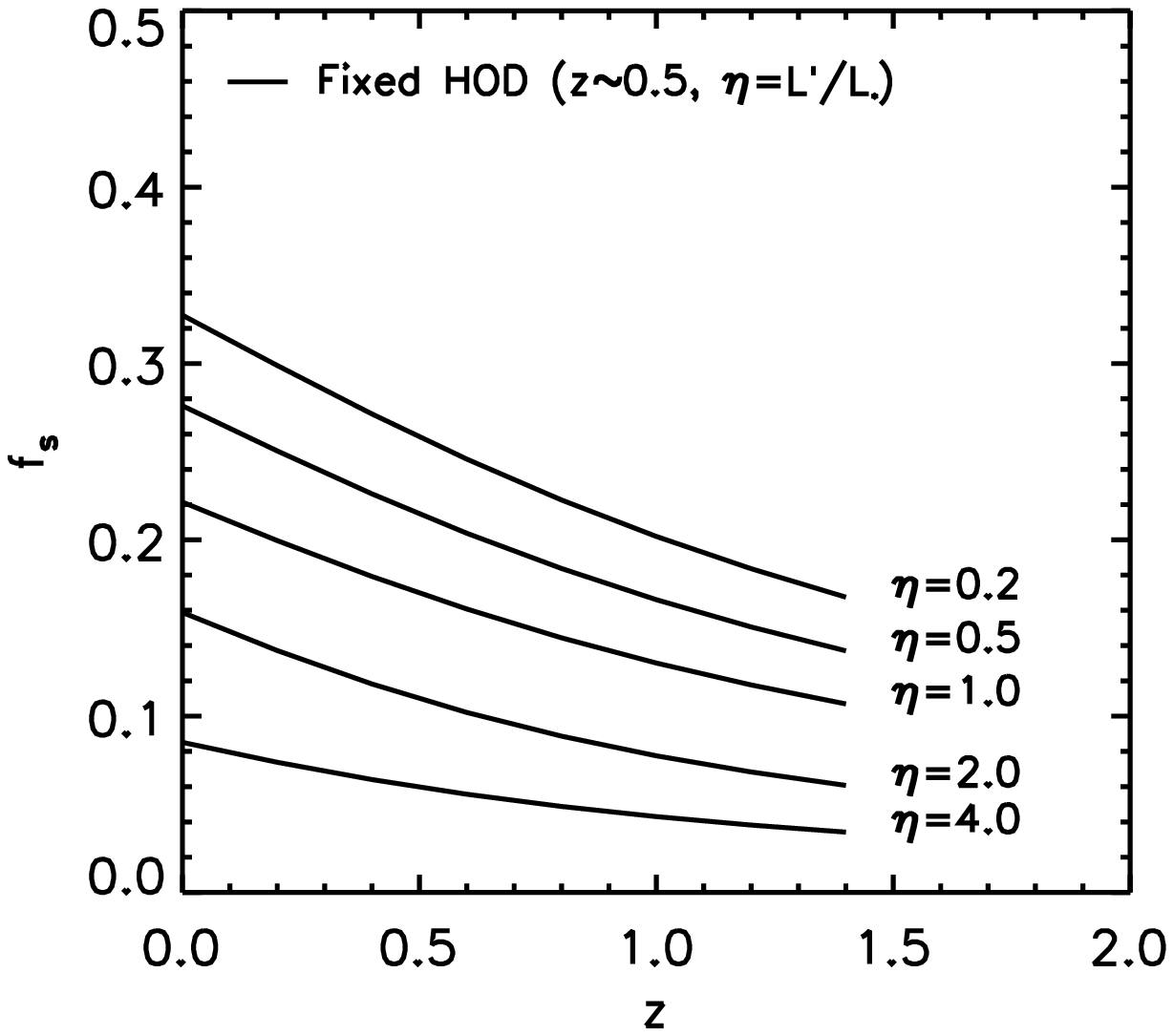}\\
       \includegraphics[width=0.33\textwidth]{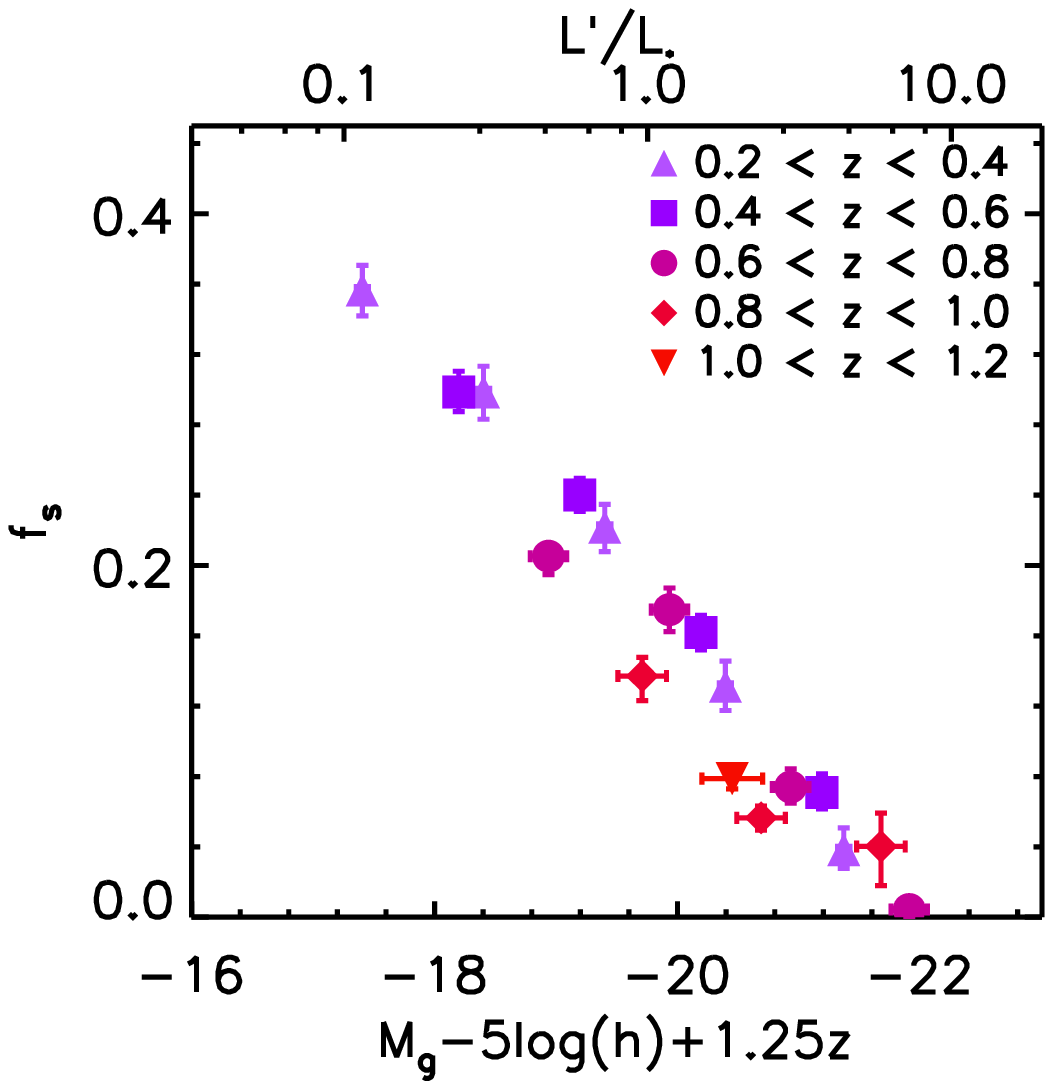}&
       \includegraphics[width=0.33\textwidth]{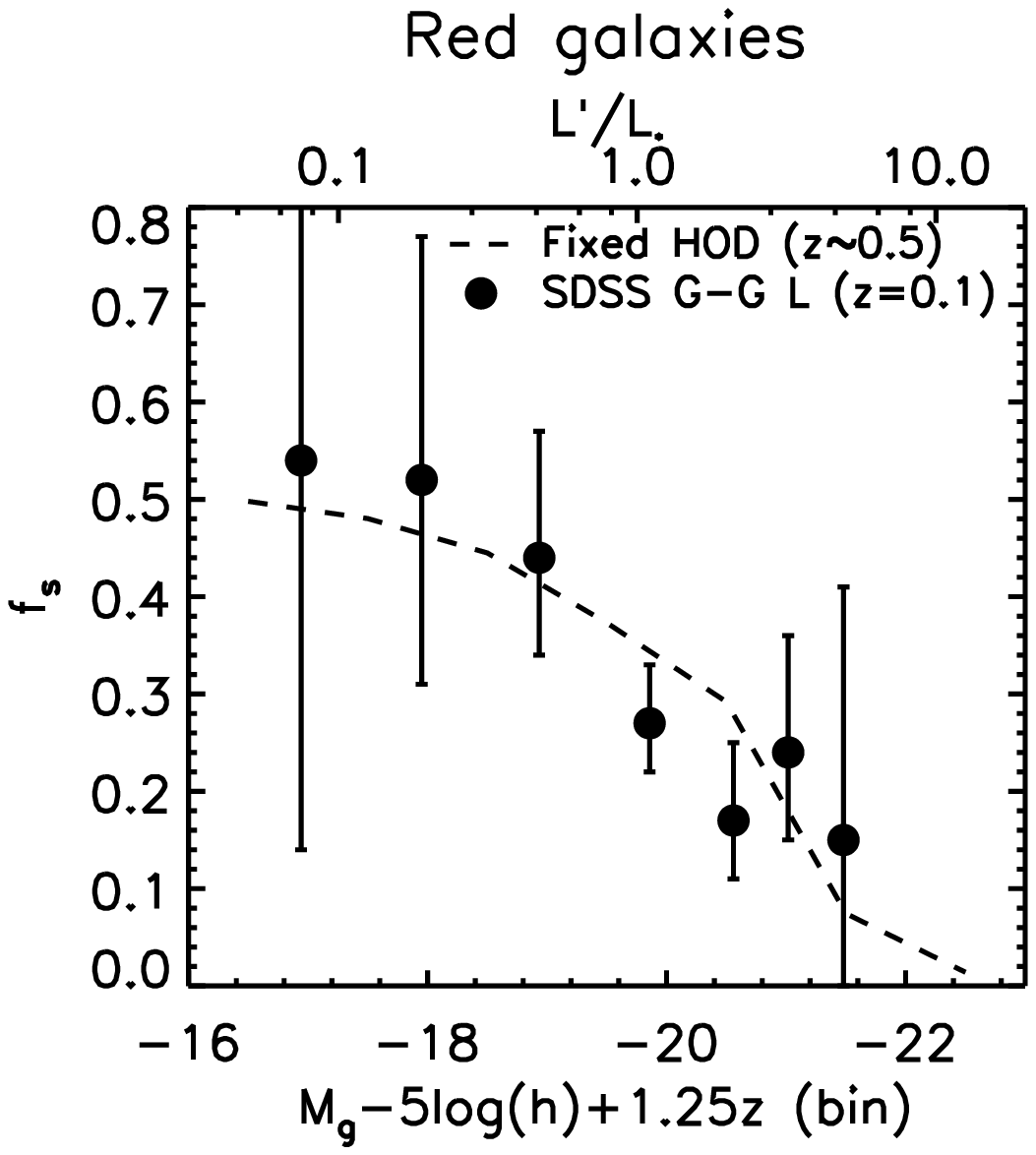}&
       \includegraphics[width=0.33\textwidth]{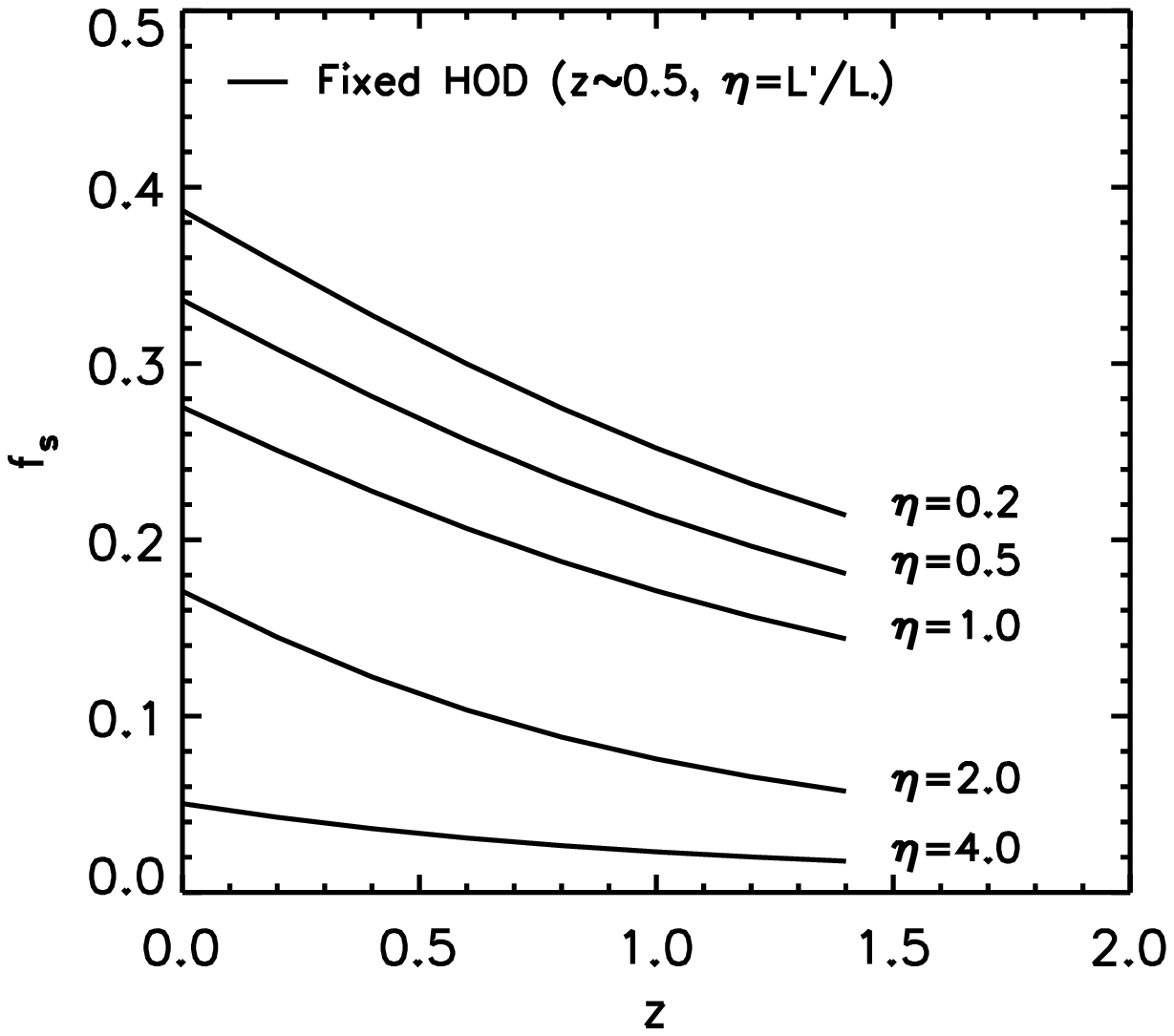}
     \end{tabular}
   \end{center}
   \caption{Satellite fractions as function of luminosity and redshift,
     for all and red samples (top and bottom panels). The left panels
    show the measured satellite fraction as function of corrected
    luminosity. Dashed lines in the middle panels show satellite
    fractions extrapolated in redshift, based on our HOD parameters
    measured in the redshift range $0.4 < z < 0.6$. We compare our
    results with the literature for all galaxies at $z=0.1$ and
    $z=1.0$ \citep[top,][]{2007ApJ...667..760Z} and with galaxy-galaxy
    lensing estimates at $z=0.1$
    \citep[bottom,][]{2006MNRAS.368..715M} for red galaxies. In the
    right panels we provide values extrapolated in redshift for
    several $L'/L_{\ast}$ ratios.}
  \label{fig:frsat}
\end{figure*}

The left panels of Figure \ref{fig:frsat} show the satellite fraction
$f_{\rm s}$, computed using Eq.~\ref{eq:fsat}. As before, each sample
is converted to an approximately mass-limited sample by ``correcting''
for passive luminosity evolution (see Sect.~\ref{sec:masstol}).
At all redshifts, as the luminosity threshold increases, corresponding
to higher mass thresholds, the satellite fractions become
progressively smaller. The similarity between the full and red samples
is simply a consequence of the dominance of red galaxies at
bright luminosity thresholds.

At fainter luminosity thresholds, the satellite fraction rises and
eventually reaches a plateau.  This effect can be understood by
considering both the halo mass function and the halo occupation
function: the increase of the latter with decreasing luminosity
threshold (see Fig.~\ref{fig:example}) is rapidly compensated by a
sharp decrease in massive halo number density, so that faint galaxies
are preferentially central galaxies in smaller haloes rather than
being satellites in more massive haloes.

We compare our results with satellite fractions found in SDSS
($z\sim0.1$) and DEEP2 ($z\sim1$).  To explore a complete range in
redshift and luminosity, we extrapolate our best-fitting HOD
parameters measured in the redshift range $0.4 < z < 0.6$; that is, at
a given luminosity, we keep the HOD parameters fixed and compute
Eq.~\ref{eq:fsat} at different redshifts.  The choice for this
intermediate redshift range is motivated by the fact that it contains
the most robust photometric redshift estimates and our measured points
span a large range in luminosity.  Each HOD parameter set is
constructed from $M_{\rm min}$ and $M_1$ interpolated in luminosity
using Eq.~\ref{eq:zehavi} with best-fitting parameters given in
Table~\ref{tab:zehavi}. For red galaxies, $M_0$ is set to $M_{\rm min}/100$,
$\sigma_{\log M}$ is fixed to 0.3 and $\alpha=1.1$ for $L < L_{\ast}$
or $\alpha=1.4$ otherwise. For all galaxies, $M_0$ is set to $M_{\rm min}/100$,
$\sigma_{\log M}=0.4$ and $\alpha=1.1$ for $L < L_{\ast}$,
and $\sigma_{\log M}=0.3$ and $\alpha=1.3$  for brighter luminosity
thresholds.

In the middle panels of Fig.~\ref{fig:frsat} we compare our
extrapolated satellite fractions for all galaxies (in dashed lines)
with results derived from SDSS and DEEP2, using a HOD analysis similar to ours
\citep{2007ApJ...667..760Z}. Our satellite fractions are consistent
with these measurements at both redshifts, although we slightly
over-predict satellite fractions at $z\sim0.1$ for
samples brighter than $\sim 0.5 L_{\ast}$.
An overall increase of a factor of two is measured
between $z\sim1$ and $z\sim0$ for samples with $L > L_{\ast}$; we
note that this value is lower than the factor of three from $z\sim0.5$
to $z\sim0$ measured in the VVDS \citep{2010MNRAS.406.1306A}.

We fit the HOD parameters at each redshift independently. Therefore,
our model provides us with ``snapshots'' of the occupation function at
various epochs. If any physical process had played a significant role
in reducing the satellite number density between $z\sim1$ and
$z\sim0$, our fixed-HOD model calibrated at $z\sim0.5$ would
over-predict the satellite fraction observed in the local Universe and
under-predict the satellite fraction at redshift one. The excellent
match found for bright luminosity threshold samples shows that our
satellite fraction measurements follow closely a redshift-independent
HOD. Since a redshift evolution of the merging rate would change the
occupation function, we therefore conclude that our measurements imply
a constant merging rate since $z\sim1$, which confirms the constant
(albeit minor) role of merging for intermediate-mass galaxies
\citep[e.g.,][]{2009ApJ...697.1369B}.  As outlined by
\cite{2007ApJ...655L..69W}, if halo merging was not followed by galaxy
merging, we would observe an increase of halo masses at constant
galaxy density. Instead, no redshift evolution of the $M_{\rm
  min}$-$n_{\rm gal}$ relationship (see Fig.~\ref{fig:Ngal}) supports
a scenario with constant galaxy merging with time.

We compare our red satellite fraction with the galaxy-galaxy lensing
analysis from SDSS \citep{2006MNRAS.368..715M}.  These samples were
constructed from luminosity slices, rather than the threshold samples
we use here. We then computed the number of satellites per luminosity
bin as the difference of satellite fractions between two threshold
samples:
\begin{equation}
  f^{\rm bin}_s = \frac{f^{\rm thres2}_s n^{\rm thres2}_{\rm
      gal}-f^{\rm thres1}_s
    n^{\rm thres1}_{\rm gal}}{n^{\rm thres2}_{\rm gal}-n^{\rm thres1}_{\rm gal}}
  \, .
\end{equation}
As we found for the full galaxy sample, the good agreement supports a
non-evolving HOD from $z=0.5$ to $z=0$.

Finally, in the right panels of Fig.~\ref{fig:frsat}, we provide
extrapolated satellite fractions as function of redshift for several
luminosity thresholds. It is interesting to examine our results in the
context of recent results studying the role of satellite galaxies
in galaxy formation and evolution
\citep{2010ApJ...721..193P,2011arXiv1106.2546P}. 
These works show measurements from numerical simulations and 
SDSS group catalogues which indicate
that the fraction of satellites \textit{at a fixed overdensity} is
independent of halo mass and redshift. In this context, the growth of
the satellite fraction we observe is simply a consequence of average
growth in overdensity in rich structures between $z\sim1$ to $z\sim0$,
reflecting the growing importance of environment at low redshift. 

\subsection{Galaxy bias}
\label{sec:bias}

\begin{figure*}
  \begin{center}
    \includegraphics[width=0.49\textwidth]{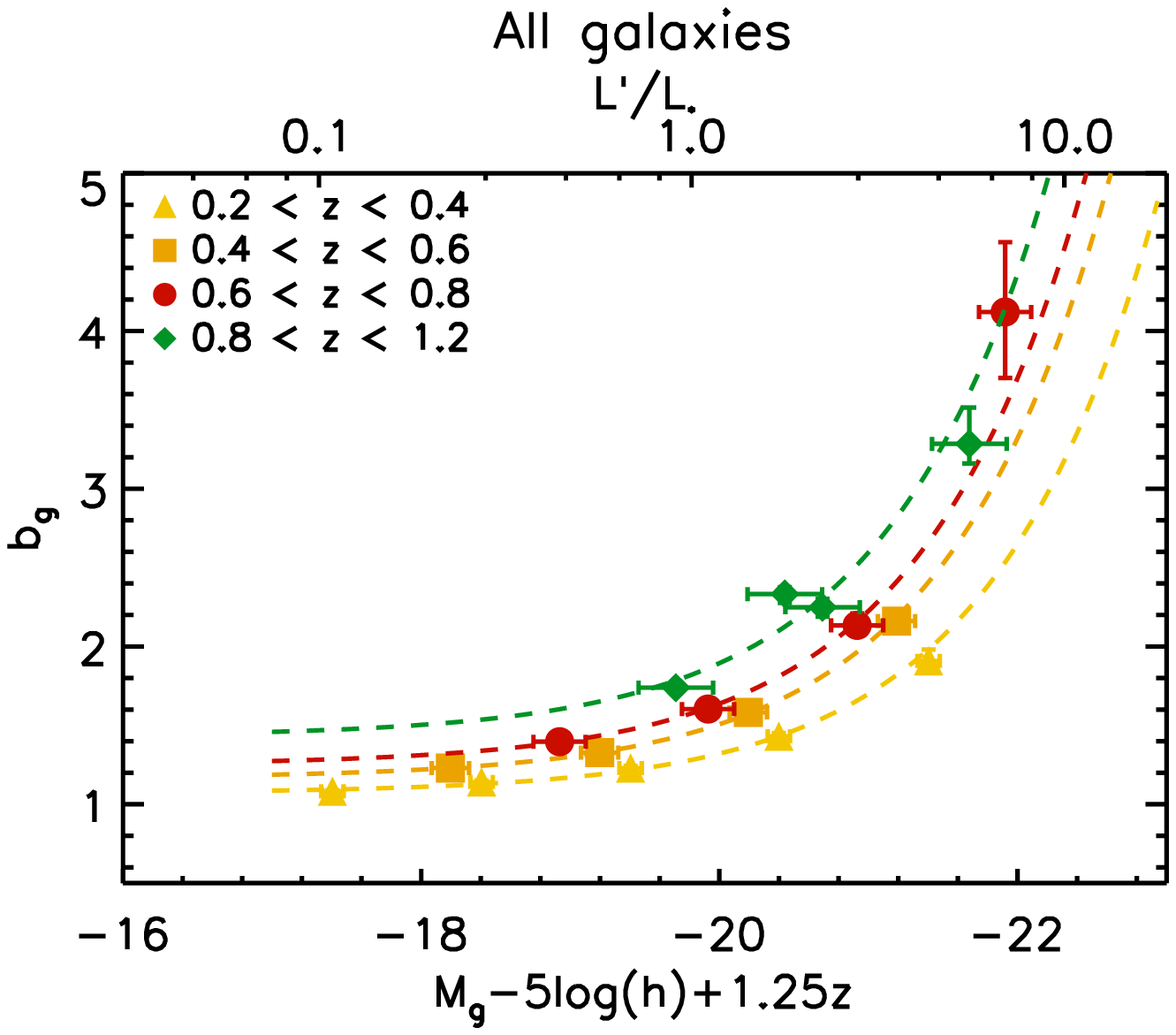} 
    \includegraphics[width=0.49\textwidth]{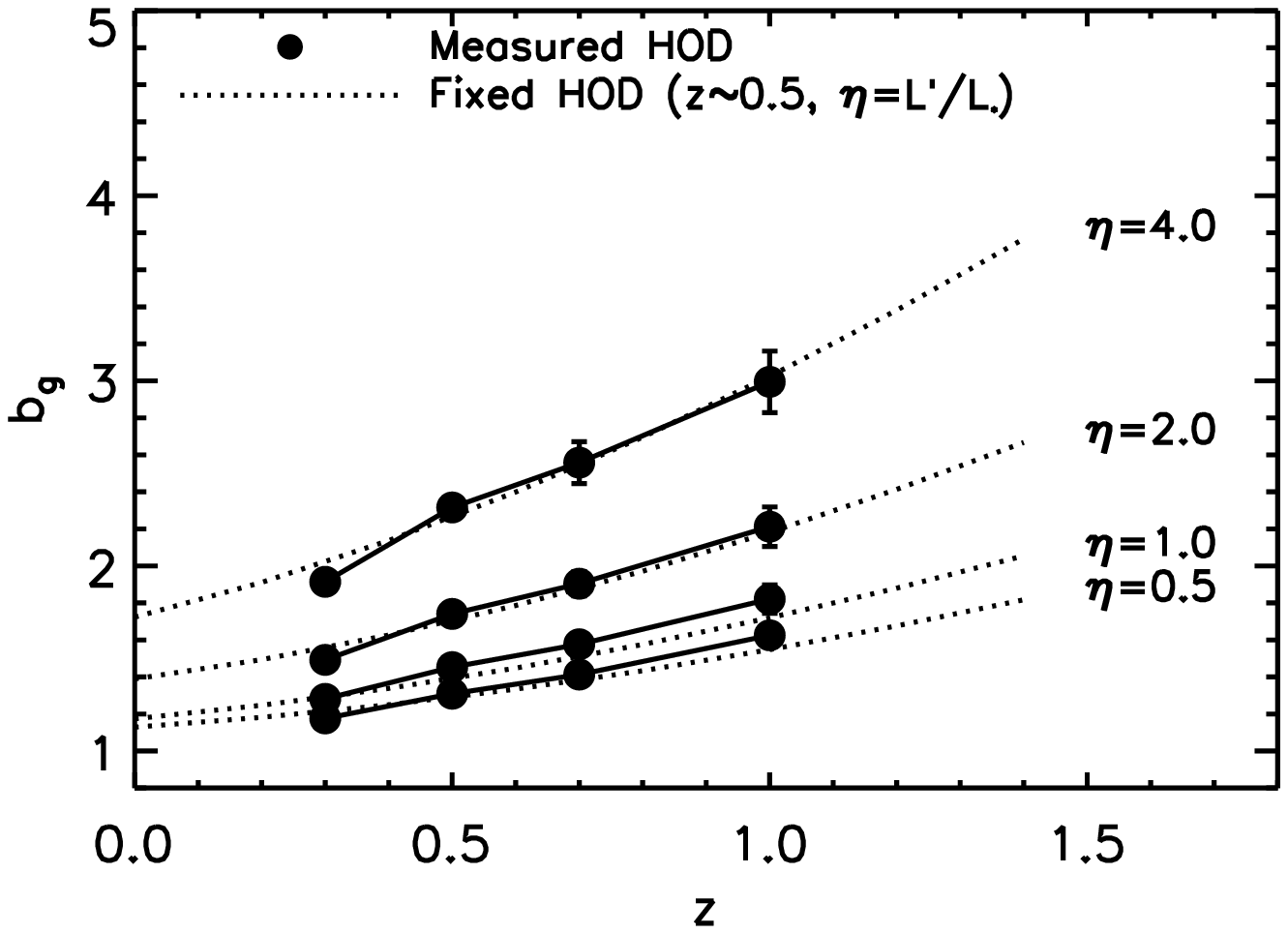} 
    \includegraphics[width=0.49\textwidth]{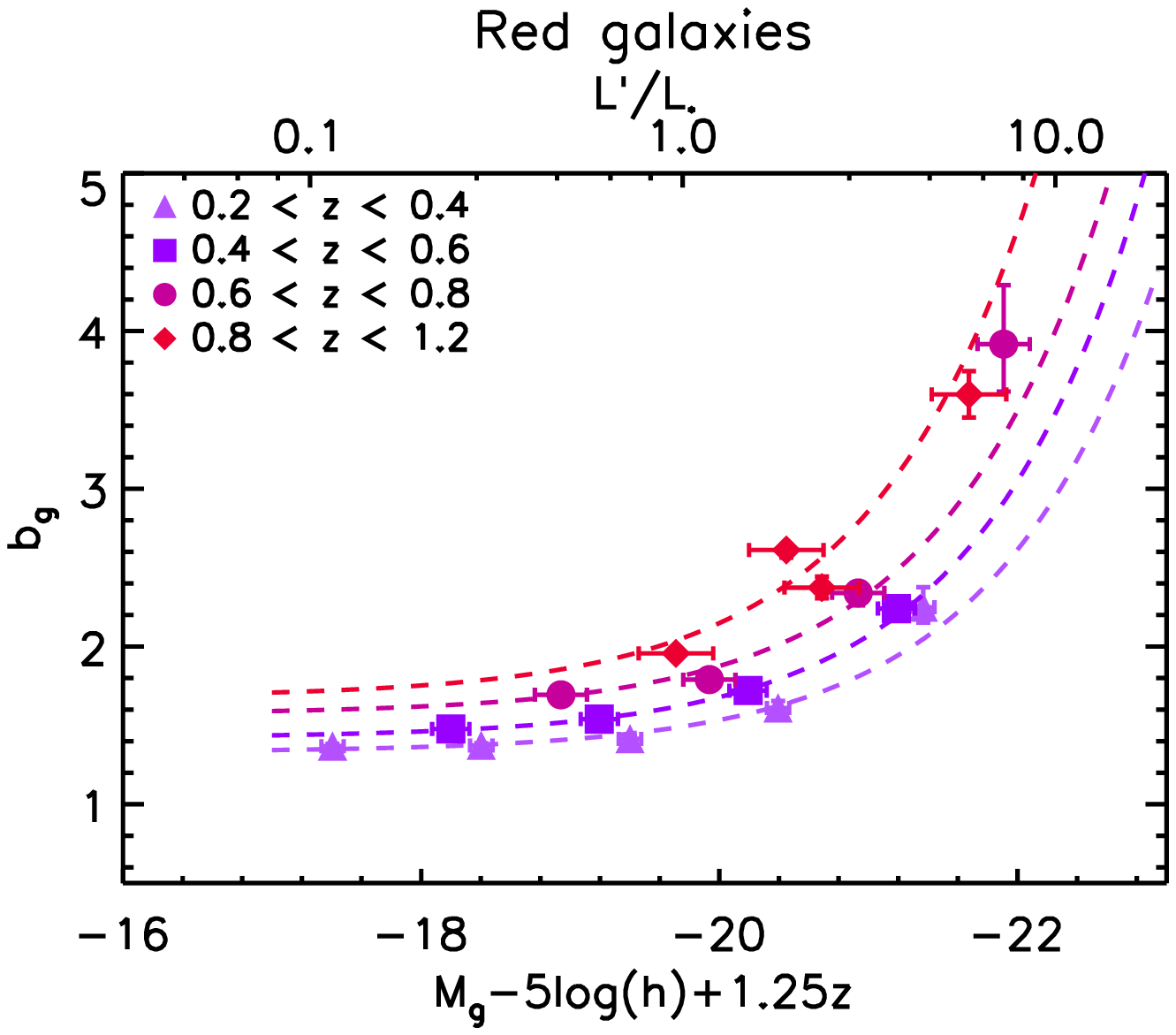}  
    \includegraphics[width=0.49\textwidth]{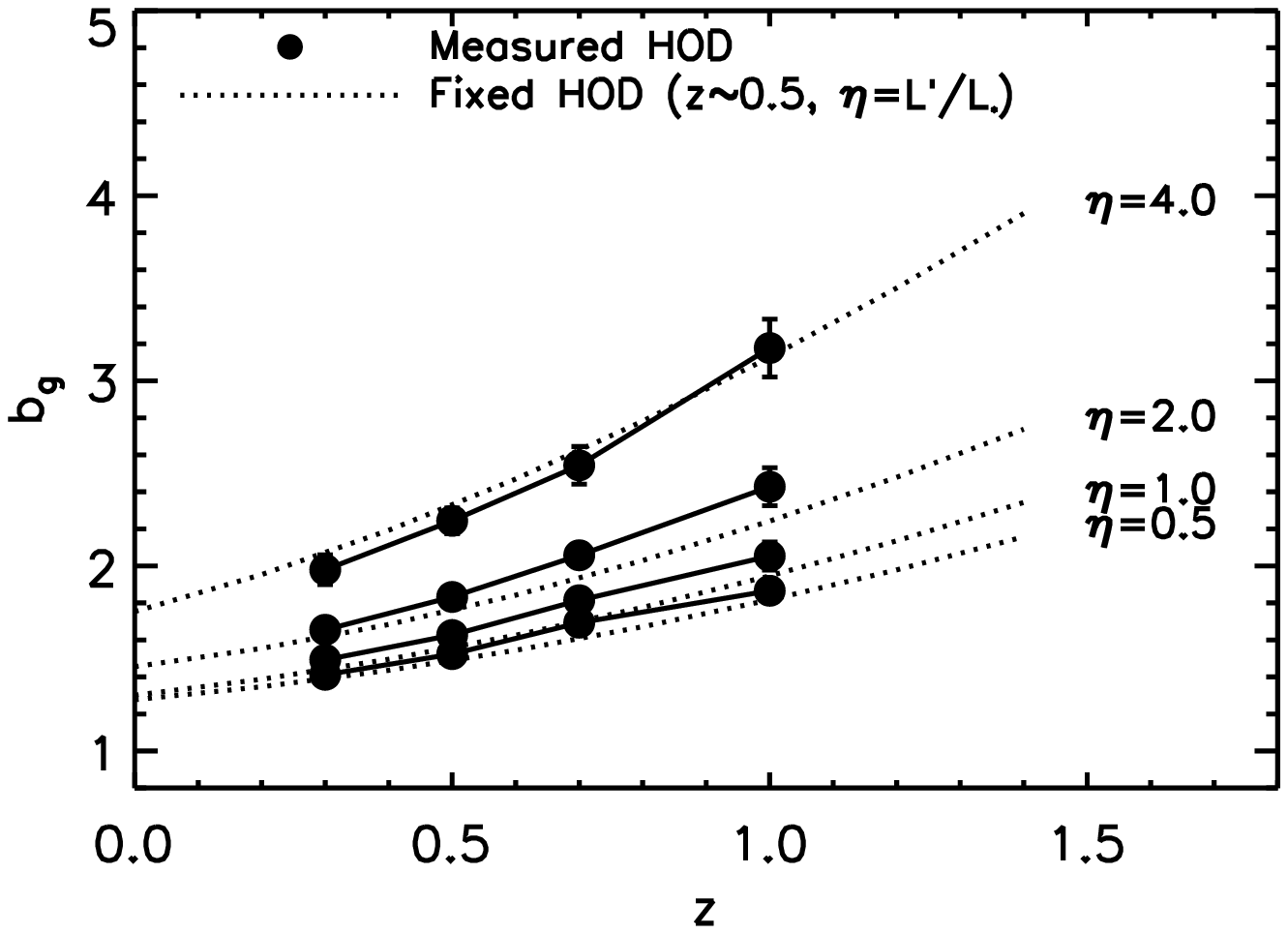} 
  \end{center}
  \caption{Galaxy bias as function of luminosity and redshift. Left
    panels: bias for full and red samples (top and bottom panels) as
    function of corrected luminosity threshold $L'$. The dashed line
    represents Eq.~\ref{eq:biasmodel}.  Right panels: computed bias for
    several luminosity threshold samples $L'_{\ast}$ as function of
    redshift. Error bars were derived from the parametric form shown
    on the left (dashed line), whereas the dotted lines show the bias
    evolution computed from our model, assuming a
    redshift-independent HOD parameter set.}
  \label{fig:bias}
\end{figure*}

We compute the galaxy bias (Eq.~\ref{eq:bias}) from our best-fitting
HOD parameters.  We fit the bias measurements as function of
luminosity and redshift, adopting a linear relationship
\citep{2001MNRAS.328...64N}:
\begin{equation}
  \label{eq:biasmodel}
  b_g(>L') = a_{\rm bias}+b_{\rm bias}
  \frac{L'}{L_{\ast}} \, ,
\end{equation}
where $L'$ represents the corrected luminosity (computed from 
Eq.~\ref{eq:corrred}  for the full and red samples).
We fit the two highest redshift bins together ($0.8 < z < 1.0$, $1.0 <
z < 1.2$), as only one point is available in the highest one. Results
are plotted in Fig.~\ref{fig:bias} and best-fitting parameters are
given in Table~\ref{tab:bgmhalo}.

This parameterised bias relation reproduces our results best at
$z<0.6$. At a fixed redshift, for $L'/L_{\ast}<1$, the bias is a very
weak function of luminosity; for $L'/L_{\ast}>1$, bias depends
strongly on luminosity. For the faint, low-redshift full-sample
galaxies, dominated by more weakly-clustered bluer galaxies, $b\sim1$;
the faint red samples have higher biases, of around $b\sim1.5$. These
results are a consequence of the higher clustering amplitudes observed
for red samples compared to the full galaxy sample for
$L'/L_{\ast}<1$, and are consistent with similar measurements made by
smaller surveys at intermediate redshifts 
\cite [see for example][]{2009A&A...505..463M}.

The right panels of Fig.~\ref{fig:bias} illustrate the redshift
evolution of the bias at several luminosity thresholds. We compare our
results (solid lines) derived from Eq.~\ref{eq:biasmodel} with a
galaxy bias evolution computed using a constant set of HOD parameters
evaluated at $z\sim0.5$ (see Sec~\ref{sec:frsat}). 
The slight difference observed between the measured HOD and the HOD
fixed model at $z=0.5$ comes from the fact that
we set $\sigma_{\log M}$ to 0.3 and $\alpha$ to 1.0. For the full galaxy
population, a model with fixed HOD parameters provides an excellent
fit to the observations, with more luminous objects undergoing a much
more rapid evolution in bias. For the red galaxy population, there is
some evidence that the bias evolution is more rapid
than the model with fixed HOD parameters. 

\subsection{Mean halo mass}
\label{sec:m_halo}

\begin{figure*}
  \begin{center}
    \includegraphics[width=0.49\textwidth]{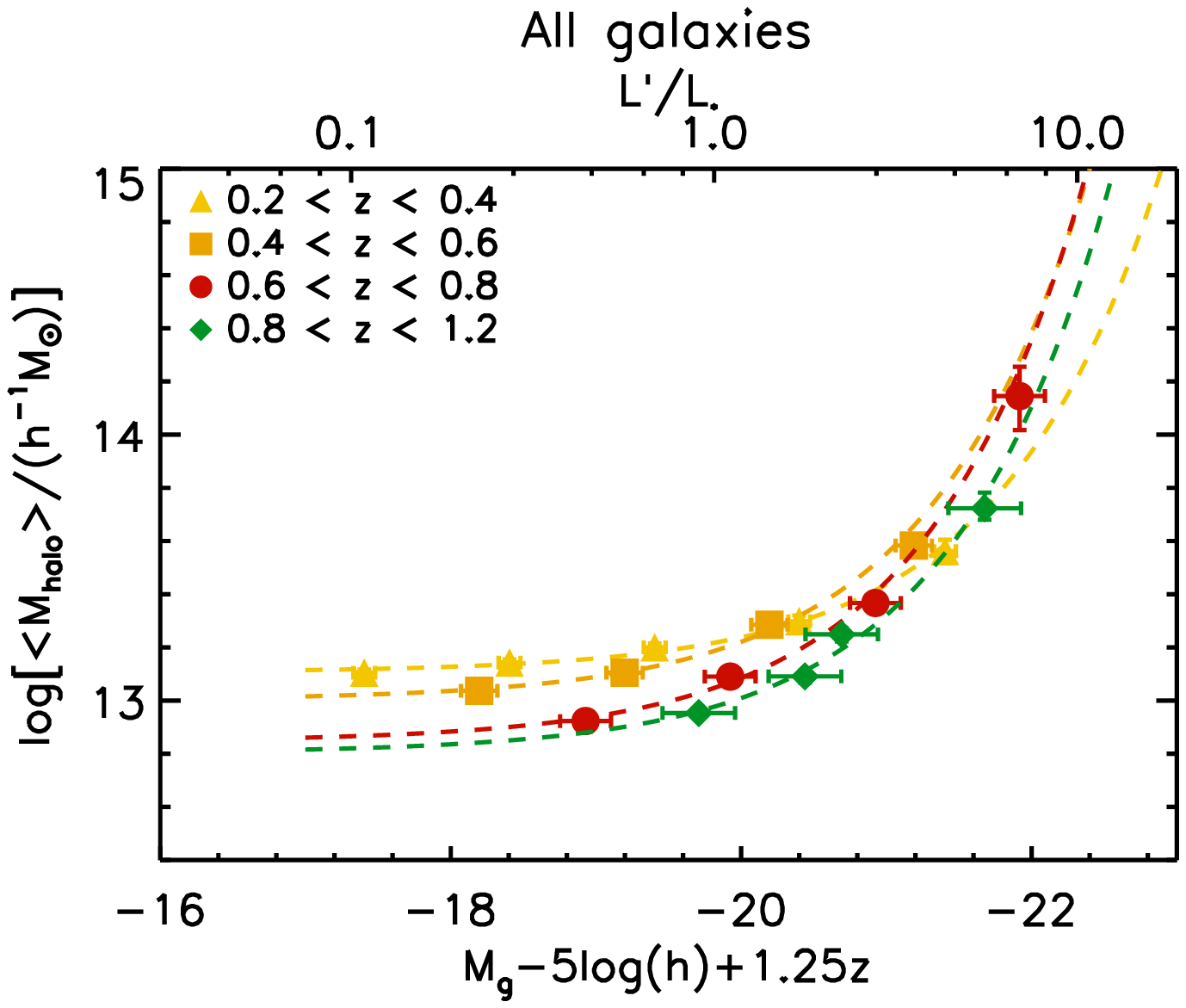}
    \includegraphics[width=0.49\textwidth]{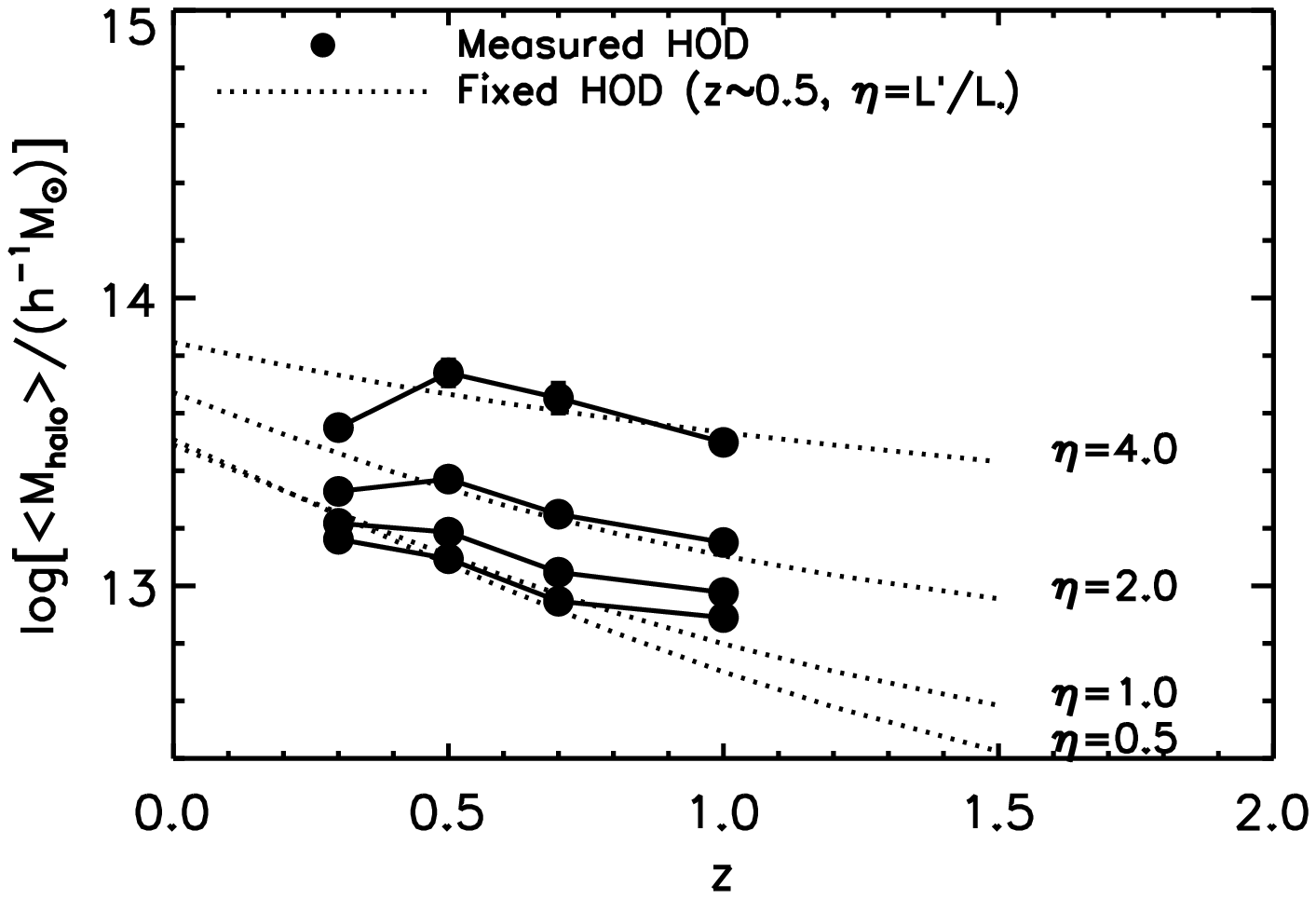}
    \includegraphics[width=0.49\textwidth]{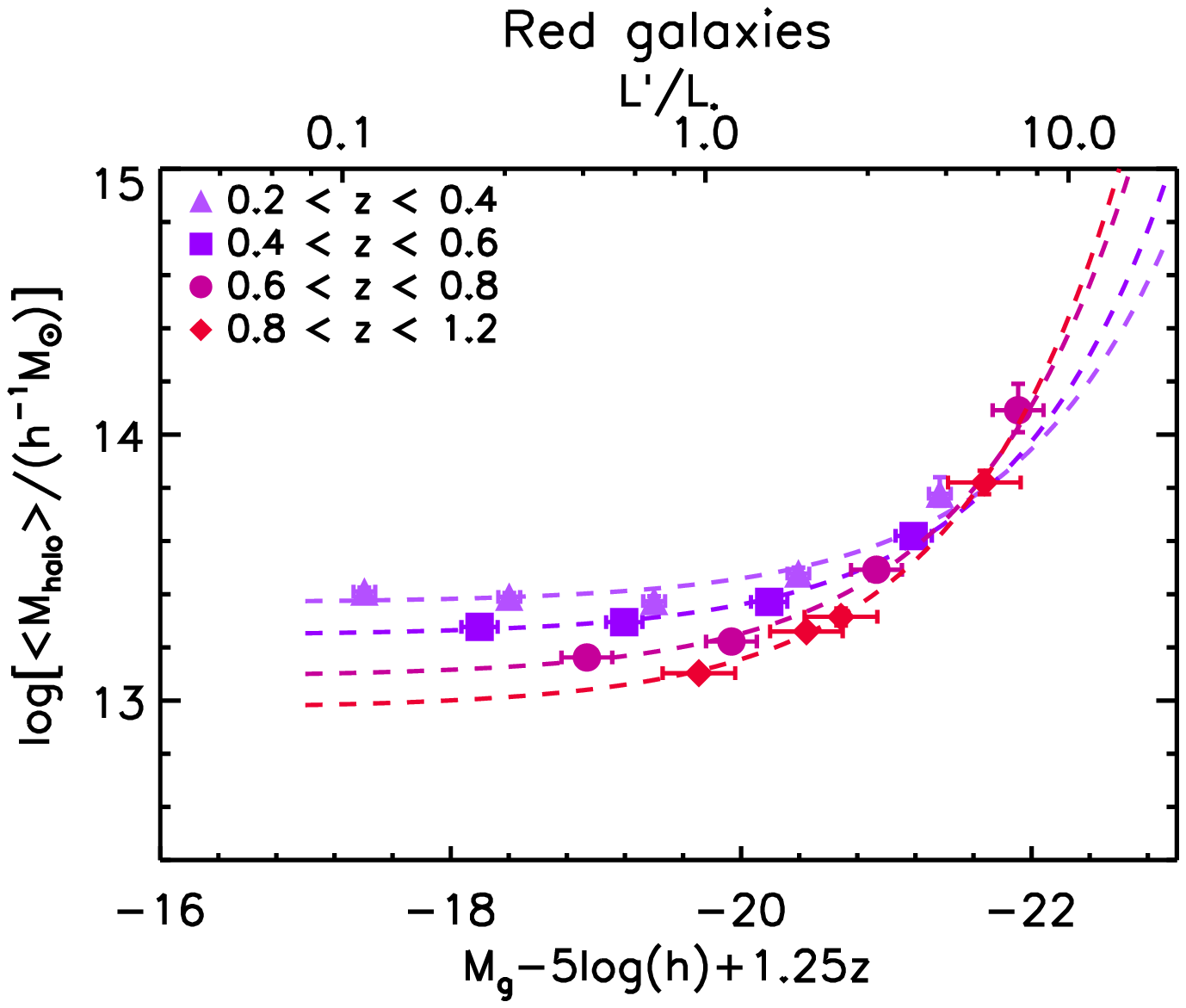}
    \includegraphics[width=0.49\textwidth]{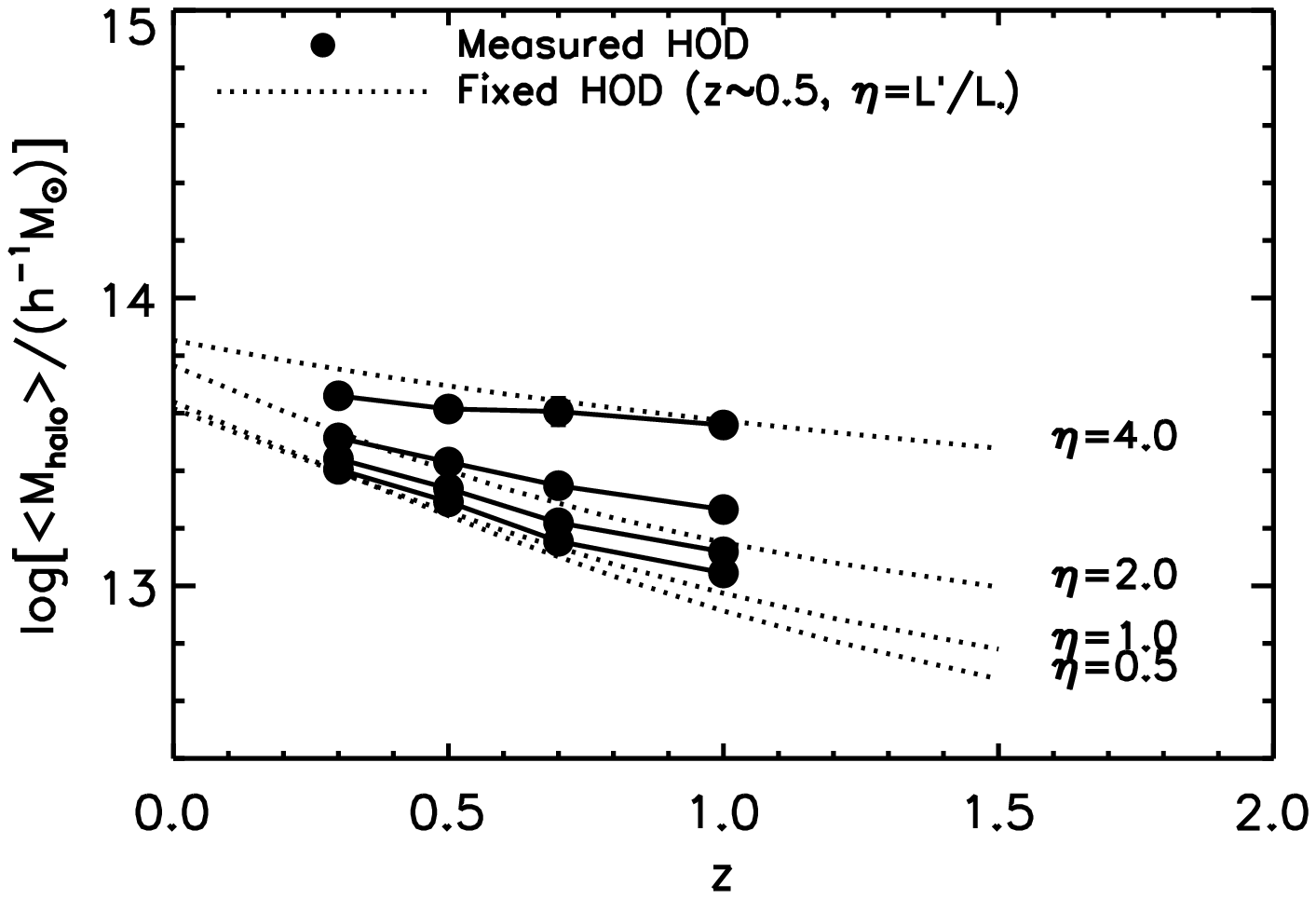}
  \end{center}

  \caption{
    Mean halo mass as function of luminosity and redshift. Left
    panels: mean halo mass for full and red samples (top and bottom panels) as
    function of corrected luminosity threshold $L'$. The dashed line
    represents Eq.~\ref{eq:mhalomodel}.  Right panels: computed mean halo mass for
    several luminosity threshold samples $L'_{\ast}$ as function of
    redshift. Error bars were derived from the parametric form shown
    on the left (dashed line), whereas the dotted lines show the mean halo mass
    evolution computed from our model, assuming a redshift-independent HOD parameter
    set.}
  \label{fig:Mhalo}
\end{figure*}

From Eq.~\ref{eq:Mhalo} we compute the average dark matter halo mass
for each of our samples. Note that this quantity represents the halo
mass averaged over samples containing galaxies above the luminosity
threshold and differs from $M_{\rm min}$ which represents the halo
mass for central galaxies whose luminosity corresponds to the
luminosity threshold. As before, after correcting luminosities to a
corresponding stellar mass threshold, we fit the mean halo mass as
function of luminosity and redshift, adopting a linear relation:

\begin{equation}
  \label{eq:mhalomodel}
  \log\langle M_{\rm halo} \rangle (>L') = a_{\rm halo}+b_{\rm halo} \frac{L'}{L_{\ast}}, \,
\end{equation}
where $L'$ is the corrected luminosity. Once again, we fit the two
highest redshift bins simultaneously. Results are displayed in
Fig.~\ref{fig:Mhalo} and best-fitting 
parameters are given in Table~\ref{tab:bgmhalo}.

For fainter samples with $L'< 2L_{\ast}$ the mean halo mass gradually
increases with luminosity and changes more rapidly for brighter
samples. Faint red galaxy samples on average have higher halo masses
than the full sample which is dominated by bluer galaxies at these
luminosities. This is consistent with a scenario in which high numbers
of red satellite galaxies in faint samples reside in more massive
haloes.

In the right panels of Fig.~\ref{fig:Mhalo} we show the redshift
evolution of the mean halo mass for several luminosity thresholds. As
before, we compare mean halo masses derived from our measured HOD
parameters (solid lines) with a model of fixed halo model parameters
set measured at redshift 0.5. At all luminosities, the mean halo mass
decreases with increasing redshift. This is an indication of the fact
that haloes merge through cosmic history and the effective mass of
haloes grows with time. In bright samples, the mean halo mass evolves
more slowly than in fainter ones. At low luminosity thresholds and
high redshifts, both the full sample and red galaxy sample deviate
slightly from our model with fixed HOD parameters.

\section{Summary and conclusions}
\label{sec:conclusion}

We have made one of the most precise measurements to date of the
angular correlation function $w$ and its dependence on luminosity and
rest-frame colour from $z\sim0$ to $z\sim1$. Our measurements were
constructed from a series of $i'<22.5$ volume-limited samples,
containing $\sim 3\times 10^6$ galaxies in four independent
fields. These samples were created using accurate five-band
photometric redshifts in the CFHTLS Wide survey calibrated with
$\sim10^4$ spectroscopic redshifts. A cross-correlation analysis
showed consistency with a relatively small contamination fraction
arising from scatter between photometric redshift bins.

We interpreted these measurements in the framework of a model in which
the number of galaxies inside each dark matter halo is parametrised by
a simple analytic function (the halo occupation distribution) with
five adjustable parameters. In this model, central and satellite
galaxies are treated separately. For each galaxy sample, we performed a
full likelihood analysis and explored model parameter space using an
efficient Population Monte Carlo analysis. The very large survey
volume probed by the CFHTLS Wide allowed us to place robust constraints
on the redshift evolution of the halo model parameters, and make
accurate estimates for halo properties of many different galaxy
samples over more than two orders of magnitude in luminosity and three
orders of magnitude in halo mass, from $10^{12}$ to $10^{15}(h^{-1}
M_{\sun})$.

Using 30-band photometric data from the COSMOS survey, we derived an
empirical relation between stellar mass, luminosity and redshift and
used this to convert our luminosity-selected samples to stellar-mass
limited samples. All of our conclusions (following below) are
independent of this correction.

These are our principal results:
\begin{enumerate}
\item For a given luminosity threshold, galaxies with redder
  rest-frame colours are more clustered than bluer ones. Clustering
  strength increases with increasing luminosity, reflecting that
  bright galaxies reside in more strongly clustered massive haloes. 

\item We consider the redshift and luminosity evolution of halo
  parameters $M_1$, representing the characteristic halo mass required
  to host at least one satellite galaxy, and $M_{\rm min}$,
  representing the halo mass where on average 50\% of haloes contain
  one galaxy. For our full sample, $M_1$ is closely approximated by
  $\sim 17\times M_{\rm min}$. This ratio becomes smaller at higher
  redshifts for lower-mass haloes. For red galaxies, $M_1/M_{\rm min}$
  is $\sim12$ at intermediate mass scale, and remains constant for all
  halo masses over $0.2 < z < 0.8$. This lower ratio for redder
  galaxies reflects the higher abundance of satellites in these haloes;
  in general, redder populations reside in more massive
  haloes.

\item By fitting a simple analytic relation between central
  galaxy luminosity and halo mass to our
  observations, we find a maximum in the stellar-to-halo mass ratio. In
  the lowest redshift bin $z\sim0.3$, this peak mass is $M^{\rm peak}_{\rm h} =
  4.5\times 10^{11} h^{-1} M_{\sun}$ for the full samples, and $M^{\rm peak}_{\rm h} =21
  \times 10^{11} h^{-1} M_{\sun}$ for the red samples,
  respectively. This transition represents the halo masses where
  baryons were most efficiently converted into stars, and is in good
  agreement with measurements from other studies.
  
\item For the full sample, $M^{\rm peak}_{\rm h}$ shifts to higher
  halo masses at higher redshifts. For the red galaxy sample, the peak
  position evolves less rapidly with redshift. These results can be
  understood qualitatively from the lack of on-going star-formation in
  the red galaxy population which means that the stellar mass content
  in these massive haloes changes very slowly. For the full galaxy
  sample, which is expected to contain galaxies still undergoing
  active star-formation, the shift of the transition mass to
  progressively more massive haloes at higher redshifts is a
  manifestation of ``anti-hierarchical'' galaxy formation. These
  results also indicate that the massive, passive galaxies in our
  survey are already fully formed by $z\sim1.2$.
  
\item At increasing luminosities, the fraction of satellite galaxies
  rapidly drops to zero. For less luminous galaxies, the satellite
  fraction is higher in our red sample than in our full sample,
  reflecting the high number of faint red satellites, consistent with
  observations in the local Universe
  \citep{2006MNRAS.368..715M,2010arXiv1005.2413Z}. For our full
  sample, the number of satellites increases by a factor of two from
  $z\sim1$ to $z\sim0$, consistent with a combined study of galaxies
  in DEEP2 and SDSS \citep{2007ApJ...667..760Z}. This result suggests
  that the occupation function for satellite galaxies inside
  intermediate-mass haloes ($\sim10^{12} h^{-1} M_{\sun}$) remains
  constant with cosmic time, and implies that the galaxy merging rate
  does not change since $z=1$.
  
\item Above the characteristic sample luminosity, the mean galaxy bias
  $b_{\rm g}$ increases rapidly. Faint red samples have $b_{\rm
    g}\sim1.5$, compared to $b_{\rm g}\sim1$ for the full
  sample. Redder samples also have higher average halo masses, which
  decrease towards higher redshifts.
 
\item We compare the measured evolution of our mean halo masses,
  biases and satellite fractions with a model assuming constant HOD
  parameters over our redshift range.  This model provides a
  remarkably good fit to our measurements for massive haloes. We note
  that lower mean halo masses (particularly in the full sample) show a
  slower decrease compared to a redshift-independent HOD model.
  
\end{enumerate}

Systematic photometric redshift errors have little impact on our faint
luminosity-threshold samples, for which we find no evidence for
incompleteness larger than the field-to-field variance estimates.  Our
very brightest samples however may be affected by incompleteness or
contamination from fainter objects. We have shown that these errors
could reduce our measured clustering amplitudes and consequently could
lead to an underestimation of halo masses. The effect will be more
important in contaminated samples, where the higher galaxy number
density will lead to lower halo masses. In the worst case, $\log
M_{\rm min}$ might be underestimated by $\sim0.5$.

Because of the lack of deep near-infrared data in the CFHTLS, we
cannot calculate reliable stellar masses and cleanly separate the
active and passive populations in mass-selected samples. Such a data
set, perhaps combined with approximate local density estimators
calibrated using spectroscopic redshifts, and a refined halo model
would allow us to understand in greater detail the physical origin of
the shift of the transition masses to higher halo masses observed
here.  Furthermore, given the fact that the peak of the star-formation
and mass assembly takes place in the redshift range $1<z<2$, it is
important to push these studies to higher redshifts by constructing
larger, wide-area surveys with deep near-infrared data.

In our model, we have assumed a halo mass function, halo density
profile, and a cosmological model, although it is well known that the
evolution of galaxy clustering depends on these parameters. Combining
galaxy-galaxy lensing, galaxy clustering and mass-function
measurements from an expanded CFHTLS including near-infrared data
should provide new insights on both cosmology and galaxy evolution.

\section*{Acknowledgements}
\label{sec:acknowledgements}

We acknowledge CFHT, TERAPIX, CADC and the CFHTLS Steering Group for
their assistance in planning, executing and reducing the CFHTLS
survey. We also acknowledge the COSMOS, VVDS and DEEP2 collaborations
for making their data publicly available.  We thank T. Hamana,
S. Colombi, M. Brown, M. White, A. Ross, A. Leauthaud 
and P. Behroozi for useful comments. 
We thank the referee for his careful and detailed report. HJMCC
acknowledges support from ANR grant ``ANR-07-BLAN-0228''. JC
is supported by the Japanese Society for the Promotion of
Science.

\bibliographystyle{aa}
\bibliography{references}

\appendix

\section{The halo model}
\label{sec:model_app}

The halo occupation distribution (HOD) model
\citep{2002ApJ...575..587B,2004ApJ...609...35K,2005ApJ...633..791Z}
allows one to derive physical information about galaxy populations and
the dark matter haloes which host them. The HOD
prescription is based on the halo model, which describes how dark
matter is distributed in space. In this framework, all of the matter
is assumed to reside in virialised haloes. The HOD parametrisation
specifies how many galaxies populate haloes, on average, as function
of halo mass. Accordingly, the number of galaxies per halo $N$ only
depends on the mass $M$ of the halo.

\subsection{Dark-matter halo model}

The three ingredients to the halo model of dark matter are the halo
mass function, the halo profile and the halo bias.  The dark matter
halo abundance can be inferred using the \cite{1974ApJ...187..425P}
approach where dark matter collapses into overdense regions above the
critical density $\delta_{\rm c}$, linearly evolved to $z=0$. The mass
function, which is the halo number density per unit mass, can be
parametrized as
\begin{equation}
  n(M, z) \, \ud M = \frac{\overline{\rho_0}}{M} f(\nu) \, \ud \nu \, , 
\end{equation}
where $\overline{\rho_0}$ is the mean density of matter at the present
day. The new mass variable $\nu$ writes
\begin{equation}
  \nu = \frac{\delta_{\rm c}(z)}{D(z)\sigma(M)} \, ,
\end{equation}
and characterizes the peak heights of the density field as function of
mass and redshift. The linear critical density $\delta_{\rm c}$
depends on the adopted cosmology and redshift; we use the fitting
formula from \citet{2003MNRAS.341..251W}; see also
\citet{1996ApJ...469..480K},
\begin{eqnarray}
  \delta_{\rm c}(z) & = & \frac 3 {20} (12 \pi)^{2/3} \left[ 1 + 0.013
    \log_{10} \Omega_{\rm m}(z) \right]; \\
  \Omega_{\rm m}(z)  & = & \Omega_{\rm m} (1+z)^3 \left[\frac{H_0}{H(z)}\right]^2.
\end{eqnarray}
 $D(z)$ is the linear growth factor at redshift $z$, and $\sigma(M)$
is  the rms of density fluctuations in a top-hat filter of width 
$R = (3M/4\pi \overline{\rho_0})^{1/3}$, computed from
linear theory,
\begin{equation} 
  \sigma^2(M) = \int^{\infty}_{0}
  \frac{\ud k}{k} \, \frac{k^3 P_{\rm lin}(k)}{2\pi^2} W^2 (kR) \, ,
\end{equation}
where $W(x) = (3/x^3) [\sin x - x \cos x]$. For the mass function
$f(\nu)$, we choose the parameterisation by \cite{1999MNRAS.308..119S}, 
calibrated on simulations:
\begin{equation}
  \nu f(\nu) = A \sqrt{\frac{2  a\nu^2}{\pi}} \left [ 1+ (a\nu^2)^{-p} \right
  ] \exp \left ( -\frac{a\nu^2}{2} \right ) \, ,
  \label{eq:nufnu}
\end{equation}
where the normalisation $A$ is fixed by imposing:
\begin{equation}
\int n(M, z) \frac{M}{\overline{\rho_0}} \ud M = \int f(\nu)
\ud \nu =1 \, ,
\end{equation}
We adopt the values $p = 0.3$ and $a = 1/\sqrt 2$.
If not indicated otherwise, all integrals over the mass 
function are performed between
$M_{\rm low} = 10^3 h^{-1} M_{\odot}$ and $M_{\rm high} = 10^{16}
h^{-1} M_{\odot}$.

We describe the halo density profile by the following form
\cite{1997ApJ...490..493N},

\begin{equation}
  \rho_{\rm h}(r|M) =
  \frac{\rho_{\rm s}}{(r/r_{\rm s})(1+r/r_{\rm s})^2} \, .
\end{equation}

The total halo mass is then written as \citep[see][]{2003MNRAS.344..857T}:
\begin{equation}
  M = \int_0^{r_{\rm vir}} 4\pi r^2 \ud r \, \rho_{\rm h}(r|M) =
  \frac{4\pi\rho_{\rm s} r_{\rm vir}^3}{c^3} \left
    [ \ln (1+c) - \frac{c}{1+c} \right ] \, ,
\end{equation}
where $c = r_{\rm vir}/r_{\rm s}$ is the ``concentration
  parameter'', for which we assume the following expression,
\begin{equation}
  c(M,z) = \frac{c_0}{1+z} \left [ \frac{M}{M_{\star}} \right
  ]^{-\beta} \, .
\end{equation}
We take $c_0 = 11$ and $\beta = 0.13$, and $M_{\star}$ is
defined such that $\nu(z=0) = 1$, i.e.\ $\delta_{c}(0) = \sigma(M_{\star})$.
The virial radius $r_{\rm vir}$ is given by the following relation
\begin{equation}
  \label{eq:delta}
  M = \frac{4\pi r^3_{\rm vir}}{3} \overline{\rho_0} \Delta_{\rm vir}(z) \, ,
\end{equation} 
with $\Delta_{\rm vir}(z)$ being the critical overdensity for
virialisation at redshift $z$ \citep{1996ApJ...469..480K,1997PThPh..97...49N,2000ApJ...534..565H}.
We take the fitting formula from \citet{2003MNRAS.341..251W}
\begin{equation}
\Delta_{\rm vir}(z) = 18 \pi^2 \left[ 1 + 0.399 \left( \Omega_{\rm
      m}^{-1}(z) - 1 \right)^{0.941} \right] .
\end{equation}
Following \cite{2005ApJ...631...41T},
we use the scale-dependent halo bias
\begin{equation}
  b^2_{\rm h}(M,z,r) = b^2_{\rm h}(M, z)
  \frac{[1+1.17\xi_{\rm m}(r, z)]^{1.49}}{[1+0.69\xi_{\rm m}(r, z)]^{2.09}}
  \, ,
\end{equation}
where $\xi_{\rm m}$ is the matter correlation function. The
large-scale halo bias $b_{\rm h}(M, z)$ is given by
\cite{2001MNRAS.323....1S} as
\begin{eqnarray}
  b_{\rm h}(M, z) = b_{\rm h}(\nu) & = & 1 + \frac{1}{\sqrt a \delta_{\rm c}} \left [ \sqrt a (a\nu^2) +
    \sqrt a b (a\nu^2)^{1-c}  \right .   \nonumber\\
  & & \left . - \frac{(a\nu^2)^c}{(a\nu^2)^c +   b(1-c)(1-c/2)} \right
  ] \, .
  \label{eq:halo_bias}
\end{eqnarray}
As in \cite{2005ApJ...631...41T}, we adopt the 
revised parameters $a = 1/\sqrt 2$, $b = 0.35$ and $c =0.8$.

\subsection{The galaxy correlation function}

We  write the correlation function as a sum of two components:
The one-halo term, to express the galaxy correlation inside a halo,
 and the two-halo term, to account for halo-to-halo correlation,
\begin{equation}
  \xi(r) = 1+\xi_1(r) + \xi_2(r) \, ,
\end{equation}
The one-halo term depends on the number of galaxy pairs $\langle N
(N-1) \rangle$
per halo. This is comprised
of the central-satellite contribution $\langle N_{\rm c} N_{\rm s} \rangle$ and the
satellite-satellite term $\langle N_{\rm s} (N_{\rm s} - 1) \rangle$. Assuming a Poisson
distribution, we write
\begin{eqnarray}
   \langle N_{\rm s}(N_{\rm s} - 1) \rangle(M) & = & N^2_{\rm s}(M) \, ;
   \nonumber \\
  \langle N_{\rm c}N_{\rm s} \rangle(M) & = & N_{\rm c}(M) N_{\rm s}(M) \, .
\end{eqnarray}
The one-halo term correlation function for central-satellite pairs is then
\begin{equation}
  1+\xi_{\rm cs}(r, z) = \int_{M_{\rm vir}(r)}^{M_{\rm high}} \ud M \,
  n(M, z)
  \frac{N_{\rm c}(M)N_{\rm s}(M)}{n_{\rm gal}^2/2} \rho_{\rm h}(r|M) \, .
\end{equation}
The lower integration limit $M_{\rm vir}(r)$ is the virial mass
contained in a halo of radius $r$, computed with
Eq.~\ref{eq:delta}. This accounts for the fact that
less-massive haloes are too small to contribute to the correlation at
separation $r$. 

The one-halo satellite-satellite contribution $\xi_{\rm ss}$ involves the halo
profile auto-convolution and is therefore easier to compute in Fourier
space. The corresponding power spectrum is written as
\begin{equation}
  P_{ss} (k) = \int_{M_{\rm low}}^{M_{\rm high}} \ud M \,  n(M)
    \frac{N^2_{\rm s}(M)}{n_{\rm gal}^2}  |u_h(k|M)|^2\, , 
\end{equation} 
where $u_{\rm h}(k|M)$ is the Fourier transform of the dark-matter halo
profile $\rho_{\rm h}(r|M)$. The correlation function $\xi_{\rm ss}$
is then obtained via a Fourier transform.
The one-halo correlation function is the sum of the two
contributions,
\begin{equation}
  \xi_1 (r) = \xi_{\rm cs} (r) + \xi_{\rm ss} (r) \, .  
\end{equation}

The two-halo term is derived from the dark-matter power spectrum and
the halo two-point correlation function:
\begin{eqnarray}
\label{eq:p2h}
  P_2(k,r) &= & P_{\rm m}(k)\nonumber \\ 
  & &\times \left [ \int_{M_{\rm low}}^{M_{\rm lim}(r)} \ud M
    n(M) \frac{N(M)}{n'_{\rm gal}(r)} b_{\rm h}(M,r) |u_{\rm h}(k|M)| \right ]^2 \, ,
\end{eqnarray}
where
\begin{equation}
  n'_{\rm gal}(r) = \int_{M_{\rm low}}^{M_{\rm lim}(r)} n(M) N(M,z) \, \ud M \, .
\end{equation}
The upper integration limit $M_{\rm lim}(r)$ takes into account the halo
exclusion \citep{2004ApJ...610...61Z}, i.e.~the fact that haloes are non-overlapping. 
We follow \cite{2005ApJ...631...41T} to compute
$M_{\rm lim}(r)$ by matching  $n'_{\rm gal}(r)$ with the
following expression,
\begin{eqnarray}
   n'^2_{\rm gal}(r) &= &\int \ud M_1 \, n(M_1) N(M_1) \nonumber \\
  & & \times  \int \ud M_2\,  n(M_2) N(M_2)  P(r,M_1,M_2) \, ,
\end{eqnarray}
where $P(r, M_1, M_2)$ is the probability that two ellipsoidal haloes
of mass $M_1$ and $M_2$, respectively, do not overlap. Defining $x =
r/[r_{\rm vir}(M_1) +r_{\rm vir}(M_2)]$ as the ratio of the halo
separation and the sum of the virial radii, and $y = (x-0.8)/0.29$,
\cite{2005ApJ...631...41T} found the probability of non-overlapping haloes to
be
\begin{equation}
P(r, M_1, M_2) = P(y) = \left\{ \begin{array}{ll}
0 & \textrm{if $y<0$}\\
(3y^2-2y^3) & \textrm{if $0 \le y \le 1$}\\
1 & \textrm{if $y > 1$}
\end{array} \right. \, .
\end{equation}
We Fourier-transform Eq.~\ref{eq:p2h} for a range of tabulated values of $r$,
to compute the two-halo term $\xi_2$ of the galaxy autocorrelation
function. Finally, we renormalise it to the total number of galaxy pairs:
\begin{equation}
  1+\xi_2 (r) = \left [ \frac{n'_{gal}(r)}{n_{\rm gal}}  \right ]
  [1+\xi_2(r)] \, .
\end{equation}

The angular two-point correlation function $w(\theta)$ is computed
from the observed photometric redshift distribution and $\xi(r)$ using
Limber's equation \citep{1954ApJ...119..655L}:
\begin{equation}
  w(\theta) = 2 \int_0^{\infty} \ud x \, f(x)^2 \int_0^{\infty} \ud u
  \, \xi(r = \sqrt{u^2 + x^2\theta^2}) \, ,
\end{equation} 
with
\begin{equation}
  f(x) = \frac{n(z)}{\ud x(z) / \ud z} \, ,
\end{equation}
and $x(z)$, the radial comoving coordinate.

\section{Best-fitting HOD parameters and deduced quantities}
\label{sec:parameters}

\begin{landscape}
  \begin{center}
    \begin{table}
      \caption{Description of all galaxy samples and best-fitting HOD
        parameters. Halo masses are given in $h^{-1} M_{\sun}$ and
        galaxy number densities in $h^{-3}\, {\rm Mpc^3}$.}
      \label{tab:HODall}
      \begin{tabular*}{\linewidth}{@{\extracolsep{\fill}}   p{4cm} c r
          c c c c c c c c c c r}
        \hline
        \hline
        \centering Redshift \T \B & $M_g - 5\log h$ & $N$ & $\log n_{\rm
          gal, obs}$ & $\log M_{\rm min}$ &  $\log M_{1}$ & $\log M_{0}$ &
      $\sigma_{{\rm log}M}$ & $\alpha$ & $\log n_{\rm gal, mod}$ & $b_g$ &
      log$M_{\rm halo}$ & $f_{\rm s}$ &  $\chi^2/{\rm dof}$\\ [0.1cm]
      \hline

0.2 $< z <$ 0.4 \dotfill\T& $< -17.8$ & $451\,203$ & -1.53$^{+0.02}_{-0.02}$ & 11.18$^{+0.06}_{-0.05}$ & 12.53$^{+0.03}_{-0.03}$ & 7.54$^{+1.72}_{-1.71}$  & 0.40$^{+0.12}_{-0.13}$ & 1.10$^{+0.01}_{-0.01}$ & -1.53$^{+0.02}_{-0.02}$ & 1.08$^{+0.01}_{-0.01}$ & 13.11$^{+0.01}_{-0.01}$ & 0.23$^{+0.01}_{-0.01}$ &  3.58\\  [0.1cm]
0.2 $< z <$ 0.4 \dotfill& $< -18.8$ & $243\,833$ & -1.80$^{+0.02}_{-0.02}$ & 11.47$^{+0.06}_{-0.06}$ & 12.76$^{+0.03}_{-0.03}$ & 7.57$^{+1.86}_{-1.78}$  & 0.42$^{+0.12}_{-0.15}$ & 1.12$^{+0.01}_{-0.01}$ & -1.79$^{+0.02}_{-0.02}$ & 1.14$^{+0.01}_{-0.01}$ & 13.14$^{+0.01}_{-0.01}$ & 0.22$^{+0.01}_{-0.01}$ &  4.93\\  [0.1cm]
0.2 $< z <$ 0.4 \dotfill& $< -19.8$ & $104\,889$ & -2.17$^{+0.03}_{-0.03}$ & 11.85$^{+0.05}_{-0.06}$ & 13.11$^{+0.04}_{-0.04}$ & 8.18$^{+2.03}_{-2.15}$  & 0.45$^{+0.10}_{-0.16}$ & 1.16$^{+0.02}_{-0.02}$ & -2.16$^{+0.03}_{-0.03}$ & 1.23$^{+0.01}_{-0.01}$ & 13.20$^{+0.01}_{-0.02}$ & 0.18$^{+0.01}_{-0.01}$ &  3.33\\  [0.1cm]
0.2 $< z <$ 0.4 \dotfill& $< -20.8$ & $28\,081$ & -2.74$^{+0.04}_{-0.04}$ & 12.36$^{+0.09}_{-0.05}$ & 13.69$^{+0.05}_{-0.05}$ & 8.35$^{+2.22}_{-2.32}$  & 0.32$^{+0.17}_{-0.13}$ & 1.28$^{+0.05}_{-0.05}$ & -2.74$^{+0.04}_{-0.05}$ & 1.43$^{+0.02}_{-0.03}$ & 13.30$^{+0.02}_{-0.02}$ & 0.12$^{+0.01}_{-0.01}$ &  4.87\\  [0.1cm]
0.2 $< z <$ 0.4 \dotfill& $< -21.8$ & $3\,679$ & -3.62$^{+0.05}_{-0.05}$ & 13.17$^{+0.09}_{-0.08}$ & 14.53$^{+0.36}_{-0.14}$ & 11.09$^{+2.32}_{-3.28}$  & 0.39$^{+0.15}_{-0.18}$ & 1.27$^{+0.33}_{-0.47}$ & -3.64$^{+0.05}_{-0.06}$ & 1.91$^{+0.07}_{-0.06}$ & 13.56$^{+0.04}_{-0.05}$ & 0.04$^{+0.01}_{-0.01}$ &  3.21\\  [0.1cm]
0.4 $< z <$ 0.6 \dotfill& $< -18.8$ & $551\,547$ & -1.80$^{+0.01}_{-0.01}$ & 11.48$^{+0.06}_{-0.06}$ & 12.66$^{+0.03}_{-0.03}$ & 10.96$^{+0.26}_{-0.72}$  & 0.43$^{+0.11}_{-0.15}$ & 1.09$^{+0.02}_{-0.02}$ & -1.80$^{+0.01}_{-0.02}$ & 1.23$^{+0.01}_{-0.01}$ & 13.04$^{+0.01}_{-0.01}$ & 0.23$^{+0.01}_{-0.00}$ &  3.71\\  [0.1cm]
0.4 $< z <$ 0.6 \dotfill& $< -19.8$ & $244\,854$ & -2.15$^{+0.02}_{-0.02}$ & 11.82$^{+0.06}_{-0.06}$ & 13.01$^{+0.04}_{-0.04}$ & 11.02$^{+0.38}_{-0.63}$  & 0.43$^{+0.11}_{-0.13}$ & 1.15$^{+0.03}_{-0.03}$ & -2.14$^{+0.02}_{-0.03}$ & 1.33$^{+0.01}_{-0.01}$ & 13.11$^{+0.01}_{-0.01}$ & 0.20$^{+0.01}_{-0.01}$ &  3.62\\  [0.1cm]
0.4 $< z <$ 0.6 \dotfill& $< -20.8$ & $65\,557$ & -2.73$^{+0.04}_{-0.04}$ & 12.33$^{+0.06}_{-0.03}$ & 13.58$^{+0.04}_{-0.04}$ & 8.49$^{+2.40}_{-2.26}$  & 0.30$^{+0.11}_{-0.09}$ & 1.37$^{+0.03}_{-0.03}$ & -2.74$^{+0.04}_{-0.03}$ & 1.58$^{+0.02}_{-0.02}$ & 13.29$^{+0.01}_{-0.01}$ & 0.14$^{+0.01}_{-0.01}$ &  3.86\\  [0.1cm]
0.4 $< z <$ 0.6 \dotfill& $< -21.8$ & $6\,724$ & -3.71$^{+0.06}_{-0.06}$ & 13.18$^{+0.09}_{-0.07}$ & 14.47$^{+0.48}_{-0.11}$ & 10.93$^{+1.69}_{-1.86}$  & 0.30$^{+0.16}_{-0.12}$ & 1.36$^{+0.24}_{-0.44}$ & -3.74$^{+0.05}_{-0.07}$ & 2.16$^{+0.06}_{-0.07}$ & 13.58$^{+0.03}_{-0.04}$ & 0.06$^{+0.01}_{-0.01}$ &  1.93\\  [0.1cm]
0.6 $< z <$ 0.8 \dotfill& $< -19.8$ & $458\,863$ & -2.08$^{+0.03}_{-0.03}$ & 11.77$^{+0.05}_{-0.05}$ & 12.83$^{+0.04}_{-0.04}$ & 11.54$^{+0.15}_{-0.18}$  & 0.50$^{+0.07}_{-0.10}$ & 1.07$^{+0.03}_{-0.03}$ & -2.07$^{+0.03}_{-0.03}$ & 1.40$^{+0.01}_{-0.01}$ & 12.92$^{+0.01}_{-0.01}$ & 0.19$^{+0.01}_{-0.01}$ &  1.85\\  [0.1cm]
0.6 $< z <$ 0.8 \dotfill& $< -20.8$ & $141\,129$ & -2.59$^{+0.04}_{-0.04}$ & 12.24$^{+0.04}_{-0.07}$ & 13.34$^{+0.05}_{-0.04}$ & 10.50$^{+0.84}_{-0.84}$  & 0.50$^{+0.06}_{-0.36}$ & 1.28$^{+0.03}_{-0.03}$ & -2.57$^{+0.04}_{-0.05}$ & 1.60$^{+0.03}_{-0.02}$ & 13.09$^{+0.01}_{-0.01}$ & 0.15$^{+0.01}_{-0.01}$ &  2.92\\  [0.1cm]
0.6 $< z <$ 0.8 \dotfill& $< -21.8$ & $17\,683$ & -3.50$^{+0.06}_{-0.06}$ & 12.96$^{+0.10}_{-0.09}$ & 14.10$^{+0.07}_{-0.09}$ & 12.47$^{+0.62}_{-0.93}$  & 0.38$^{+0.14}_{-0.21}$ & 1.28$^{+0.23}_{-0.51}$ & -3.49$^{+0.05}_{-0.06}$ & 2.13$^{+0.08}_{-0.07}$ & 13.37$^{+0.03}_{-0.03}$ & 0.06$^{+0.01}_{-0.01}$ &  4.76\\  [0.1cm]
0.6 $< z <$ 0.8 \dotfill& $< -22.8$ & $656$ & -4.93$^{+0.18}_{-0.18}$ & 14.12$^{+0.32}_{-0.16}$ & 16.09$^{+0.58}_{-0.57}$ & 9.72$^{+3.15}_{-3.11}$  & 0.31$^{+0.18}_{-0.12}$ & 1.31$^{+0.44}_{-0.36}$ & -5.37$^{+0.29}_{-0.43}$ & 4.12$^{+0.44}_{-0.42}$ & 14.14$^{+0.11}_{-0.13}$ & 0.01$^{+0.01}_{-0.00}$ &  10.17\\  [0.1cm]
0.8 $< z <$ 1.0 \dotfill& $< -20.8$ & $200\,516$ & -2.57$^{+0.04}_{-0.04}$ & 12.14$^{+0.09}_{-0.06}$ & 13.21$^{+0.08}_{-0.09}$ & 12.23$^{+0.18}_{-0.27}$  & 0.35$^{+0.14}_{-0.13}$ & 1.12$^{+0.09}_{-0.13}$ & -2.58$^{+0.04}_{-0.05}$ & 1.74$^{+0.03}_{-0.03}$ & 12.95$^{+0.02}_{-0.02}$ & 0.13$^{+0.01}_{-0.01}$ &  4.85\\  [0.1cm]
0.8 $< z <$ 1.0 \dotfill& $< -21.8$ & $31\,001$ & -3.38$^{+0.05}_{-0.05}$ & 12.80$^{+0.10}_{-0.07}$ & 13.94$^{+0.06}_{-0.36}$ & 12.15$^{+0.91}_{-1.01}$  & 0.33$^{+0.14}_{-0.15}$ & 1.52$^{+0.20}_{-0.88}$ & -3.39$^{+0.05}_{-0.06}$ & 2.25$^{+0.05}_{-0.06}$ & 13.25$^{+0.02}_{-0.03}$ & 0.06$^{+0.01}_{-0.01}$ &  8.69\\  [0.1cm]
0.8 $< z <$ 1.0 \dotfill& $< -22.8$ & $1\,411$ & -4.72$^{+0.15}_{-0.15}$ & 13.76$^{+0.10}_{-0.16}$ & 15.13$^{+0.47}_{-0.36}$ & 9.11$^{+2.48}_{-2.91}$  & 0.48$^{+0.08}_{-0.37}$ & 0.94$^{+0.60}_{-0.20}$ & -4.64$^{+0.10}_{-0.14}$ & 3.28$^{+0.23}_{-0.12}$ & 13.72$^{+0.06}_{-0.04}$ & 0.05$^{+0.02}_{-0.02}$ &  7.35\\  [0.1cm]
1.0 $< z <$ 1.2 \dotfill\B& $< -21.8$ & $52\,310$ & -3.24$^{+0.04}_{-0.04}$ & 12.62$^{+0.08}_{-0.04}$ & 13.79$^{+0.05}_{-0.04}$ & 8.67$^{+2.54}_{-2.44}$  & 0.30$^{+0.14}_{-0.09}$ & 1.50$^{+0.09}_{-0.11}$ & -3.25$^{+0.04}_{-0.04}$ & 2.33$^{+0.04}_{-0.06}$ & 13.09$^{+0.02}_{-0.02}$ & 0.08$^{+0.01}_{-0.01}$ &  3.24\\  [0.1cm]

      \hline
    \end{tabular*}
  \end{table}
\end{center}
\end{landscape}
\begin{landscape}
  \begin{center}
    \begin{table}
      \caption{Description of red galaxy samples and best-fitting HOD
        parameters. Halo masses are given in $h^{-1} M_{\sun}$ and
        galaxy number densities in $h^{-3}\, {\rm Mpc^3}$.}
      \label{tab:HODred}
      \begin{tabular*}{\linewidth}{@{\extracolsep{\fill}}   p{4cm} c r
          c c c c c c c c c c r}
        \hline
        \hline
        \centering Redshift \T \B & $M_g - 5\log h$ & $N$ & $\log n_{\rm
          gal, obs}$ & $\log M_{\rm min}$ &  $\log M_{1}$ & $\log M_{0}$ &
      $\sigma_{{\rm log}M}$ & $\alpha$ & $\log n_{\rm gal, mod}$ & $b_g$ &
      log$M_{\rm halo}$ & $f_{\rm s}$ &  $\chi^2/{\rm dof}$\\ [0.1cm]
     \hline
 0.2 $< z <$ 0.4 \dotfill\T& $< -17.8$ & $153\,729$ & -2.00$^{+0.04}_{-0.04}$ & 11.77$^{+0.06}_{-0.07}$ & 12.70$^{+0.05}_{-0.05}$ & 9.69$^{+1.65}_{-3.45}$  & 0.41$^{+0.12}_{-0.22}$ & 1.13$^{+0.02}_{-0.04}$ & -1.99$^{+0.03}_{-0.04}$ & 1.37$^{+0.01}_{-0.01}$ & 13.41$^{+0.02}_{-0.02}$ & 0.36$^{+0.01}_{-0.01}$ &  4.77\\  [0.1cm]
0.2 $< z <$ 0.4 \dotfill& $< -18.8$ & $100\,829$ & -2.19$^{+0.04}_{-0.04}$ & 11.90$^{+0.08}_{-0.06}$ & 12.94$^{+0.06}_{-0.05}$ & 8.04$^{+2.09}_{-2.02}$  & 0.37$^{+0.14}_{-0.15}$ & 1.17$^{+0.02}_{-0.02}$ & -2.17$^{+0.04}_{-0.05}$ & 1.37$^{+0.01}_{-0.02}$ & 13.39$^{+0.01}_{-0.02}$ & 0.30$^{+0.01}_{-0.02}$ &  5.64\\  [0.1cm]
0.2 $< z <$ 0.4 \dotfill& $< -19.8$ & $54\,393$ & -2.45$^{+0.04}_{-0.04}$ & 12.15$^{+0.08}_{-0.06}$ & 13.25$^{+0.07}_{-0.07}$ & 10.58$^{+1.28}_{-2.42}$  & 0.36$^{+0.15}_{-0.15}$ & 1.19$^{+0.04}_{-0.13}$ & -2.46$^{+0.04}_{-0.05}$ & 1.42$^{+0.02}_{-0.02}$ & 13.37$^{+0.02}_{-0.02}$ & 0.22$^{+0.01}_{-0.01}$ &  3.63\\  [0.1cm]
0.2 $< z <$ 0.4 \dotfill& $< -20.8$ & $17\,200$ & -2.95$^{+0.05}_{-0.05}$ & 12.60$^{+0.07}_{-0.07}$ & 13.82$^{+0.09}_{-0.07}$ & 8.56$^{+2.46}_{-2.48}$  & 0.31$^{+0.15}_{-0.10}$ & 1.42$^{+0.06}_{-0.06}$ & -2.99$^{+0.06}_{-0.08}$ & 1.61$^{+0.04}_{-0.03}$ & 13.48$^{+0.02}_{-0.02}$ & 0.13$^{+0.01}_{-0.01}$ &  3.00\\  [0.1cm]
0.2 $< z <$ 0.4 \dotfill& $< -21.8$ & $1\,545$ & -3.85$^{+0.08}_{-0.08}$ & 13.40$^{+0.16}_{-0.09}$ & 14.51$^{+0.26}_{-0.30}$ & 13.47$^{+0.43}_{-1.52}$  & 0.30$^{+0.18}_{-0.11}$ & 1.24$^{+0.51}_{-0.46}$ & -3.98$^{+0.10}_{-0.17}$ & 2.25$^{+0.13}_{-0.10}$ & 13.78$^{+0.06}_{-0.05}$ & 0.04$^{+0.01}_{-0.01}$ &  2.92\\  [0.1cm]
0.4 $< z <$ 0.6 \dotfill& $< -18.8$ & $209\,469$ & -2.22$^{+0.03}_{-0.03}$ & 11.94$^{+0.07}_{-0.05}$ & 12.90$^{+0.04}_{-0.04}$ & 10.04$^{+1.29}_{-2.37}$  & 0.40$^{+0.13}_{-0.12}$ & 1.15$^{+0.02}_{-0.03}$ & -2.21$^{+0.03}_{-0.03}$ & 1.48$^{+0.01}_{-0.01}$ & 13.28$^{+0.01}_{-0.01}$ & 0.30$^{+0.01}_{-0.01}$ &  4.96\\  [0.1cm]
0.4 $< z <$ 0.6 \dotfill& $< -19.8$ & $122\,073$ & -2.46$^{+0.03}_{-0.03}$ & 12.10$^{+0.04}_{-0.03}$ & 13.19$^{+0.04}_{-0.03}$ & 8.37$^{+1.99}_{-1.97}$  & 0.26$^{+0.12}_{-0.07}$ & 1.22$^{+0.02}_{-0.02}$ & -2.46$^{+0.03}_{-0.03}$ & 1.54$^{+0.01}_{-0.01}$ & 13.29$^{+0.01}_{-0.01}$ & 0.24$^{+0.01}_{-0.01}$ &  3.07\\  [0.1cm]
0.4 $< z <$ 0.6 \dotfill& $< -20.8$ & $42\,303$ & -2.92$^{+0.03}_{-0.03}$ & 12.52$^{+0.05}_{-0.05}$ & 13.61$^{+0.05}_{-0.11}$ & 11.79$^{+0.65}_{-1.03}$  & 0.29$^{+0.12}_{-0.08}$ & 1.26$^{+0.07}_{-0.27}$ & -2.93$^{+0.03}_{-0.04}$ & 1.72$^{+0.02}_{-0.02}$ & 13.37$^{+0.01}_{-0.02}$ & 0.16$^{+0.01}_{-0.01}$ &  3.88\\  [0.1cm]
0.4 $< z <$ 0.6 \dotfill& $< -21.8$ & $5\,210$ & -3.82$^{+0.05}_{-0.05}$ & 13.27$^{+0.08}_{-0.05}$ & 14.70$^{+0.42}_{-0.22}$ & 9.48$^{+2.65}_{-2.83}$  & 0.32$^{+0.14}_{-0.11}$ & 1.03$^{+0.28}_{-0.26}$ & -3.84$^{+0.05}_{-0.05}$ & 2.24$^{+0.07}_{-0.07}$ & 13.62$^{+0.04}_{-0.04}$ & 0.07$^{+0.01}_{-0.01}$ &  1.14\\  [0.1cm]
0.6 $< z <$ 0.8 \dotfill& $< -19.8$ & $216\,675$ & -2.41$^{+0.05}_{-0.05}$ & 12.14$^{+0.07}_{-0.02}$ & 13.21$^{+0.04}_{-0.04}$ & 11.65$^{+0.19}_{-0.16}$  & 0.25$^{+0.11}_{-0.06}$ & 1.22$^{+0.02}_{-0.03}$ & -2.56$^{+0.02}_{-0.03}$ & 1.69$^{+0.01}_{-0.01}$ & 13.16$^{+0.01}_{-0.01}$ & 0.21$^{+0.01}_{-0.01}$ &  4.35\\  [0.1cm]
0.6 $< z <$ 0.8 \dotfill& $< -20.8$ & $83\,687$ & -2.82$^{+0.05}_{-0.05}$ & 12.40$^{+0.09}_{-0.07}$ & 13.46$^{+0.05}_{-0.05}$ & 8.25$^{+2.20}_{-2.22}$  & 0.36$^{+0.14}_{-0.14}$ & 1.31$^{+0.03}_{-0.03}$ & -2.80$^{+0.04}_{-0.05}$ & 1.79$^{+0.02}_{-0.03}$ & 13.22$^{+0.01}_{-0.01}$ & 0.17$^{+0.01}_{-0.01}$ &  2.76\\  [0.1cm]
0.6 $< z <$ 0.8 \dotfill& $< -21.8$ & $11\,982$ & -3.67$^{+0.07}_{-0.07}$ & 13.06$^{+0.11}_{-0.06}$ & 14.19$^{+0.07}_{-0.06}$ & 9.43$^{+2.38}_{-3.31}$  & 0.28$^{+0.16}_{-0.13}$ & 1.54$^{+0.12}_{-0.19}$ & -3.67$^{+0.06}_{-0.07}$ & 2.34$^{+0.06}_{-0.08}$ & 13.49$^{+0.03}_{-0.03}$ & 0.07$^{+0.01}_{-0.01}$ &  2.93\\  [0.1cm]
0.6 $< z <$ 0.8 \dotfill& $< -22.8$ & $385$ & -5.16$^{+0.16}_{-0.16}$ & 14.08$^{+0.21}_{-0.13}$ & 16.27$^{+0.48}_{-0.61}$ & 9.87$^{+3.43}_{-3.28}$  & 0.34$^{+0.17}_{-0.15}$ & 1.35$^{+0.44}_{-0.46}$ & -5.24$^{+0.17}_{-0.35}$ & 3.92$^{+0.37}_{-0.30}$ & 14.09$^{+0.10}_{-0.08}$ & 0.00$^{+0.00}_{-0.00}$ &  6.24\\  [0.1cm]
0.8 $< z <$ 1.0 \dotfill& $< -20.8$ & $111\,158$ & -2.83$^{+0.06}_{-0.06}$ & 12.37$^{+0.07}_{-0.05}$ & 13.48$^{+0.07}_{-0.05}$ & 11.76$^{+0.37}_{-0.34}$  & 0.26$^{+0.11}_{-0.09}$ & 1.31$^{+0.05}_{-0.10}$ & -2.88$^{+0.05}_{-0.07}$ & 1.96$^{+0.03}_{-0.03}$ & 13.10$^{+0.02}_{-0.02}$ & 0.14$^{+0.01}_{-0.01}$ &  4.65\\  [0.1cm]
0.8 $< z <$ 1.0 \dotfill& $< -21.8$ & $21\,494$ & -3.54$^{+0.06}_{-0.06}$ & 12.93$^{+0.11}_{-0.06}$ & 14.07$^{+0.07}_{-0.06}$ & 10.86$^{+1.57}_{-1.52}$  & 0.34$^{+0.14}_{-0.14}$ & 1.61$^{+0.14}_{-0.40}$ & -3.55$^{+0.06}_{-0.06}$ & 2.37$^{+0.07}_{-0.07}$ & 13.31$^{+0.03}_{-0.03}$ & 0.06$^{+0.01}_{-0.01}$ &  4.65\\  [0.1cm]
0.8 $< z <$ 1.0 \dotfill& $< -22.8$ & $955$ & -4.89$^{+0.12}_{-0.12}$ & 13.74$^{+0.05}_{-0.10}$ & 15.62$^{+0.43}_{-0.35}$ & 9.27$^{+1.95}_{-2.67}$  & 0.34$^{+0.07}_{-0.16}$ & 0.82$^{+0.78}_{-0.15}$ & -4.77$^{+0.08}_{-0.10}$ & 3.60$^{+0.15}_{-0.15}$ & 13.82$^{+0.04}_{-0.04}$ & 0.04$^{+0.02}_{-0.02}$ &  5.21\\  [0.1cm]
1.0 $< z <$ 1.2 \dotfill\B& $< -21.8$ & $30\,637$ & -3.47$^{+0.04}_{-0.04}$ & 12.77$^{+0.07}_{-0.03}$ & 13.84$^{+0.03}_{-0.05}$ & 11.72$^{+0.70}_{-0.59}$  & 0.21$^{+0.16}_{-0.07}$ & 1.72$^{+0.08}_{-0.13}$ & -3.49$^{+0.04}_{-0.04}$ & 2.61$^{+0.04}_{-0.05}$ & 13.26$^{+0.02}_{-0.02}$ & 0.08$^{+0.01}_{-0.01}$ &  3.37\\  [0.1cm]
\hline
\end{tabular*}
\end{table}
\end{center}
\end{landscape}
\begin{table*}
  \caption{Description of blue galaxy samples. Galaxy number
      densities are given in $h^{-3}\, {\rm Mpc^3}$.}
  \label{tab:HODblue}
  \begin{center}
    \begin{tabular*}{0.5\textwidth}{@{\extracolsep{\fill}}  p{4cm} c c
        r}
      \hline
      \hline
      \centering Redshift \T \B & $M_g - 5\log h$ & $N$ & $\log n_{\rm
        gal, obs}$\\ [0.1cm]
      \hline
      0.2 $< z <$ 0.4 \dotfill\T& $< -17.8$ & $297\,474$ & -1.72$^{+0.01}_{-0.01}$ \\ [0.1cm]
      0.2 $< z <$ 0.4 \dotfill   & $< -18.8$ & $143\,004$ &-2.03$^{+0.02}_{-0.02}$ \\ [0.1cm]
      0.2 $< z <$ 0.4 \dotfill   & $< -19.8$ & $50\,496$ & -2.49$^{+0.04}_{-0.04}$\\ [0.1cm]
      0.2 $< z <$ 0.4 \dotfill   & $< -20.8$ & $10\,881$ & -3.15$^{+0.06}_{-0.06}$\\ [0.1cm]
      0.4 $< z <$ 0.6 \dotfill   & $< -18.8$ & $342\,078$ & -2.01$^{+0.02}_{-0.02}$\\ [0.1cm]
      0.4 $< z <$ 0.6 \dotfill   & $< -19.8$ & $122\,781$ & -2.45$^{+0.04}_{-0.04}$\\ [0.1cm]
      0.4 $< z <$ 0.6 \dotfill   & $< -20.8$ & $23\,254$ & -3.18$^{+0.08}_{-0.08}$\\ [0.1cm]
      0.6 $< z <$ 0.8 \dotfill   & $< -19.8$ & $242\,188$ & -2.36$^{+0.02}_{-0.02}$\\ [0.1cm]
      0.6 $< z <$ 0.8 \dotfill   & $< -20.8$ & $57\,442$ & -2.99$^{+0.03}_{-0.03}$\\ [0.1cm]
      0.6 $< z <$ 0.8 \dotfill   & $< -21.8$ & $5\,701$ & -3.99$^{+0.06}_{-0.06}$\\ [0.1cm]
      0.8 $< z <$ 1.0 \dotfill   & $< -20.8$ & $89\,358$  & -2.92$^{+0.03}_{-0.03}$\\ [0.1cm]
      0.8 $< z <$ 1.0 \dotfill   & $< -21.8$ & $9\,507$ & -3.89$^{+0.05}_{-0.05}$\\ [0.1cm]
      1.0 $< z <$ 1.2 \dotfill\B& $< -21.8$ & $21\,673$ & -3.62$^{+0.04}_{-0.04}$\\ [0.1cm]
      \hline
    \end{tabular*}
  \end{center}
\end{table*}
\section{Two-point correlation function measurements}
\label{sec:wtheta}
\begin{table*}
  \caption{Two-point correlation function measurements in the range $0.2 < z < 0.4$.}
  \label{tab:wtheta02}
  \begin{center}
    \begin{tabular}{c c c c c c}
      \multicolumn{6}{c}{All galaxies\B}\\
      \hline
      \hline
         \centering  $\theta({\rm deg})$ \T \B & $M_g - 5\log h < -17.8$
      &  $M_g - 5\log h < -18.8$ & $M_g - 5\log h < -19.8$  & $M_g - 5\log h < -20.8$ & $M_g - 5\log h < -21.8$  \\ [0.1cm]
      \hline
      0.0013 \T& $1.4818\pm0.0322$  	      & $2.3199\pm0.0670$  	      & $3.5549\pm0.1685$  	      & $7.3561\pm0.9543$  	      & $10.3172\pm6.7590$  \\
      0.0024 & $0.8895\pm0.0213$  	      & $1.3596\pm0.0377$  	      & $1.9627\pm0.0878$  	      & $3.8350\pm0.2778$  	      & $7.1094\pm2.8470$  \\
      0.0042 & $0.5787\pm0.0132$  	      & $0.8173\pm0.0222$  	      & $1.1466\pm0.0466$  	      & $2.0245\pm0.1489$  	      & $2.0728\pm1.2725$  \\
      0.0075 & $0.3949\pm0.0127$  	      & $0.5195\pm0.0180$  	      & $0.7166\pm0.0307$  	      & $1.0055\pm0.0949$  	      & $1.2243\pm0.4861$  \\
      0.0133 & $0.2819\pm0.0106$  	      & $0.3630\pm0.0140$  	      & $0.4808\pm0.0213$  	      & $0.7403\pm0.0500$  	      & $0.9581\pm0.2822$  \\
      0.0237 & $0.1883\pm0.0091$  	      & $0.2340\pm0.0130$  	      & $0.2893\pm0.0182$  	      & $0.4364\pm0.0403$  	      & $0.5257\pm0.1548$  \\
      0.0422 & $0.1243\pm0.0072$  	      & $0.1501\pm0.0103$  	      & $0.1861\pm0.0140$  	      & $0.2800\pm0.0268$  	      & $0.4263\pm0.0780$  \\
      0.0750 & $0.0776\pm0.0060$  	      & $0.0928\pm0.0081$  	      & $0.1179\pm0.0111$  	      & $0.1700\pm0.0238$  	      & $0.1334\pm0.0493$  \\
      0.1334 & $0.0499\pm0.0052$  	      & $0.0593\pm0.0068$  	      & $0.0761\pm0.0086$  	      & $0.1172\pm0.0173$  	      & $0.1055\pm0.0393$  \\
      0.2371 & $0.0342\pm0.0052$  	      & $0.0426\pm0.0067$  	      & $0.0536\pm0.0088$  	      & $0.0820\pm0.0169$  	      & $0.0888\pm0.0225$  \\
      0.4217 & $0.0241\pm0.0046$  	      & $0.0297\pm0.0060$  	      & $0.0362\pm0.0083$  	      & $0.0624\pm0.0138$  	      & $0.0620\pm0.0176$  \\
      0.7499 & $0.0135\pm0.0033$  	      & $0.0165\pm0.0045$  	      & $0.0212\pm0.0056$  	      & $0.0327\pm0.0076$  	      & $0.0330\pm0.0121$  \\
      1.3335 \B& $0.0047\pm0.0046$  	      & $0.0060\pm0.0047$  	      & $0.0082\pm0.0047$  	      & $0.0148\pm0.0068$  	      & $0.0210\pm0.0104$  \\
      \hline\\
    \end{tabular}
    \begin{tabular}{c c c c c c}
      \multicolumn{6}{c}{Red galaxies\B}\\
      \hline
      \hline
      \centering  $\theta({\rm deg})$ \T \B & $M_g - 5\log h < -17.8$
      &  $M_g - 5\log h < -18.8$ & $M_g - 5\log h < -19.8$  & $M_g - 5\log h < -20.8$ & $M_g - 5\log h < -21.8$  \\ [0.1cm]
      \hline
      0.0013 \T& $4.7974\pm0.1698$  	      & $6.0040\pm0.2551$  	      & $7.4692\pm0.5074$  	      & $12.0464\pm2.1742$  	      & $37.0212\pm31.1297$\\  
      0.0024 & $3.2359\pm0.0904$  	      & $3.7019\pm0.1303$  	      & $4.2512\pm0.2044$  	      & $7.2855\pm0.8348$  	      & $14.9207\pm9.6253$  \\
      0.0042 & $2.1016\pm0.0635$  	      & $2.2614\pm0.0780$  	      & $2.5039\pm0.1200$  	      & $3.5656\pm0.3051$  	      & $4.8226\pm3.5344$  \\
      0.0075 & $1.4436\pm0.0540$  	      & $1.4913\pm0.0594$  	      & $1.5948\pm0.0767$  	      & $1.8603\pm0.1843$  	      & $4.4201\pm1.7669$  \\
      0.0133 & $0.9918\pm0.0456$  	      & $1.0089\pm0.0472$  	      & $1.0138\pm0.0489$  	      & $1.3918\pm0.0982$  	      & $3.3844\pm0.9315$  \\
      0.0237 & $0.6053\pm0.0357$  	      & $0.6219\pm0.0381$  	      & $0.5959\pm0.0363$  	      & $0.7927\pm0.0671$  	      & $2.1519\pm0.4732$  \\
      0.0422 & $0.3411\pm0.0230$  	      & $0.3528\pm0.0257$  	      & $0.3319\pm0.0278$  	      & $0.4282\pm0.0418$  	      & $1.1750\pm0.2607$  \\
      0.0750 & $0.1768\pm0.0146$  	      & $0.1953\pm0.0173$  	      & $0.1961\pm0.0178$  	      & $0.2743\pm0.0344$  	      & $0.5895\pm0.1460$  \\
      0.1334 & $0.0950\pm0.0098$  	      & $0.1094\pm0.0125$  	      & $0.1199\pm0.0130$  	      & $0.1749\pm0.0205$  	      & $0.3728\pm0.1074$  \\
      0.2371 & $0.0596\pm0.0092$  	      & $0.0728\pm0.0116$  	      & $0.0807\pm0.0127$  	      & $0.1151\pm0.0196$  	      & $0.2964\pm0.0609$  \\
      0.4217 & $0.0409\pm0.0077$  	      & $0.0506\pm0.0096$  	      & $0.0560\pm0.0115$  	      & $0.0847\pm0.0153$  	      & $0.1940\pm0.0458$  \\
      0.7499 & $0.0239\pm0.0062$  	      & $0.0301\pm0.0087$  	      & $0.0328\pm0.0090$  	      & $0.0462\pm0.0114$  	      & $0.1166\pm0.0292$  \\
      1.3335 \B& $0.0101\pm0.0062$  	      & $0.0142\pm0.0086$  	      & $0.0142\pm0.0083$  	      & $0.0220\pm0.0119$  	      & $0.0572\pm0.0309$ \\
      \hline\\
    \end{tabular}
    \begin{tabular}{c c c c c c}
      \multicolumn{6}{c}{Blue galaxies\B}\\
      \hline
      \hline
      \centering  $\theta({\rm deg})$ \T \B & $M_g - 5\log h < -17.8$
      &  $M_g - 5\log h < -18.8$ & $M_g - 5\log h < -19.8$  & $M_g - 5\log h < -20.8$ & $M_g - 5\log h < -21.8$  \\ [0.1cm]
      \hline
      0.0013 \T& $1.0697\pm0.0363$  	      & $1.6188\pm0.0912$  	      & $2.4395\pm0.3425$  	      & $5.9775\pm2.1846$  	      & $33.1298\pm40.8099$  \\
      0.0024 & $0.5225\pm0.0225$  	      & $0.7530\pm0.0501$  	      & $1.0018\pm0.1375$  	      & $1.0833\pm0.5121$  	      & $5.9415\pm4.5196$  \\
      0.0042 & $0.3040\pm0.0145$  	      & $0.4330\pm0.0260$  	      & $0.5566\pm0.0585$  	      & $1.0238\pm0.2918$  	      & $1.3497\pm1.4446$  \\
      0.0075 & $0.1801\pm0.0086$  	      & $0.2253\pm0.0130$  	      & $0.3712\pm0.0356$  	      & $0.6845\pm0.1501$  	      & $0.8682\pm0.7125$  \\
      0.0133 & $0.1339\pm0.0067$  	      & $0.1661\pm0.0087$  	      & $0.2189\pm0.0229$  	      & $0.1885\pm0.0979$  	      & $0.2393\pm0.3912$  \\
      0.0237 & $0.0934\pm0.0049$  	      & $0.1144\pm0.0078$  	      & $0.1490\pm0.0161$  	      & $0.1619\pm0.0526$  	      & $0.2913\pm0.2202$  \\
      0.0422 & $0.0681\pm0.0045$  	      & $0.0799\pm0.0062$  	      & $0.1063\pm0.0125$  	      & $0.1598\pm0.0318$  	      & $0.1634\pm0.1230$  \\
      0.0750 & $0.0507\pm0.0046$  	      & $0.0598\pm0.0056$  	      & $0.0753\pm0.0105$  	      & $0.1043\pm0.0234$  	      & $0.0250\pm0.0562$  \\
      0.1334 & $0.0369\pm0.0039$  	      & $0.0414\pm0.0050$  	      & $0.0560\pm0.0085$  	      & $0.0872\pm0.0190$  	      & $-0.0117\pm0.0385$  \\
      0.2371 & $0.0263\pm0.0039$  	      & $0.0301\pm0.0048$  	      & $0.0366\pm0.0079$  	      & $0.0602\pm0.0165$  	      & $0.0452\pm0.0231$  \\
      0.4217 & $0.0183\pm0.0035$  	      & $0.0214\pm0.0044$  	      & $0.0279\pm0.0079$  	      & $0.0443\pm0.0160$  	      & $0.0169\pm0.0182$  \\
      0.7499 & $0.0098\pm0.0026$  	      & $0.0113\pm0.0030$  	      & $0.0178\pm0.0052$  	      & $0.0276\pm0.0086$  	      & $0.0178\pm0.0104$  \\
      1.3335 \B& $0.0029\pm0.0037$  	      & $0.0032\pm0.0035$  	      & $0.0070\pm0.0038$  	      & $0.0133\pm0.0063$  	      & $0.0112\pm0.0116$ \\
      \hline
    \end{tabular}
  \end{center}
\end{table*}
\begin{table*}
  \caption{Two-point correlation function measurements in the range $0.4 < z < 0.6$.}
  \label{tab:wtheta04}
  \begin{center}
    \begin{tabular}{ c c c c c}
      \multicolumn{5}{c}{All galaxies\B}\\
      \hline
      \hline
      \centering  $\theta({\rm deg})$ \T \B  &  $M_g - 5\log h < -18.8$ & $M_g - 5\log h < -19.8$  & $M_g - 5\log h < -20.8$ & $M_g - 5\log h < -21.8$  \\ [0.1cm]
      \hline
      0.0013\T & $1.3933\pm0.0262$  	      & $2.2367\pm0.0615$  	      & $4.1633\pm0.2645$  	      & $9.4751\pm3.9229$  \\
      0.0024 & $0.8180\pm0.0148$  	      & $1.2711\pm0.0286$  	      & $2.3764\pm0.1307$  	      & $7.9475\pm1.7637$  \\
      0.0042 & $0.5276\pm0.0104$  	      & $0.7745\pm0.0193$  	      & $1.4069\pm0.0628$  	      & $3.8402\pm0.7920$  \\
      0.0075 & $0.3481\pm0.0073$  	      & $0.4912\pm0.0125$  	      & $0.8450\pm0.0375$  	      & $1.8830\pm0.3531$  \\
      0.0133 & $0.2218\pm0.0055$  	      & $0.3019\pm0.0090$  	      & $0.5299\pm0.0206$  	      & $0.8854\pm0.1511$  \\
      0.0237 & $0.1398\pm0.0047$  	      & $0.1862\pm0.0069$  	      & $0.3035\pm0.0148$  	      & $0.6038\pm0.0930$  \\
      0.0422 & $0.0876\pm0.0035$  	      & $0.1167\pm0.0055$  	      & $0.1787\pm0.0106$  	      & $0.2470\pm0.0497$  \\
      0.0750 & $0.0563\pm0.0029$  	      & $0.0749\pm0.0044$  	      & $0.1150\pm0.0085$  	      & $0.2270\pm0.0288$  \\
      0.1334 & $0.0378\pm0.0025$  	      & $0.0504\pm0.0035$  	      & $0.0800\pm0.0052$  	      & $0.1409\pm0.0165$  \\
      0.2371 & $0.0254\pm0.0021$  	      & $0.0339\pm0.0029$  	      & $0.0499\pm0.0046$  	      & $0.0948\pm0.0146$  \\
      0.4217 & $0.0145\pm0.0017$  	      & $0.0201\pm0.0025$  	      & $0.0293\pm0.0044$  	      & $0.0497\pm0.0108$  \\
      0.7499 & $0.0066\pm0.0018$  	      & $0.0101\pm0.0027$  	      & $0.0153\pm0.0041$  	      & $0.0197\pm0.0081$  \\
      1.3335 \B& $0.0022\pm0.0044$  	      & $0.0043\pm0.0041$  	      & $0.0088\pm0.0040$  	      & $0.0124\pm0.0079$  \\
      \hline\\
    \end{tabular}
  \begin{tabular}{ c c c c c}
      \multicolumn{5}{c}{Red galaxies\B}\\
      \hline
      \hline
      \centering  $\theta({\rm deg})$ \T \B  &  $M_g - 5\log h < -18.8$ & $M_g - 5\log h < -19.8$  & $M_g - 5\log h < -20.8$ & $M_g - 5\log h < -21.8$  \\ [0.1cm]
      \hline
      0.0013 \T& $4.0250\pm0.1139$  	      & $5.0502\pm0.1719$  	      & $7.2406\pm0.5425$  	      & $19.5919\pm9.0027$  \\
      0.0024 & $2.4772\pm0.0594$  	      & $2.9115\pm0.0828$  	      & $4.0217\pm0.2380$  	      & $10.5452\pm2.8842$  \\
      0.0042 & $1.5544\pm0.0337$  	      & $1.7346\pm0.0467$  	      & $2.2900\pm0.1250$  	      & $6.5579\pm1.4118$  \\
      0.0075 & $1.0081\pm0.0254$  	      & $1.0944\pm0.0255$  	      & $1.3993\pm0.0555$  	      & $3.1910\pm0.4906$  \\
      0.0133 & $0.6123\pm0.0191$  	      & $0.6484\pm0.0204$  	      & $0.8565\pm0.0352$  	      & $1.3759\pm0.2206$  \\
      0.0237 & $0.3545\pm0.0142$  	      & $0.3719\pm0.0141$  	      & $0.4487\pm0.0231$  	      & $0.7683\pm0.1319$  \\
      0.0422 & $0.1899\pm0.0100$  	      & $0.2063\pm0.0112$  	      & $0.2154\pm0.0189$  	      & $0.2399\pm0.0625$  \\
      0.0750 & $0.1101\pm0.0063$  	      & $0.1223\pm0.0077$  	      & $0.1460\pm0.0111$  	      & $0.3053\pm0.0422$  \\
      0.1334 & $0.0662\pm0.0049$  	      & $0.0749\pm0.0059$  	      & $0.0961\pm0.0070$  	      & $0.1346\pm0.0256$  \\
      0.2371 & $0.0459\pm0.0051$  	      & $0.0500\pm0.0052$  	      & $0.0630\pm0.0058$  	      & $0.1122\pm0.0198$  \\
      0.4217 & $0.0261\pm0.0040$  	      & $0.0287\pm0.0042$  	      & $0.0366\pm0.0054$  	      & $0.0466\pm0.0119$  \\
      0.7499 & $0.0124\pm0.0037$  	      & $0.0139\pm0.0041$  	      & $0.0174\pm0.0051$  	      & $0.0242\pm0.0099$  \\
      1.3335 \B& $0.0053\pm0.0047$  	      & $0.0054\pm0.0049$  	      & $0.0075\pm0.0050$  	      & $0.0124\pm0.0090$ \\
      \hline\\
    \end{tabular}
    \begin{tabular}{ c c c c c}
      \multicolumn{5}{c}{Blue galaxies\B}\\
      \hline
      \hline
      \centering  $\theta({\rm deg})$ \T \B  &  $M_g - 5\log h < -18.8$ & $M_g - 5\log h < -19.8$  & $M_g - 5\log h < -20.8$ & $M_g - 5\log h < -21.8$  \\ [0.1cm]
      \hline
      0.0013 \T& $0.9171\pm0.0314$  	      & $1.4191\pm0.0867$  	      & $2.4257\pm0.5698$  	      & $-2.0388\pm2.8547$  \\
      0.0024 & $0.5111\pm0.0180$  	      & $0.6902\pm0.0447$  	      & $0.8761\pm0.2250$  	      & $4.0406\pm5.0154$  \\
      0.0042 & $0.2934\pm0.0088$  	      & $0.4165\pm0.0244$  	      & $0.8823\pm0.1442$  	      & $1.9021\pm2.6653$  \\
      0.0075 & $0.1813\pm0.0068$  	      & $0.2587\pm0.0172$  	      & $0.4201\pm0.0649$  	      & $-0.0745\pm0.8562$  \\
      0.0133 & $0.1165\pm0.0047$  	      & $0.1636\pm0.0106$  	      & $0.2602\pm0.0462$  	      & $-0.3840\pm0.4747$  \\
      0.0237 & $0.0863\pm0.0027$  	      & $0.1086\pm0.0069$  	      & $0.1781\pm0.0322$  	      & $0.0085\pm0.3271$  \\
      0.0422 & $0.0610\pm0.0026$  	      & $0.0759\pm0.0043$  	      & $0.1388\pm0.0175$  	      & $0.2799\pm0.1861$  \\
      0.0750 & $0.0439\pm0.0025$  	      & $0.0580\pm0.0034$  	      & $0.0942\pm0.0142$  	      & $0.2588\pm0.0975$  \\
      0.1334 & $0.0317\pm0.0020$  	      & $0.0425\pm0.0035$  	      & $0.0851\pm0.0117$  	      & $0.2357\pm0.0651$  \\
      0.2371 & $0.0206\pm0.0016$  	      & $0.0317\pm0.0026$  	      & $0.0593\pm0.0096$  	      & $0.0646\pm0.0473$  \\
      0.4217 & $0.0117\pm0.0014$  	      & $0.0202\pm0.0033$  	      & $0.0396\pm0.0094$  	      & $0.0974\pm0.0322$  \\
      0.7499 & $0.0058\pm0.0015$  	      & $0.0119\pm0.0027$  	      & $0.0289\pm0.0094$  	      & $0.0396\pm0.0289$  \\
      1.3335 \B& $0.0025\pm0.0040$  	      & $0.0071\pm0.0039$  	      & $0.0219\pm0.0104$  	      & $0.0591\pm0.0314$     \\
      \hline\\
    \end{tabular}
  \end{center}
\end{table*}
\begin{table*}
  \caption{Two-point correlation function measurements in the range $0.6 < z < 0.8$.}
  \label{tab:wtheta06}
  \begin{center}
    \begin{tabular}{ c c c c c}
      \multicolumn{5}{c}{All galaxies\B}\\
      \hline
      \hline
      \centering  $\theta({\rm deg})$ \T \B  & $M_g - 5\log h < -19.8$  & $M_g - 5\log h < -20.8$ & $M_g - 5\log h < -21.8$  & $M_g - 5\log h < -22.8$  \\ [0.1cm]
      \hline
      0.0013 \T& $1.5217\pm0.0356$  	      & $3.0868\pm0.1218$  	      & $8.0221\pm1.5873$ &     N/A\\
      0.0024 & $0.8792\pm0.0158$  	      & $1.5273\pm0.0550$  	      & $3.5010\pm0.4573$  &	    N/A\\
      0.0042 & $0.5453\pm0.0097$  	      & $0.9487\pm0.0349$     & $1.7392\pm0.2452$        &  N/A\\
      0.0075 & $0.3274\pm0.0074$  	      & $0.5640\pm0.0190$  	      & $0.9377\pm0.1134$ 	   & $4.8069\pm4.4791$  \\
      0.0133 & $0.1943\pm0.0049$  	      & $0.3223\pm0.0113$  	      & $0.6846\pm0.0694$      & $1.0014\pm1.4080$ \\ 
      0.0237 & $0.1227\pm0.0039$  	      & $0.1870\pm0.0076$  	      & $0.3379\pm0.0336$        & $1.4934\pm0.8316$ \\ 
      0.0422 & $0.0789\pm0.0036$  	      & $0.1129\pm0.0070$  	      & $0.1957\pm0.0269$         & $0.7017\pm0.3584$  \\
      0.0750 & $0.0537\pm0.0029$  	      & $0.0772\pm0.0058$  	      & $0.1499\pm0.0178$      & $0.8865\pm0.2413$  \\
      0.1334 & $0.0375\pm0.0028$  	      & $0.0543\pm0.0053$  	      & $0.0992\pm0.0151$     & $0.5009\pm0.1653$ \\
      0.2371 & $0.0252\pm0.0026$  	      & $0.0372\pm0.0047$  	      & $0.0675\pm0.0129$       & $0.2492\pm0.0983$  \\
      0.4217 & $0.0152\pm0.0029$  	      & $0.0236\pm0.0048$  	      & $0.0478\pm0.0130$   & $0.2689\pm0.0748$  \\
      0.7499 & $0.0070\pm0.0027$  	      & $0.0128\pm0.0043$  	      & $0.0251\pm0.0098$  & $0.1509\pm0.0683$ \\
      1.3335 \B& $0.0037\pm0.0046$  	      & $0.0053\pm0.0052$  	      & $0.0150\pm0.0078$   & $0.1576\pm0.0655$  \\
      \hline\\
    \end{tabular}
   \begin{tabular}{ c c c c c}
      \multicolumn{5}{c}{Red galaxies\B}\\
      \hline
      \hline
      \centering  $\theta({\rm deg})$ \T \B  & $M_g - 5\log h < -19.8$  & $M_g - 5\log h < -20.8$ & $M_g - 5\log h < -21.8$  & $M_g - 5\log h < -22.8$  \\ [0.1cm]
      \hline
      0.0013 \T & $3.6793\pm0.1127$  	      & $5.4089\pm0.3033$  	      & $15.8452\pm4.5991$  	 &     N/A\\
      0.0024 & $2.1018\pm0.0547$  	      & $2.9129\pm0.1258$  	      & $6.9176\pm1.0528$  	   & N/A\\
      0.0042 & $1.2737\pm0.0330$  	      & $1.7428\pm0.0633$  	      & $3.6702\pm0.4636$  	   & N/A\\
      0.0075 & $0.7465\pm0.0223$  	      & $0.9808\pm0.0354$  	      & $1.4866\pm0.2029$  	    & $7.6277\pm8.1243$\\  
      0.0133 & $0.4112\pm0.0125$  	      & $0.5170\pm0.0179$  	      & $1.1126\pm0.1101$  	     & $1.7428\pm3.0541$ \\ 
      0.0237 & $0.2301\pm0.0087$  	      & $0.2942\pm0.0116$  	      & $0.4482\pm0.0544$  	      & $1.9737\pm1.5615$  \\
      0.0422 & $0.1334\pm0.0057$  	      & $0.1669\pm0.0086$  	      & $0.2866\pm0.0401$  	       & $-0.1086\pm0.5148$ \\
      0.0750 & $0.0861\pm0.0045$  	      & $0.1104\pm0.0071$  	      & $0.1700\pm0.0267$  	       & $0.5358\pm0.4698$  \\
      0.1334 & $0.0590\pm0.0042$  	      & $0.0755\pm0.0066$  	      & $0.1255\pm0.0243$  	  & $0.5715\pm0.2197$        \\
      0.2371 & $0.0407\pm0.0039$  	      & $0.0548\pm0.0061$  	      & $0.0871\pm0.0164$  	      & $0.0901\pm0.1194$   \\
      0.4217 & $0.0244\pm0.0040$  	      & $0.0332\pm0.0059$  	      & $0.0552\pm0.0146$  	  & $0.2193\pm0.1365$ \\
      0.7499 & $0.0120\pm0.0042$  	      & $0.0175\pm0.0055$  	      & $0.0260\pm0.0101$  & $0.1040\pm0.0817$ 	\\
      1.3335 \B& $0.0083\pm0.0061$  	      & $0.0096\pm0.0060$  	      & $0.0182\pm0.0097$  & $0.0749\pm0.0386$  \\
      \hline\\
    \end{tabular}
    \begin{tabular}{ c c c c c}
      \multicolumn{5}{c}{Blue galaxies\B}\\
      \hline
      \hline
      \centering  $\theta({\rm deg})$ \T \B  & $M_g - 5\log h < -19.8$  & $M_g - 5\log h < -20.8$ & $M_g - 5\log h < -21.8$  & $M_g - 5\log h < -22.8$  \\ [0.1cm]
      \hline
      0.0013 \T& $0.8487\pm0.0496$  	      & $1.4324\pm0.2268$  	      & $4.1554\pm3.3984$  	   &     N/A\\
      0.0024 & $0.4539\pm0.0264$  	      & $0.4523\pm0.1133$  	      & $0.8359\pm0.9450$  	    &    N/A\\
      0.0042 & $0.2618\pm0.0144$  	      & $0.3612\pm0.0618$  	      & $-0.2525\pm0.5016$  	  &      N/A\\
      0.0075 & $0.1553\pm0.0083$  	      & $0.2681\pm0.0291$  	      & $0.0214\pm0.2699$  	  &      N/A\\
      0.0133 & $0.1038\pm0.0055$  	      & $0.1476\pm0.0189$  	      & $0.3495\pm0.1536$  	      & $-1.4572\pm0.5388$  \\
      0.0237 & $0.0771\pm0.0038$  	      & $0.1075\pm0.0114$  	      & $0.1912\pm0.0780$  	      & $-1.1570\pm0.3078$  \\
      0.0422 & $0.0563\pm0.0031$  	      & $0.0834\pm0.0077$  	      & $0.1621\pm0.0520$  	      & $1.9834\pm1.5364$  \\
      0.0750 & $0.0398\pm0.0024$  	      & $0.0624\pm0.0065$  	      & $0.1584\pm0.0294$  	& $1.5887\pm0.7844$  \\
      0.1334 & $0.0279\pm0.0023$  	      & $0.0431\pm0.0045$  	      & $0.1002\pm0.0178$  	& $0.5221\pm0.4101$  \\
      0.2371 & $0.0194\pm0.0022$  	      & $0.0322\pm0.0042$  	      & $0.0503\pm0.0160$  	& $0.4081\pm0.1886$  \\
      0.4217 & $0.0110\pm0.0020$  	      & $0.0187\pm0.0037$  	      & $0.0539\pm0.0144$  	& $0.4225\pm0.1605$  \\
      0.7499 & $0.0047\pm0.0015$  	      & $0.0094\pm0.0032$  	      & $0.0298\pm0.0122$  &	  N/A\\
      1.3335 \B& $0.0010\pm0.0032$  	      & $0.0013\pm0.0034$  	      & $0.0136\pm0.0078$   & N/A\\
      \hline
    \end{tabular}
  \end{center}
\end{table*}
\begin{table*}
  \caption{Two-point correlation function measurements in the range $0.8 < z < 1.0$.}
  \label{tab:wtheta08}
  \begin{center}
    \begin{tabular}{ c c c c}
      \multicolumn{4}{c}{All galaxies\B}\\
      \hline
      \hline
      \centering  $\theta({\rm deg})$ \T \B  & $M_g - 5\log h < -20.8$ & $M_g - 5\log h < -21.8$  & $M_g - 5\log h < -22.8$  \\ [0.1cm]
      \hline
      0.0013\T & $1.9297\pm0.0672$  	      & $6.5369\pm0.8511$  	      & $61.7234\pm41.9290$  \\
      0.0024 & $1.1410\pm0.0340$  	      & $2.2368\pm0.2615$  	      & $28.1877\pm19.7251$  \\ 
      0.0042 & $0.6653\pm0.0252$  	      & $1.5711\pm0.1320$  	      & $22.8965\pm9.3422$   \\
      0.0075 & $0.4075\pm0.0142$  	      & $0.8834\pm0.0785$  	      & $7.9313\pm3.7334$   \\
      0.0133 & $0.2298\pm0.0084$  	      & $0.4598\pm0.0453$  	      & $1.5075\pm1.4777$   \\
      0.0237 & $0.1446\pm0.0068$  	      & $0.2568\pm0.0267$  	      & $0.5253\pm0.3927$   \\
      0.0422 & $0.1004\pm0.0052$  	      & $0.1774\pm0.0162$  	      & $0.4819\pm0.1835$   \\
      0.0750 & $0.0715\pm0.0043$  	      & $0.1192\pm0.0108$  	      & $0.5160\pm0.1432$   \\
      0.1334 & $0.0525\pm0.0038$  	      & $0.0933\pm0.0076$  	      & $0.2072\pm0.0657$   \\
      0.2371 & $0.0355\pm0.0041$  	      & $0.0574\pm0.0067$  	      & $0.1539\pm0.0430$   \\
      0.4217 & $0.0231\pm0.0041$  	      & $0.0400\pm0.0066$  	      & $0.1461\pm0.0376$   \\
      0.7499 & $0.0156\pm0.0039$  	      & $0.0269\pm0.0066$  	      & $0.1218\pm0.0368$   \\
      1.3335 \B& $0.0112\pm0.0056$  	      & $0.0195\pm0.0075$  	      & $0.1075\pm0.0371$   \\
      \hline\\
    \end{tabular}
 \begin{tabular}{ c c c c}
      \multicolumn{4}{c}{Red galaxies\B}\\
      \hline
      \hline
      \centering  $\theta({\rm deg})$ \T \B  & $M_g - 5\log h < -20.8$ & $M_g - 5\log h < -21.8$  & $M_g - 5\log h < -22.8$  \\ [0.1cm]
      \hline
      0.0013 \T& $4.0563\pm0.1830$  	      & $8.0337\pm1.1816$  	      & $62.4523\pm72.7031$  \\ 
      0.0024 & $2.1944\pm0.0796$  	      & $3.5455\pm0.4726$  	      & $62.2007\pm48.4134$   \\
      0.0042 & $1.2422\pm0.0465$  	      & $2.1323\pm0.2013$  	      & $39.5324\pm16.5788$   \\
      0.0075 & $0.6819\pm0.0246$  	      & $1.1544\pm0.1257$  	      & $11.4103\pm6.7419$   \\
      0.0133 & $0.3588\pm0.0178$  	      & $0.6279\pm0.0633$  	      & $3.2474\pm3.0772$   \\
      0.0237 & $0.2091\pm0.0127$  	      & $0.3088\pm0.0311$  	      & $0.4151\pm0.5025$   \\
      0.0422 & $0.1384\pm0.0083$  	      & $0.2302\pm0.0206$  	      & $0.3642\pm0.2609$  \\ 
      0.0750 & $0.0970\pm0.0058$  	      & $0.1410\pm0.0155$  	      & $0.5988\pm0.1924$ \\
      0.1334 & $0.0678\pm0.0049$  	      & $0.1041\pm0.0101$  	      & $0.1791\pm0.0805$   \\
      0.2371 & $0.0460\pm0.0057$  	      & $0.0590\pm0.0087$  	      & $0.1322\pm0.0458$   \\
      0.4217 & $0.0295\pm0.0057$  	      & $0.0424\pm0.0068$  	      & $0.0712\pm0.0370$   \\
      0.7499 & $0.0205\pm0.0057$  	      & $0.0267\pm0.0074$  	      & $0.1058\pm0.0316$   \\
      1.3335 \B& $0.0157\pm0.0072$  	      & $0.0197\pm0.0080$  	      & $0.0573\pm0.0251$   \\
      \hline\\
    \end{tabular}
    \begin{tabular}{ c c c c}
      \multicolumn{4}{c}{Blue galaxies\B}\\
      \hline
      \hline
      \centering  $\theta({\rm deg})$ \T \B  & $M_g - 5\log h < -20.8$ & $M_g - 5\log h < -21.8$  & $M_g - 5\log h < -22.8$  \\ [0.1cm]
      \hline
      0.0013 \T & $1.0353\pm0.1344$  	      & $0.8847\pm1.3392$  	      & $320.2051\pm433.1487$   \\
      0.0024 & $0.5535\pm0.0699$  	      & $-0.0151\pm0.7771$  	      & $123.6391\pm153.4200$   \\
      0.0042 & $0.2876\pm0.0413$  	      & $0.6398\pm0.3317$  	      & $21.6725\pm22.7273$   \\
      0.0075 & $0.2044\pm0.0218$  	      & $0.5090\pm0.1797$  	      & $5.3233\pm5.7498$   \\
      0.0133 & $0.1323\pm0.0117$  	      & $0.3050\pm0.1021$  	      & $1.7491\pm2.0265$   \\
      0.0237 & $0.0975\pm0.0086$  	      & $0.2763\pm0.0572$  	      & $2.4051\pm1.5689$   \\
      0.0422 & $0.0841\pm0.0057$  	      & $0.1246\pm0.0291$  	      & $1.5988\pm0.7454$   \\
      0.0750 & $0.0543\pm0.0046$  	      & $0.0981\pm0.0227$  	      & $0.5858\pm0.3418$   \\
      0.1334 & $0.0430\pm0.0046$  	      & $0.0743\pm0.0126$  	      & $0.2787\pm0.1904$   \\
      0.2371 & $0.0304\pm0.0043$  	      & $0.0546\pm0.0106$  	      & $0.3135\pm0.1561$   \\
      0.4217 & $0.0209\pm0.0041$  	      & $0.0355\pm0.0068$  	      & $0.2642\pm0.0839$   \\
      0.7499 & $0.0135\pm0.0036$  	      & $0.0262\pm0.0077$  	& N/A \\
      1.3335 \B& $0.0086\pm0.0044$  	      & $0.0182\pm0.0071$  	& N/A   \\
      \hline
    \end{tabular}
  \end{center}
\end{table*}
\begin{table*}
  \caption{Two-point correlation function measurements in the range $1.0 < z < 1.2$.}
  \label{tab:wtheta10}
  \begin{center}
    \begin{tabular}{ c c c c }
      \multicolumn{4}{c}{$M_g - 5\log h < -21.8$\B}\\
      \hline
      \hline
      \centering  $\theta({\rm deg})$ \T \B  & All galaxies & Red
      galaxies & Blue galaxies\\ [0.1cm]
      \hline
      0.0013 \T& $3.7142\pm0.2962$  & $6.3867\pm0.7254$  & $2.2515\pm0.4971$   \\
      0.0024 & $1.6094\pm0.1283$  & $2.9118\pm0.3126$ & $0.3135\pm0.3126$   \\
      0.0042 & $0.8737\pm0.0657$ & $1.7832\pm0.1386$ & $0.2647\pm0.1136$   \\
      0.0075 & $0.6176\pm0.0433$ & $1.1093\pm0.0762$& $0.2090\pm0.0738$   \\
      0.0133 & $0.2992\pm0.0273$ & $0.5005\pm0.0425$ & $0.1765\pm0.0437$   \\
      0.0237 & $0.1926\pm0.0158$ & $0.3243\pm0.0254$  & $0.1098\pm0.0302$   \\
      0.0422 & $0.1403\pm0.0130$ & $0.2180\pm0.0208$   & $0.0851\pm0.0162$   \\
      0.0750 & $0.1003\pm0.0095$ & $0.1545\pm0.0171$ & $0.0516\pm0.0137$   \\
      0.1334 & $0.0727\pm0.0074$  & $0.1027\pm0.0162$ & $0.0515\pm0.0089$   \\
      0.2371 & $0.0520\pm0.0072$ & $0.0742\pm0.0135$ & $0.0401\pm0.0081$   \\
      0.4217 & $0.0297\pm0.0056$  & $0.0428\pm0.0101$  & $0.0213\pm0.0068$   \\
      0.7499 & $0.0137\pm0.0057$ & $0.0190\pm0.0064$ & $0.0110\pm0.0068$   \\
      1.3335 \B& $0.0059\pm0.0050$ & $0.0076\pm0.0055$ & $0.0058\pm0.0054$   \\
      \hline
    \end{tabular}
  \end{center}
\end{table*}
\begin{figure*}
  \begin{center}
    \includegraphics[width=\textwidth]{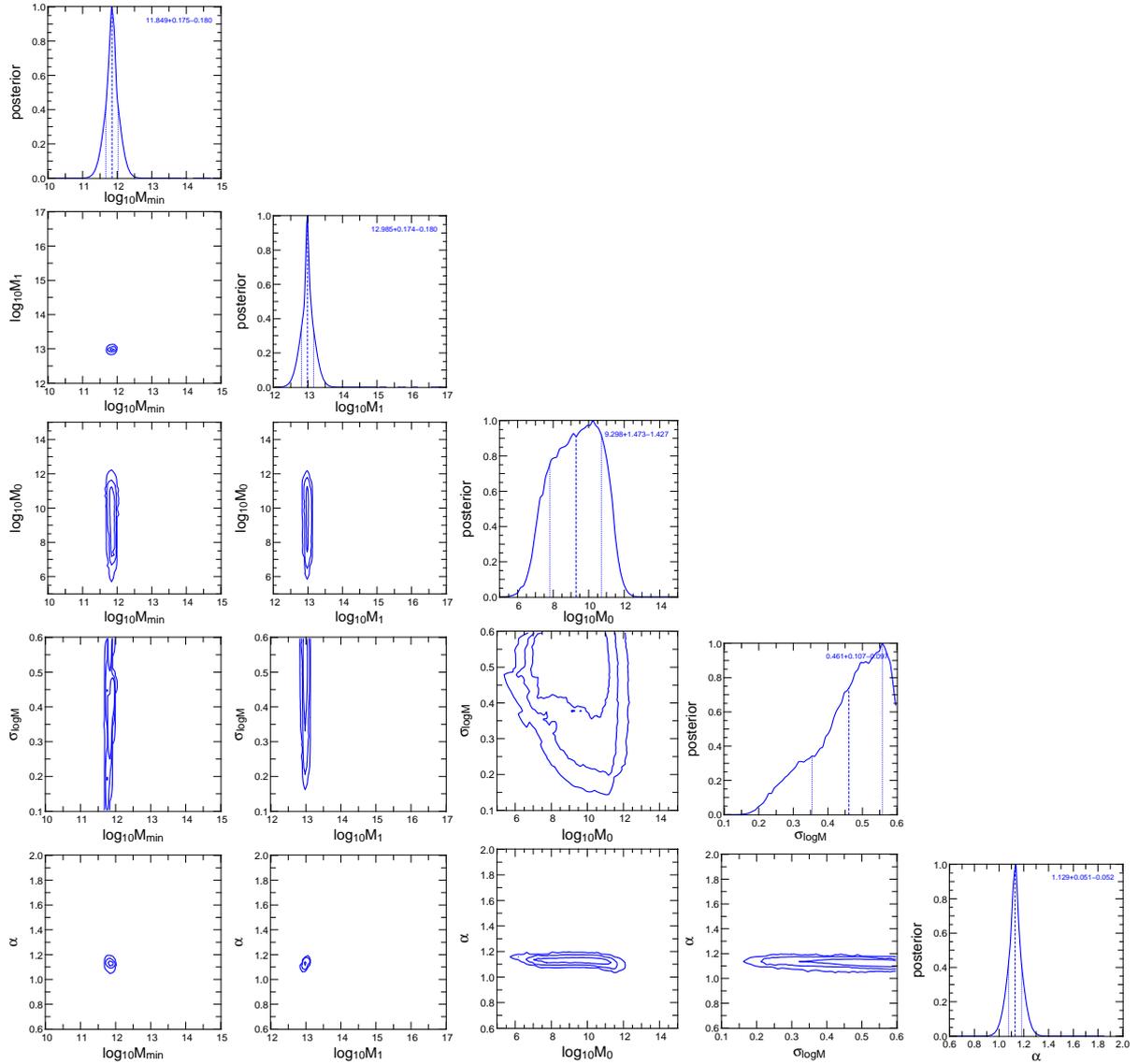}
  \end{center}
  \caption{1D (diagonal) and 2D likelihood distributions of best-fitting HOD parameters for the full sample,
    in the range $0.4 < z < 0.6$, and $M_g -5 \log h < -19.8$.}
  \label{fig:contours_HOD}
\end{figure*}
\begin{table*}
  \caption{Best-fitting parameters (from Eq.~\ref{eq:zehavi}) 
    of $M_{\rm min}$ and $M_1$ as function of
    luminosity threshold corrected for passive redshift evolution to
    approximate stellar mass selected samples. Results are given for
    all and red samples, as function of redshift bins. Halo masses are given in $h^{-1} M_{\sun}$}.
  \label{tab:zehavi}
  \begin{center}
    \begin{tabular*}{\textwidth}{@{\extracolsep{\fill}}   p{4cm}   c c c c c c}
      \multicolumn{7}{c}{All galaxies \B}\\
      \hline
      \hline
      &  \multicolumn{3}{c}{$M_{\rm min}$} \T& \multicolumn{3}{c}{$M_{1}$} \\
      \centering Redshift \B& $A$   &   $\log(M_{\rm t})$ &  $\alpha_M$ & $A$   &
      $\log(M_{\rm t})$ &  $\alpha_M$ \\ 
      \hline  
      $  0.2< z <  0.4$\dotfill \T& $ 0.203 \pm  0.005$ & $11.354 \pm  0.012$ & $ 0.503 \pm  0.010$ & $ 0.274 \pm  0.017$ & 
      $12.756 \pm  0.021$ & $ 0.439 \pm  0.035$ \\ 
      $  0.4< z <  0.6$\dotfill & $ 0.301 \pm  0.195$ & $11.570 \pm  0.224$ & $ 0.404 \pm  0.149$ & $ 0.269 \pm  0.181$ & 
      $12.715 \pm  0.211$ & $ 0.399 \pm  0.191$ \\ 
      $  0.6< z <  0.8$\dotfill & $ 0.377 \pm  0.044$ & $11.717 \pm  0.063$ & $ 0.352 \pm  0.039$ & $ 0.514 \pm  0.002$ & 
      $12.898 \pm  0.002$ & $ 0.233 \pm  0.002$ \\ 
      $  0.8< z <  1.2$\dotfill \B& $ 0.345 \pm  0.006$ & $11.731 \pm  0.013$ & $ 0.383 \pm  0.006$ & $ 0.334 \pm  0.002$ & 
      $12.759 \pm  0.005$ & $ 0.335 \pm  0.002$ \\ 
      \hline\\
      \multicolumn{7}{c}{Red galaxies \B}\\
      \hline
      \hline
      &  \multicolumn{3}{c}{$M_{\rm min}$}\T & \multicolumn{3}{c}{$M_{1}$} \\
      \centering Redshift \B& $A$   &   $\log(M_{\rm t})$ &  $\alpha_M$ & $A$   &
      $\log(M_{\rm t})$ &  $\alpha_M$ \\
      \hline  
      $  0.2< z <  0.4$\dotfill\T & $ 0.706 \pm  0.341$ & $12.154 \pm  0.129$ & $ 0.307 \pm  0.161$ & $ 0.288 \pm  0.009$ & 
      $12.930 \pm  0.011$ & $ 0.477 \pm  0.020$ \\ 
      $  0.4< z <  0.6$\dotfill & $ 0.776 \pm  0.531$ & $12.209 \pm  0.198$ & $ 0.267 \pm  0.241$ & $ 0.552 \pm  0.007$ & 
      $13.141 \pm  0.004$ & $ 0.268 \pm  0.008$ \\ 
      $  0.6< z <  0.8$\dotfill & $ 0.761 \pm  0.553$ & $12.270 \pm  0.233$ & $ 0.302 \pm  0.176$ & $ 1.033 \pm  0.114$ & 
      $13.435 \pm  0.043$ & $ 0.144 \pm  0.036$ \\ 
      $  0.8< z <  1.2$\dotfill \B& $ 0.405 \pm  0.429$ & $12.037 \pm  0.502$ & $ 0.433 \pm  0.187$ & $ 0.839 \pm  0.114$ & 
      $13.426 \pm  0.083$ & $ 0.188 \pm  0.054$ \\ 
      \hline
    \end{tabular*}
  \end{center}
\end{table*}

\begin{table*}
  \caption{Best-fitting parameters (from Eqs.~\ref{eq:biasmodel} and \ref{eq:mhalomodel})  
    of $b_g$ and $\log \langle M_{\rm halo} \rangle$ as function of redshift. 
    Results are given for all and red samples.}
  \label{tab:bgmhalo}
  \begin{center}
    \begin{tabular*}{\textwidth}{@{\extracolsep{\fill}}   p{6cm} c c c c}
      \multicolumn{5}{c}{All galaxies\B}\\
      \hline
      \hline
      &  \multicolumn{2}{c}{Galaxy bias $b_{\rm g}$} \T& \multicolumn{2}{c}{Mean halo mass $\log \langle M_{\rm halo} \rangle$} \\
      \centering Redshift \B& $a_{\rm bias}$   &   $b_{\rm bias}$ &  $a_{\rm h}$ & $b_{\rm h}$ \\
      \hline
      $0.2 < z < 0.4$ \dotfill\T& $ 1.071 \pm  0.006$ & $ 0.211 \pm  0.011$ & $13.106 \pm  0.002$ & $ 0.111 \pm  0.002$\\ 
      $0.4 < z < 0.6$ \dotfill& $ 1.166 \pm  0.002$ & $ 0.288 \pm  0.003$ & $13.002 \pm  0.009$ & $ 0.185 \pm  0.009$\\ 
      $0.6 < z < 0.8$ \dotfill& $ 1.250 \pm  0.018$ & $ 0.327 \pm  0.024$ & $12.846 \pm  0.011$ & $ 0.202 \pm  0.010$\\ 
      $0.8 < z < 1.2$ \dotfill\B& $ 1.430 \pm  0.047$ & $ 0.391 \pm  0.030$ & $12.803 \pm  0.001$ & $ 0.174 \pm  0.001$\\ 
      \hline\\
      \multicolumn{5}{c}{Red galaxies\B}\\
      \hline
      \hline
      &  \multicolumn{2}{c}{Galaxy bias $b_{\rm g}$} \T& \multicolumn{2}{c}{Mean halo mass $\log \langle M_{\rm halo} \rangle$} \\
      \centering  Redshift \B& $a_{\rm bias}$   &   $b_{\rm bias}$ &  $a_{\rm h}$ & $b_{\rm h}$ \\
      \hline
      $0.2 < z < 0.4$ \dotfill\T& $ 1.331 \pm  0.011$ & $ 0.162 \pm  0.017$ & $13.368 \pm  0.001$ & $ 0.073 \pm  0.001$\\ 
      $0.4 < z < 0.6$ \dotfill& $ 1.421 \pm  0.013$ & $ 0.206 \pm  0.014$ & $13.246 \pm  0.002$ & $ 0.092 \pm  0.002$\\ 
      $0.6 < z < 0.8$ \dotfill& $ 1.572 \pm  0.016$ & $ 0.243 \pm  0.021$ & $13.090 \pm  0.011$ & $ 0.129 \pm  0.009$\\ 
      $0.8 < z < 1.2$ \dotfill\B& $ 1.679 \pm  0.049$ & $ 0.375 \pm  0.027$ & $12.971 \pm  0.003$ & $ 0.147 \pm  0.002$\\ 
      \hline
    \end{tabular*}
  \end{center}
\end{table*}



 \end{document}